\documentstyle[aps,epsfig]{revtex}

\begin{document}

\preprint{SUNYSB-NTG-97-24}

\draft

\title{Kaons in dense matter, kaon production in heavy-ion 
collisions, and kaon condensation in neutron stars}

\bigskip
\author{G.Q. Li, C.-H. Lee, and G.E. Brown}
\address{Department of Physics, State University of New
York at Stony Brook, Stony Brook, NY 11794, USA}

\maketitle
  
\begin{abstract}
The recent past witnesses the growing interdependence
between the physics of hadrons, the physics of relativistic
heavy-ion collisions, and the physics of compact objects in
astrophysics. A notable example is the kaon which plays special
roles in all the three fields. In this paper, we
first review the various theoretical investigations
of kaon properties in nuclear medium, focusing on
possible uncertainties in each model. We then present
a detailed transport model study of kaon production in
heavy-ion collisions at SIS energies. We shall discuss 
especially the elementary kaon and antikaon production
cross sections in hadron-hadron interactions, that
represent one of the most serious uncertainties in
the transport model study of particle production in 
heavy-ion collisions. The main purpose of such a study 
is to constrain kaon in-medium properties from the heavy-ion 
data. This can provide useful guidances for the development 
of theoretical models of the kaon in medium. In the last 
part of the paper, we apply the kaon in-medium properties 
extracted from heavy-ion data to the study of neutron star 
properties. Based on a conventional equation of state of 
nuclear matter that can be considered as one of the best 
constrained by available experimental data on finite nuclei, 
we find that the maximum mass of neutron stars is about 
2$M_\odot$, which is reduced to about 1.5$M_\odot$ once kaon 
condensation as constrained by heavy-ion data is introduced. 

pacs: 25.75.Dw, 97.60.Jd, 26.60.+c, 24.10.Lx


\end{abstract}

\section{INTRODUCTION}

There is currently growing interplay between physics of
hadrons (especially the properties of hadrons in 
dense matter which might reflect spontaneous
chiral symmetry breaking and its restoration), the physics
of relativistic heavy-ion collisions (from which one might 
extract hadron properties in dense matter), and the
physics of compact objects in astrophysics (which needs
as inputs the information gained from the first two fields).
A notable example is the kaon ($K$ and ${\bar K}$), which, 
being a Goldstone boson with strangeness, 
plays a special role in all the three fields mentioned.

Ever since the pioneering work of Kaplan and Nelson \cite{kap86,kap87}
on the possibility of kaon condensation in nuclear matter,
a huge amount of theoretical effort has been devoted to
the study of kaon properties in dense matter, based mostly on the 
SU(3) chiral perturbation theory 
\cite{brown87,wise91,muto92,yabu93,lee94,lutz94,sch94,lee95,kai95,kai96,waas96,lee96,kai97,waas97}. 
Kaons, as Goldstone bosons with strangeness, play a special 
role in the development of hadron models. The mass of the strange 
quark is about 150 MeV, which is, on the one hand, considerably 
larger than the mass of light (up and down) quarks 
($\sim 5$ MeV), but on the other hand, much smaller than that 
of the charmed quark ($\sim 1.5$ GeV). In the limit of vanishing quark 
mass the chiral symmetry is good, and systematic studies can 
be carried out using chiral perturbation theory for hadrons 
made of light quarks. Since its mass is not negligibly small as
compared with the typical QCD chiral symmetry breaking scale
$\Lambda _\chi\sim 1 $ GeV, the expansion in terms of
the strange quark mass is much more delicate. In spite of this,
chiral perturbation calculations have been extensively and 
quite successfully carried out in the recent past for the 
study of kaon-nucleon ($KN$) and antikaon-nucleon 
(${\bar K}N$) scattering.

Part of the reason for this success, at least in free
space, can be attributed to the availability of a large body
of experimental data which can be used to constrain the
various parameters in the chiral Lagrangian.
In studying kaon properties in nuclear matter, a new scale,
namely the nuclear Fermi momentum $k_F$, is introduced.
This renders the chiral perturbation calculation in matter
much more subtle than that in free space. 
It is very important to have experimental data available so
that the predictions of chiral perturbation calculations 
can be tested. Kaonic atom data provide information on
this, but is restricted to very low densities \cite{gal94}. 
For the study of kaon condensation, densities much 
higher than that accessible by kaonic atoms are involved.
This can only be obtained by analysing heavy-ion collision 
data on kaon spectra and flow. 

Measurements of kaon spectra and flow have been systematically
carried out in heavy-ion collisions at SIS (1-2 AGeV), 
AGS (10 AGeV),  and SPS (200 AGeV) energies \cite{qm96}. 
By comparing transport model predictions with experimental 
data, one can learn not only the global reaction dynamics,
but more importantly, the kaon properties in dense
matter. Of special interest is kaon production in
heavy-ion collisions at SIS energies, as it has been shown
that particle production at subthreshold energies 
is sensitive to its properties in dense matter
\cite{cass90,koli96,kkl97}. Recently, high quality data 
concerning $K^+$ and $K^-$ production in heavy-ion collisions at 
SIS energies have been published by the KaoS collaboration 
at GSI \cite{kaos}. The KaoS data show that the $K^-$ yield 
at 1.8 AGeV agrees roughly with the $K^+$ yield at 1.0 AGeV. 
This is a nontrivial observation. These beam energies were 
purposely chosen, such that the Q-values (the difference
between the available center-of-mass energy and the 
threshold) for $NN\rightarrow NK\Lambda$ and 
$NN\rightarrow NNK{\bar K}$ are identical and both are 
about $-230$ MeV.  Near their respective production thresholds,
the cross section for the $K^-$ production in proton-proton 
interaction is one to two orders of magnitude smaller than 
that for $K^+$ production \cite{data}. In addition, antikaons
are strongly absorbed in heavy-ion collisions, which should further
reduce the $K^-$ yield. The KaoS results of $K^-/K^+\sim 1$
indicate thus the importance of kaon medium effects which 
act oppositely on the $K^+$ and $K^-$ production in nuclear
medium, and can thus provide useful information on kaon
properties in dense matter.
 
Studies of neutron star properties also have a long history, both
observationally and theoretically. A recent 
compilation by Thorsett quoted by Brown \cite{nsmass}
shows that well-measured neutron star masses are all less
than 1.5$M_\odot$. On the other hand, most of the theoretical 
calculations based on conventional nuclear equation of state 
(EOS) predict a maximum neutron state mass above 2$M_\odot$
\cite{eng96,serot,lee97}. The EOS can, therefore, 
be substantially softened without running into 
contradiction with observation. Various scenarios have been 
proposed that can lead to a soft EOS. Brown and collaborators 
suggested that kaon condensation might happen at a critical 
density of 2-4$\rho_0$ \cite{brown94}. The existence of many 
$K^-$ mesons implies the coexistence of a large number of protons 
that are needed to neutralize the negative charge of $K^-$ mesons
\cite{thor94}. Based on this, Brown and Bethe also proposed the 
interesting possibility of the existence of a large number of 
low mass black holes in the galaxy \cite{bethe}.

In the first part of this paper, Section 2, we review
various theoretical approaches for studying kaon 
properties in nuclear medium. We shall focus on
the limitations and possible uncertainties in each
approach. In the second part, Section 3, we present
a detailed analysis of kaon production in
heavy-ion collisions at SIS energies. We will 
discuss in detail the elementary kaon and antikaon
production cross sections in hadron-hadron interactions,
which are the major sources of uncertainties in the
transport model study of particle production
in heavy-ion collisions. By comparing transport model
predictions with experimental data, we try to extract 
kaon in-medium properties at high densities. This information
is very useful for the development of chiral perturbation
theory for finite density. This also provides guidance
for the study of kaon condensation in neutron stars, which is
the focus of Section 4. We calculate neutron star masses based on
the nuclear equation of state (EOS) of Furnstahl, Tang and Serot 
\cite{fst}. Without kaon condensation, the maximum neutron star mass
is predicted to be about 2.0$M_\odot$, which reduces to about
1.5$M_\odot$ once kaon condensation is introduced. 
The paper ends with summary and outlook in Section 5.

\section{kaon in dense matter: a review}

Ever since the pioneering work of Kaplan and Nelson \cite{kap86}
on the possibility of kaon condensation in dense nuclear
matter, many works have been devoted to the study of kaon
properties in nuclear matter. There are two typical
approaches to this problem. One was initiated by Kaplan and Nelson, 
and is based on the chiral perturbation theory. The other
has been pursued by Schaffer {\it et al.} and Knorren
{\it et al.} based on the extension of the Walecka-type
mean field model from SU(2) to SU(3). In other words,
it follows the traditional meson-exchange idea for
hadron interactions. This latter approach
has also been developed by the J\"ulich group for
the description of kaon-nucleon scattering in free
space \cite{speth90a,speth90b}. In addition,
kaon properties in dense matter have been
studied using the Nambu$-$Jona-Lasinio model \cite{lutz94},
treating the kaon as quark-antiquark excitation,
and phenomenological off-shell meson-nucleon
interaction \cite{yabu93}.  Although quantitatively,
the results from these different models are not
completely identical, qualitatively, a consistent 
picture, namely in nuclear matter the kaon feels a weak
repulsive potential and the antikaon feels a strong
attractive potential, has emerged. In this section we
will discuss mainly the results of the chiral perturbation 
theory and the meson-exchange model.

\subsection{Chiral perturbation theory}

The interactions between pseudoscalar mesons (pion, kaon, and eta meson)
and baryons (nucleon and hyperon) are described by
the SU(3)$_L\times$SU(3)$_R$ nonlinear chiral Lagrangian
which can be written as
\begin{eqnarray}
{\cal L}&=&{1\over 4}f^2{\rm Tr}\partial^\mu\Sigma\partial_\mu\Sigma^\dagger
+{1\over 2}f^2\Lambda[{\rm Tr}M_q(\Sigma-1)+{\rm h.c.}]
+{\rm Tr}{\bar B}(i\gamma^\mu\partial_\mu-m_B)B\nonumber\\
&+&i{\rm Tr}{\bar B}\gamma^\mu[V_\mu, B]
+D{\rm Tr}{\bar B}\gamma^\mu\gamma^5\{A_\mu, B\}
+F{\rm Tr}{\bar B}\gamma^\mu\gamma^5[A_\mu, B]\nonumber\\
&+&a_1{\rm Tr}{\bar B}(\xi M_q\xi+{\rm h.c.})B
+a_2{\rm Tr}{\bar B}B(\xi M_q\xi+{\rm h.c.})\nonumber\\
&+&a_3[{\rm Tr}M_q\Sigma+{\rm h.c.}]{\rm Tr}{\bar B}B.\label{LAG}
\end{eqnarray}
In the above, $B$ is the baryon octet with a degenerate mass $m_B$, and
\begin{equation}
\Sigma=\exp (2i\pi/f),~~{\rm and}~~
\xi=\sqrt\Sigma=\exp(i\pi/f),
\end{equation}
with $\pi$ being the pseudoscalar meson octet. The pseudoscalar meson 
decay constants are equal in the SU(3)$_V$ limit and are denoted by 
$f=f_\pi \simeq 93$ MeV.  The meson vector $V_\mu$ and axial vector 
$A_\mu$ currents are defined as
\begin{equation}
V_\mu={1\over 2}(\xi^\dagger\partial_\mu\xi
   +\xi\partial_\mu\xi^\dagger)~~{\rm and}~~
A_\mu={i\over 2}(\xi^\dagger\partial_\mu\xi-\xi\partial_\mu\xi^\dagger),
\end{equation}
respectively. The current quark mass matrix is given by 
$M_q={\rm diag}\{m_q,m_q,m_s\}$, where we neglect the small difference 
between the up and down quark masses.
 
Expanding $\Sigma$ to order of $1/f^2$ and
keeping explicitly only the kaon field, the first two terms in Eq. 
(\ref{LAG}) can be written as
\begin{equation}
\partial^\mu\bar K\partial_\mu K-\Lambda(m_q+m_s)\bar KK+\cdots,
\end{equation}
where
\begin{equation}
K=\left(\matrix{K^+ \cr
                K^0 \cr}\right)~~
{\rm and} ~~\bar K=(K^- ~~\bar {K^0}),
\end{equation}
and the ellipsis denotes terms containing other mesons.
 
Keeping explicitly only the nucleon and kaon, the third and fourth terms in 
Eq. (\ref{LAG}) become
\begin{equation}
{\bar N}(i\gamma^\mu\partial_\mu-m_B)N
-{3i\over 8f^2}{\bar N}\gamma^0 N 
\bar K \buildrel \leftrightarrow\over \partial_t K+\cdots,
\end{equation}
where
\begin{equation}
N=\left(\matrix{p \cr
                n \cr}\right)~~{\rm and}~~\bar N=(\bar p~~ \bar n),
\end{equation}
and the ellipsis denotes terms involving other baryons and mesons.
 
The last three terms in Eq. (\ref{LAG}) can be similarly worked out, and
the results are
\begin{eqnarray}
{\rm Tr}{\bar B}(\xi M_q\xi+{\rm h.c.})B&=&2m_q{\bar N}N-{{\bar N}N\over 2f^2}
(m_q+m_s){\bar K}K+\cdots,\nonumber\\
{\rm Tr}{\bar B}B(\xi M_q\xi+{\rm h.c.})&=&2m_s{\bar N}N-{{\bar N}N\over f^2}
(m_q+m_s){\bar K}K+\cdots,\nonumber\\
\,[{\rm Tr}M_q\Sigma+{\rm h.c.}]{\rm Tr}{\bar B}B&=&2(2m_q+m_s){\bar N}N
-{2{\bar N}N\over f^2}(m_q+m_s){\bar K}K+\cdots.
\end{eqnarray}
Combining above expressions, one arrives at the following Lagrangian,
\begin{eqnarray}
{\cal L}&=&{\bar N}(i\gamma^\mu\partial_\mu-m_N)N
+\partial^\mu{\bar K}\partial_\mu K
-(m_K^2-{\Sigma_{KN}\over f^2}{\bar N}N){\bar K}K\nonumber\\
&-&{3i\over 8f^2}{\bar N}\gamma^0 N
\bar K \buildrel \leftrightarrow\over \partial_t K
+\cdots.\label{LAG2} ,
\end{eqnarray}
where the kaon mass is given by
\begin{equation}
m_K^2=\Lambda(m_q+m_s),
\end{equation}
and the nucleon mass by
\begin{eqnarray}\label{mn}
m_N=m_B-2[a_1m_q+a_2m_s+a_3(2m_q+m_s)].
\end{eqnarray}
Also, the $KN$ sigma term can be expressed as
\begin{eqnarray}\label{skn}
\Sigma_{KN}&\equiv&{1\over 2}(m_q+m_s)\langle N|{\bar u}u+{\bar s}s|
N\rangle\nonumber\\
&=&-{1\over 2}(m_q+m_s)(a_1+2a_2+4a_3).
\end{eqnarray}

We note that the last term in Eq. (\ref{LAG2}) is the usual
Weinberg-Tomozawa term; it gives rise to a repulsive 
vector potential for $K^+$, and an attractive vector 
potential for $K^-$. The one that goes with the kaon mass
is usually called Kaplan-Nelson term; it provides
scalar attraction for both $K^+$ and $K^-$, and thus reduces
their mass. The amount of scalar attraction is linearly
proportional to the kaon-nucleon sigma term, $\Sigma _{KN}$.
From Eq. (\ref{skn}) we see that $\Sigma _{KN}$ depends
on three coefficients, $a_1m_s$, $a_2m_s$, and $a_3m_s$.
While the first two coefficients are relatively well determined
from the baryon mass splitting \cite{wise91}, the 
dominant term, $a_3m_s$, is only poorly known, mainly because
of the large uncertainties in the nucleon strangeness content.

On the other hand, the kaon-nucleon sigma term can be related to
pion-nucleon sigma term, $\Sigma _{\pi N}$ \cite{br96}. 
The latter is relatively well determined to be about 
45 MeV \cite{leut91},
\begin{eqnarray}
\Sigma _{KN} &= &
{(m_q+m_s)\langle N|{\bar u}u + {\bar s}s |N\rangle\
\over 2m_q \langle N|{\bar u}u + {\bar d}d |N\rangle }
\Sigma _{\pi N}\nonumber\\
&\approx &{1\over 4} (1+{m_s\over m_q})(1+y)\Sigma _{\pi N},
\end{eqnarray}
where the strangeness content of nucleon is defined as
\begin{eqnarray}
y={2\langle N| {\bar s}s |N\rangle \over
\langle N| {\bar u}u + {\bar d} d | N\rangle}.
\end{eqnarray}
From the particle data book \cite{pdata}, we know that
the average mass of the light quark is $m_q\approx 7.5$ MeV.
There is a large uncertainty in the strange quark mass
$m_s$ which ranges from 100 to 300 MeV \cite{pdata}. 
There is also a large uncertainty in the nucleon strangeness 
content, which ranges from 0 to about 0.3 \cite{wise91,liu94}. 
  
In the mean-field approximation and including only the
Weinberg-Tomozawa and the Kaplan-Nelson terms,
the kaon dispersion relation in nuclear matter is then given by
\begin{eqnarray}
\omega^2({\bf p},\rho _B) =m_K^2+{\bf p}^2 -{\Sigma_{KN}\over f^2}
\rho_S +{3\over 4}{\omega\over f^2}\rho_N,\label{DIS}
\end{eqnarray}
where {\bf p} is the three-momentum of the kaon. 
From the dispersion relation, the kaon energy in medium can be 
obtained, i.e.,
\begin{eqnarray}\label{omek}
\omega_K=\left[m_K^2+{\bf k}^2-{\Sigma_{KN}\over f^2}\rho_S
+\left({3\over 8}{\rho_N\over f^2}\right)^2\right]^{1/2}
+{3\over 8}{\rho_N\over f^2},
\end{eqnarray}
and similarly for the antikaon,
\begin{eqnarray}\label{omeak}
\omega_{\bar K}=\left[m_K^2+{\bf k}^2-{\Sigma_{KN}\over f^2}\rho_S
+\left({3\over 8}{\rho_N\over f^2}\right)^2\right]^{1/2}
-{3\over 8}{\rho_N\over f^2}.
\end{eqnarray}
Note that the scalar attraction depends on nucleon scalar
density $\rho_S$, which is model dependent \cite{sch94}.

There are a number of corrections to these simple mean-field results.
Below we discuss some of them.

{\it Range term:} At the same order in chiral perturbation theory 
as the Kaplan-Nelson term is the energy-dependent range term. 
This has been studied in great detail in Refs. \cite{lee95,lee96}.
It has been shown that the range term
reduces the scalar attraction and can be approximately
included, with $\Sigma_{KN}\approx 400$ MeV, 
by multiplying the Kaplan-Nelson term by 
a factor $1-0.37\omega_{K, {\bar K}}^2/m_K^2$ \cite{lee95}. 
Note that in the nuclear medium, the kaon effective mass increases,
while that of the antikaon decreases. Therefore, while the range term
is very important for $K^+$, it is less so for $K^-$.
In other words, the scalar attraction for $K^+$ is different
from that for $K^-$. 

{\it Brown-Rho scaling:} In the nuclear medium, the pion decay constant
$f_\pi$ may decrease \cite{br91}. At finite density, this can be easily
seen from the Gell-Mann$-$Oakes$-$Renner relation.
In Ref. \cite{br96}, Brown and Rho has shown that
the scaling in $f_\pi$ is instructive in linking the 
chiral Lagrangian in free space to Walecka mean-field
model for finite nuclei. The enhancement of low-mass
dileptons in heavy-ion collisions might be considered
as empirical evidence for this scaling \cite{ceres,likobrown95,cass95}.
Including this scaling, the vector attraction for $K^-$ increases
significantly.

{\it Short-range correlations:} At high densities,
a simple mean-field type treatment of kaon-nucleon
and nucleon-nucleon interactions might be inadequate.
The effects of short-range correlations in 
nucleon-nucleon and kaon-nucleon interactions
were studied by Pandharipande, Pethick, and Thorsson
\cite{pand95}, and were found to reduce the scalar attraction
significantly. There will, however, be a tendency for
the effects from scaling in $f_\pi$ and short range correlation 
to cancel each other. 
 
{\it $\Lambda (1405)$ and coupled-channel effects:}
The isospin averaged $K^-N$ scattering length is negative in free space 
\cite{martin81}, implying a repulsive $K^-$ optical potential in the 
simple impulse approximation. However, a systematic analysis of the 
kaonic atom data shows that the $K^-$ optical potential is deeply 
attractive, with a value of about 200$\pm$ 20 MeV at normal nuclear 
matter density \cite{gal94}. Unlike the kaon which interacts with 
nucleons relatively weakly so that the impulse approximation is reasonable, 
the antikaon interacts strongly with nucleons so we do not expect the 
impulse approximation to be reliable. The antikaon-nucleon (${\bar K}N$) 
interaction at low-energy is strongly affected by the $\Lambda (1405)$ 
which is a quasi bound state of an antikaon and a nucleon in the isospin 
$I=0$ channel and can decay into $\Sigma \pi$ channel 
\cite{waas96}. Thus, in principle one needs to carry 
out a coupled-channel calculation for ${\bar K}N$, $\Lambda \pi$ 
and $\Sigma \pi$ including the effects of $\Lambda (1405)$ in both 
free space and in nuclear medium \cite{waas96,toki94,koch94}.
This has been studied in detail by
Weise and collaborators \cite{kai97,waas97}.
In effective chiral Lagrangian, however, one can also 
introduce $\Lambda(1405)$ as a fundamental field, which is
consistent with phenomenological ${\bar K}N$ scattering amplitudes 
and branching ratios \cite{rho96}. 
Because of Pauli blocking effects on the intermediate states, 
the possibility of forming a bound ${\bar K}N$ state 
($\Lambda (1405)$) decreases with increasing 
density, leading to a dissociation of $\Lambda(1405)$ in nuclear medium.  
This induces a transition of the $K^-$ potential from repulsion at very 
low densities to attraction at higher densities \cite{waas96,koch94}. 
It was shown in Refs. \cite{lee95,lee96}, that, because of the diminishing 
role of $\Lambda (1405)$ in dense matter, the prediction for the kaon 
condensation threshold is relatively robust with respect to different 
treatments of $\Lambda (1405)$. 

In view of large uncertainties in $\Sigma _{KN}$ and
difficulties in treating systematically high-order
corrections, we adopt in this paper a more
phenomenological approach. We assume that the effects from 
the scaling in $f_\pi$ and short-range correlations
approximately cancel each other. Furthermore, we
introduce two free parameters, $a_K$ and $a_{\bar K}$, 
which determine the scalar attractions for $K^+$ and
$K^-$. We assume that these are density independent, although
in principle they should be density dependent, since 
the range term, the scaling in $f_\pi$, and the short-range
corrections are all density dependent. The kaon and
antikaon energy in the nuclear medium can then be written as
\begin{eqnarray}
\omega _K=\left[m_K^2+{\bf k}^2-a_K\rho_S
+(b_K \rho_N )^2\right]^{1/2} + b_K \rho_N  ,
\end{eqnarray}
\begin{eqnarray}
\omega _{\bar K}=\left[m_K^2+{\bf k}^2-a_{\bar K}\rho_S
+(b_K \rho_N )^2\right]^{1/2} - b_K \rho_N  ,
\end{eqnarray}
where $b_K=3/(8f_\pi^2)\approx 0.333$ GeVfm$^3$. 
We try to determine $a_K$ and $a_{\bar K}$ from the 
experimental observables in heavy-ion collisions.

Since the kaon-nucleon interaction is relatively weak as
compared to other hadron-nucleon interactions,
impulse approximation is considered to be reasonable
for the kaon potential in nuclear matter, at least at 
low densities. This provides some constraint on $a_K$.
Using $a_K=0.22$ GeV$^2$fm$^3$, we find that at normal
nuclear matter density $\rho_0 =0.16$ fm$^{-3}$,
the $K^+$ feels a repulsive potential of about 20 MeV.
This is in rough agreement with what is expected from
the impulse approximation using the $KN$ scattering length
in free space. Note that as $\rho$ approaches
zero,  
\begin{eqnarray}
a_K \approx {0.63 \Sigma _{KN} \over f_\pi^2}.
\end{eqnarray}
Therefore $a_K=0.22$ GeV$^2$fm$^3$ implies
$\Sigma _{KN} \approx 400 $ MeV. This is exactly
the value determined in Ref. \cite{lee95} by
fitting the $KN$ scattering length. 

Determination of the $a_{\bar K}$ is more delicate, as
impulse approximation does not apply. The $a_{\bar K}$ should
show stronger density dependence than $a_K$, as
the effects of $\Lambda (1450)$ and coupled-channels
are both strongly density dependent. We will try to determine
this value by fitting to heavy-ion data. We find that, 
when using the model of Furnstahl, Tang, and Serot \cite{fst}
for dense matter,  $a_{\bar K}=0.45$ GeV$^2$fm$^3$
provides a good fit to the $K^-$ data in heavy-ion
collisions at SIS energies. Since the $K^-$ production
chiefly proceeds at the high densities
($2\rho_0$-$3\rho_0$), the value of $a_{\bar K}$ determined
here needs not apply to lower densities, e.g., those 
sampled in kaonic atoms.

\begin{figure}
\begin{center}
\epsfig{file=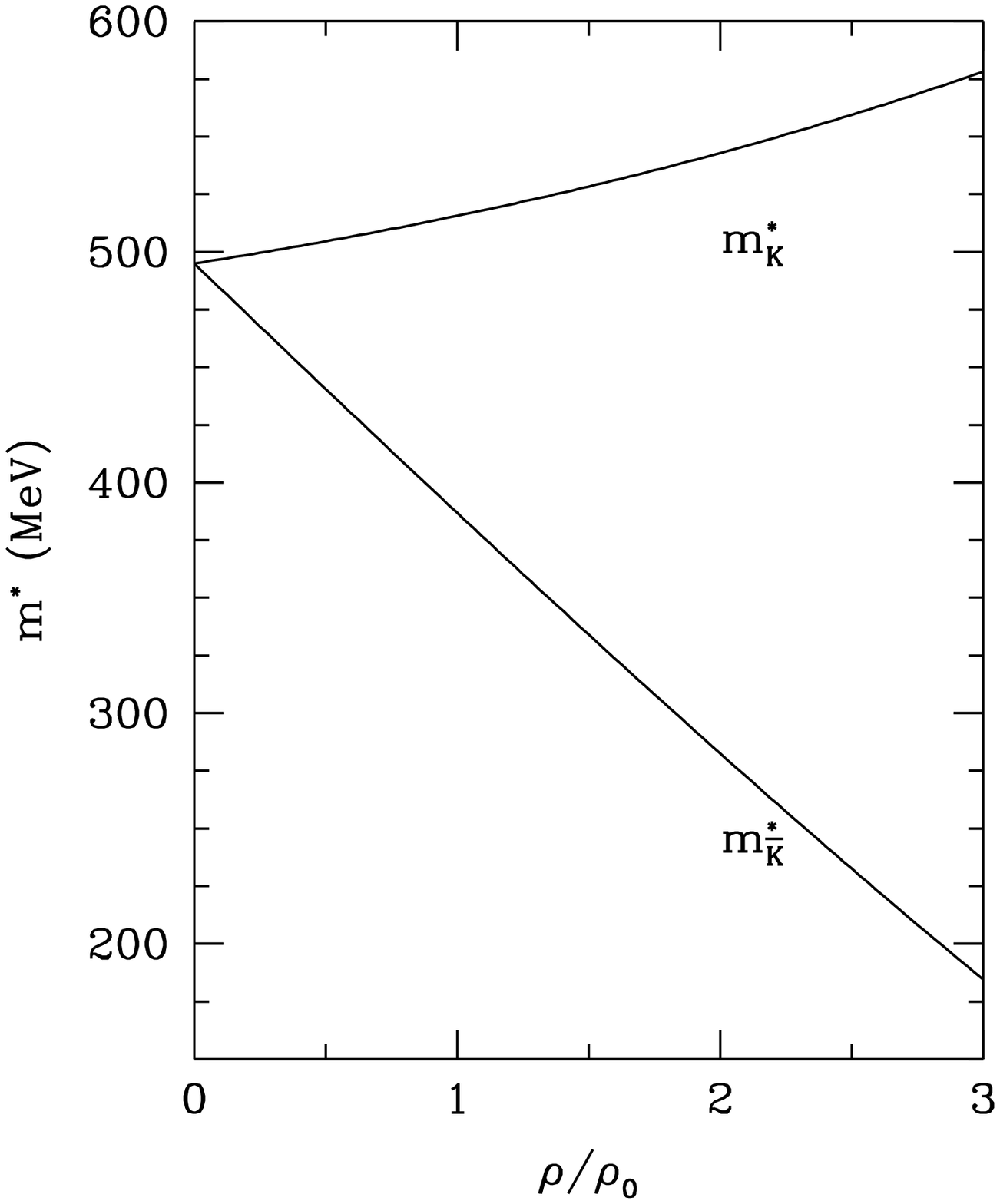,height=7.0in,width=7.0in}
\caption{Effective mass of kaon and antikaon in
nuclear medium. 
\label{kmass}} 
\end{center}
\end{figure}

With these two parameters we show in Fig. \ref{kmass} the effective 
masses of kaon and antikaon defined as their energies at zero momentum. 
It is seen that the kaon mass increases slightly with density, 
resulting from near cancellation of the attractive scalar and 
repulsive vector potential. The mass of the antikaon drops substantially. 
At normal nuclear matter density, the kaon mass increases about
4\%, in rough agreement with the prediction of impulse approximation
based on $KN$ scattering length. The antikaon mass drops by about 22\%,
which is somewhat smaller than what has been inferred from
the kaonic atom data {\cite{gal94}}, namely, an attractive
$K^-$ potential of $200\pm 20$ MeV at $\rho_0$.

\begin{figure}
\begin{center}
\epsfig{file=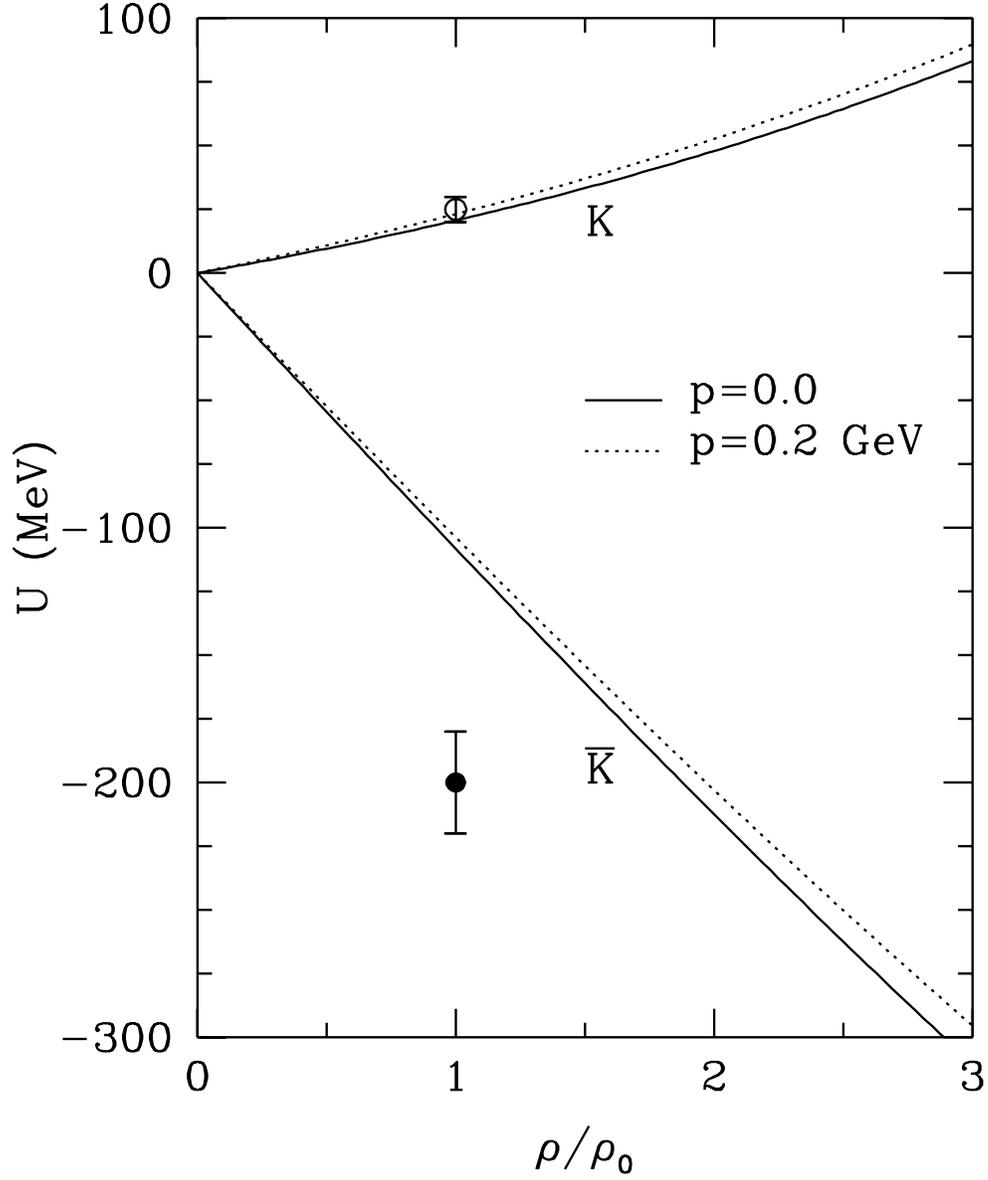,height=7.0in,width=7.0in}
\caption{Kaon and antikaon potential in
nuclear medium at two different momenta. 
The open circle in the
figure is the $K^+$ potential expected from the
impulse approximation using the kaon-nucleon scattering
length in free space \protect\cite{koch95a,dover82}. The
solid circle is the $K^-$ potential extracted
from the kaonic atom data \protect\cite{gal94}.
\label{kpot}} 
\end{center}
\end{figure}

From their in-medium dispersion relations, we can define the 
kaon and antikaon potential as the difference
between its energies in the medium and in free space \cite{shu92},
\begin{eqnarray}
U_{K,{\bar K}} = \omega _{K,{\bar K}} -\omega _0,
\end{eqnarray}
with $\omega _0 = \sqrt {m_K^2+{\bf p}^2}$.
Kaon and antikaon potentials at two different momenta are
shown in Fig. \ref{kpot}. The open circle in the
figure is the $K^+$ potential expected from the
impulse approximation using the kaon-nucleon scattering
length in free space \cite{koch95a,dover82}. The
solid circle is the $K^-$ potential extracted
from the kaonic atom data \cite{gal94}. 

\subsection{Meson-exchange model}

The chiral Lagrangian as given above does not describe properly 
the nuclear matter properties. Therefore in most of 
the studies about kaon condensation in density matter, 
the nuclear matter properties are obtained based on 
some form of Walecka-type mean-field model; e.g., in this
work we use the model of Ref. \cite{fst}, which has imposed
chiral constraints on the mean field. The question then is,
as put forward in Ref. \cite{knor95}, 
whether it is possible to describe the kaon-nucleon
interaction using the idea of meson-exchange.
This has been addressed by Schaffer {\it et al.} \cite{sch94}
for symmetric nuclear matter, and by Knorren {\it et al.}
\cite{knor95} for more general case as encountered in
neutron stars.

In the meson-exchange picture, the scalar and vector interaction 
between kaon and nucleon are mediated by the exchange
of $\sigma$ and $\omega$ meson, respectively. 
Although in chiral Lagrangians one cannot attach mean 
fields to Goldstone bosons, in the sense of Ref. \cite{br96},
one might try to describe the effects of explicit
chiral symmetry breaking by a scalar field $\sigma$.
The Lagrangian is,
\begin{eqnarray}
{\cal L} _K = \partial _\mu {\bar K} \partial _\mu K
-(m_K^2 - g_{\sigma K} m_K \sigma ) K{\bar K} 
+i g_{\omega K} \omega_0
\bar K \buildrel \leftrightarrow\over \partial_t K,
\end{eqnarray}
where $\sigma $ and $\omega _0$ are the scalar and the (time-component
of) vector fields, respectively. $g_{\sigma K}$ and 
$g_{\omega K}$ are the coupling constants between the kaon
and the scalar and the vector fields, respectively.
The energies of kaon and antikaon in nuclear medium are then
given by
\begin{eqnarray}\label{omkm}
\omega_K=\left[m_K^2+{\bf k}^2-g_{\sigma K} m_K \sigma 
+(g_{\omega K} \omega _0)^2\right]^{1/2}
+g_{\omega K} \omega _0,
\end{eqnarray}
\begin{eqnarray}\label{omakm}
\omega_{\bar K}=\left[m_K^2+{\bf k}^2-g_{\sigma K} m_K \sigma 
+(g_{\omega K} \omega _0)^2\right]^{1/2}
-g_{\omega K} \omega _0.
\end{eqnarray}
The results in the meson-exchange picture also have model
dependence. First, they depend on the strength of the
scalar and vector fields, which might be constrained
from the nuclear matter properties. Second, they
depend on the coupling constants $g_{\sigma K}$ and
$g_{\omega K}$. In the simple quark model, 
$g_{\omega K}/g_{\omega N}\approx 1/3$, since there is
only one light quark in kaon. The relation for scalar
couplings is more subtle, since there is no fundamental
scalar meson; it represent correlated s-wave two-pion
exchange. 

By analysing kaon-nucleon scattering in the meson-exchange
model, and by fitting to experimental data, one may be
able to determine these coupling constants. This is the
approach adopted in Refs. \cite{speth90a,speth90b} by
the J\"ulich group. The coupling constants so determined,
however, cannot be used directly in Eqs. (\ref{omkm}) and
(\ref{omakm}), as these expressions are obtained in the
simple mean-field approximation, while in Refs. 
\cite{speth90a,speth90b}, more much complete diagram 
sets and infinite summation of ladder diagrams are
involved. This is analogous to the case of nuclear matter
saturation, namely, one cannot apply directly the coupling
constants in the Bonn model to the Walecka-type mean-field
model for nuclear matter saturation.

Starting from the Bonn model for nucleon-nucleon interaction,
one can carry out Dirac-Brueckner-Hartree-Fock calculation
to study nuclear matter properties. In the same sense,
one can study kaon properties in nuclear matter starting
from the J\"ulich kaon-nucleon (and antikaon-nucleon) interactions,
by solving in-medium scattering equation (the so-called G-matrix).
This kind of approach should provide a consistent way
of study hadron systems containing strangeness degrees of
freedom.  

\section{kaon production in heavy-ion collisions}

One of the most important ingredients in the transport
model study of particle production in heavy-ion
collisions is the elementary particle production
cross sections in hadron-hadron interactions. At SIS
energies, the colliding system consists mainly of
nucleons, delta resonances, and pions. We need thus
kaon and antikaon production cross sections from nucleon-nucleon
($NN$), nucleon-delta ($N\Delta$), delta-delta ($\Delta\Delta$),
pion-nucleon ($\pi N$), and pion-delta ($\pi\Delta$)
interactions. In addition, the antikaon can also be produced
from strangeness-exchange processes such as $\pi Y\rightarrow 
{\bar K} N$. In the next two subsections, we will discuss these
cross sections. Because of the lack of experimental data,
especially near production threshold that are important
for heavy-ion collisions at SIS energies, we have to adopt
the strategy that combines the parameterization of experimental
data with some theoretical investigations and reasonable assumptions
and prescriptions. 
 
\subsection{Elementary kaon production cross sections}

In this subsection we discuss kaon production cross sections
in pion-baryon and baryon-baryon collisions.

\subsubsection{kaon production in pion-baryon interactions}

Two major processes for kaon production in $\pi N$
interaction, namely, $\pi N\rightarrow \Lambda K$ and 
$\pi N\rightarrow \Sigma K$, have actually been 
measured quite extensively in the literature \cite{data}.
Cugnon and Lombard proposed the following parameterizations
for experimental data in some specific channels \cite{cugnon84},
\begin{eqnarray}
\sigma _{\pi^- p\rightarrow \Lambda K^0} & = & \left\{
\begin{array}{lll}
9.89 (\sqrt s -\sqrt {s_0}) \; {\rm mb} & {\rm if} & 
\sqrt {s_0}=m_\Lambda +m_K < \sqrt s < 1.7 \; {\rm GeV}\\
{0.09 \over \sqrt s -1.6} \; {\rm mb}   & {\rm if} &
\sqrt s > 1.7  \; {\rm GeV}\\
\end{array}\right.  ,
\end{eqnarray}
\begin{eqnarray}
\sigma _{\pi^+ p\rightarrow \Sigma^+ K^+} & = & \left\{
\begin{array}{lll}
3.21 (\sqrt s -\sqrt {s_0}) \; {\rm mb} & {\rm if} & 
\sqrt {s_0}=m_\Sigma +m_K < \sqrt s < 1.9 \; {\rm GeV}\\
{0.14 \over \sqrt s -1.7} \; {\rm mb}   & {\rm if} &
\sqrt s > 1.9  \; {\rm GeV}\\
\end{array}\right.  ,
\end{eqnarray}
\begin{eqnarray}
{1\over 2}\left[\sigma _{\pi^-p\rightarrow \Sigma ^-K^+}
+\sigma_{\pi^-p\rightarrow \Sigma^0K^0}\right]
=0.25 \left[1.0-0.75(\sqrt s-1.682)\right] \; {\rm mb}
\; \sqrt s>1.682 \; {\rm GeV}.
\end{eqnarray}
Furthermore, under some assumptions, they obtained the isospin
averaged cross sections
\begin{eqnarray}
\sigma _{\pi N\rightarrow \Lambda K} = 
{1\over 2} \sigma _{\pi^-p\rightarrow \Lambda K^0},
\end{eqnarray}
\begin{eqnarray}
\sigma _{\pi N\rightarrow \Sigma K} = 
{1\over 2} \left[\sigma _{\pi^-p\rightarrow \Sigma^- K^+}+
\sigma_{\pi^-p\rightarrow \Sigma^0K^0}+
\sigma_{\pi^+p\rightarrow \Sigma ^+ K^+}\right].
\end{eqnarray}
These parameterizations have been used by several groups
in the study of kaon production in heavy-ion collisions at 1-2
AGeV region \cite{xiong90}.

\begin{figure}
\begin{center}
\epsfig{file=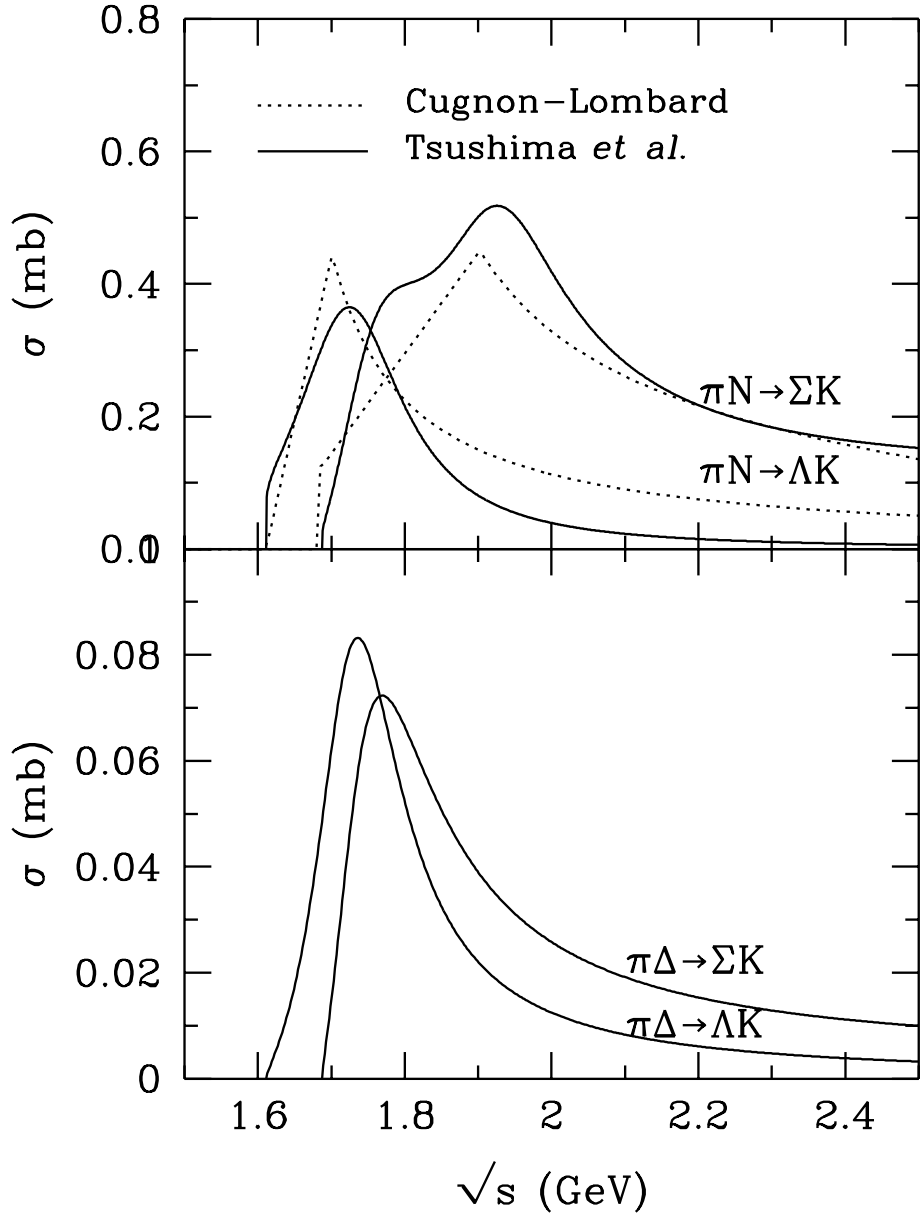,height=7.0in,width=7.0in}
\caption{Isospin-averaged cross sections $\sigma _{\pi N\rightarrow 
YK}$. Solid and dotted lines are from Refs. \protect\cite{fae94}
and \protect\cite{cugnon84}, respectively. 
\label{isopinyk}} 
\end{center}
\end{figure}

Recently, Tsushima {\it et al.} \cite{fae94} studied kaon
production in $\pi N$ and $\pi\Delta$ interactions
in the resonance model. The kaon is produced mainly in the
s-channel processes from the decay of baryon resonances
such as $N(1650)$, $N(1710)$, $N(1720)$, and $\Delta (1920)$.
The $\Delta (1920)$ in their model is treated as an effective
description of the contributions from six $\Delta$-resonances 
between 1.9 and 1.94 GeV that have decay possibilities into
$\Sigma K$ final state. Overall, very good agreement with
experimental data has been achieved. One of the advantages of 
a model calculation is that the isospin average can be done explicitly. 
In addition to kaon production in the $\pi N$ interaction, these
authors have also calculated kaon production cross sections
in $\pi\Delta$ interactions, within the same model.
Generally, these cross sections are substantially smaller
than those in the $\pi N$ interaction, because the branching ratios
of these resonances into $\pi\Delta$ are smaller than those
into $\pi N$. In Fig. \ref{isopinyk} we show the 
isospin-averaged cross sections from Ref. \cite{fae94}, 
which will be used in this study. For the $\pi N$ interaction
we also show those from Cugnon and Lombard \cite{cugnon84}.

In addition to processes without the pion in the final state, 
the kaon can also be produced together with one or more pions.
At SIS energies, these processes might not be very important,
since they have a higher production threshold and are thus
suppressed as compared to those without pions in the final
states. They, nevertheless, may not be totally negligible,
and become increasingly important as beam energy increases.
Here we will consider processes with one pion in the final
state.

\begin{figure}
\begin{center}
\epsfig{file=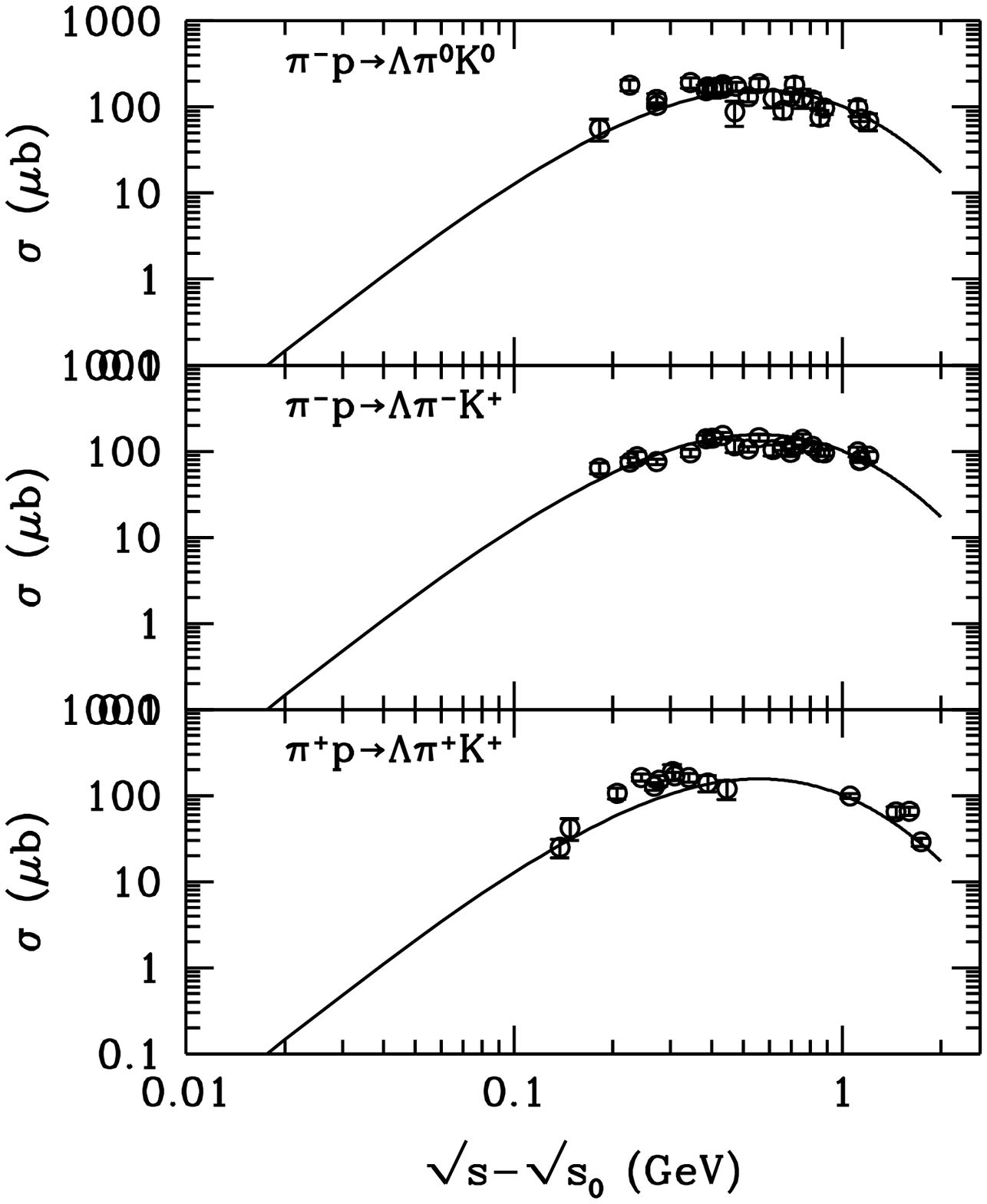,height=7.0in,width=7.0in}
\caption{Kaon production cross sections $\sigma_{\pi N\rightarrow
\Lambda \pi K}$. Open circles are experimental data, and the lines
are parameterizations.
\label{pinlop}} 
\end{center}
\end{figure}

For the $\Lambda \pi K$ channel, there are three sets of data 
available, and they can be approximately fitted by
the following expression,
\begin{eqnarray}
\sigma_{\pi^-p\rightarrow \Lambda \pi^0K^0}
\approx \sigma_{\pi^-p\rightarrow \Lambda \pi^-K^+}
\approx \sigma_{\pi^+p\rightarrow \Lambda \pi^+K^+}
\approx 24.0\left(1-{s_0\over s}\right)^{3.16}
\left({s_0\over s}\right)^{4.24} ~{\rm mb} ,
\end{eqnarray} 
where $\sqrt s$ is the available energy and $\sqrt {s_0}
=m_\Lambda +m_\pi + m_K$. 
The comparison of this parameterization with
experimental data is shown in Fig. \ref{pinlop}.  
Assuming that all the other channels have the similar cross 
sections, the isospin-averaged cross section is then obtained,
\begin{eqnarray}
\sigma _{\pi N\rightarrow \Lambda \pi K}
\approx 40.0\left(1-{s_0\over s}\right)^{3.16}
\left({s_0\over s}\right)^{4.24} ~{\rm mb}.
\end{eqnarray} 

\begin{figure}
\begin{center}
\epsfig{file=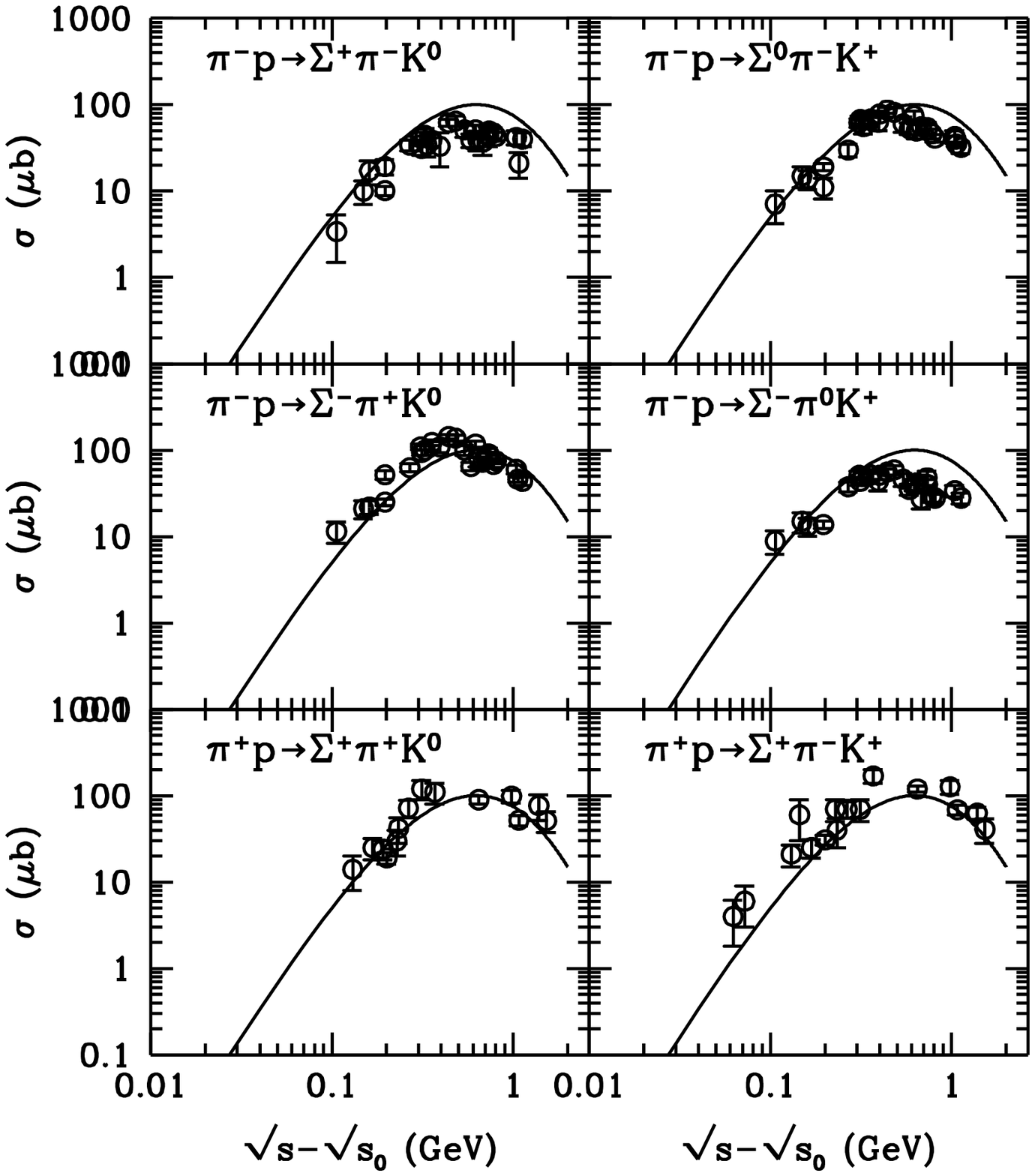,height=7.0in,width=7.0in}
\caption{Same as Fig. \protect\ref{pinlop}, for 
$\sigma_{\pi N\rightarrow \Sigma \pi K}$.
\label{pinsop}}
\end{center}
\end{figure}

Similarly, the six sets of available experimental data 
for $\Sigma \pi K$ final state can be approximately fitted by the
same expression
\begin{eqnarray}
\sigma_{\pi^-p\rightarrow \Sigma^+ \pi^-K^0}
\approx \sigma_{\pi^-p\rightarrow \Sigma^0 \pi^-K^+}
\approx \cdot\cdot\cdot
\approx 19.4\left(1-{s_0\over s}\right)^{3.41}
\left({s_0\over s}\right)^{4.25} ~{\rm mb},
\end{eqnarray} 
where $\sqrt {s_0}=m_\Sigma +m_\pi + m_K$. 
The comparison of this parameterization with data is shown 
in Fig. \ref{pinsop}. Under the same assumption that all
other channels have the similar cross sections, we get the
isospin-averaged cross section
\begin{eqnarray}
\sigma _{\pi N\rightarrow \Sigma \pi K}
\approx 84.1\left(1-{s_0\over s}\right)^{3.41}
\left({s_0\over s}\right)^{4.25} ~{\rm mb}.
\end{eqnarray} 

\subsubsection{kaon production in baryon-baryon collisions}

The situation of experimental measurement of kaon production
in nucleon-nucleon collisions is much worse than for pion-nucleon
interactions. Until very recently, there are basically no data
available near the threshold. The first detailed analysis
of the experimental data was carried out by Randrup and Ko
\cite{rk80}. They proposed the following parameterizations
for the available data
\begin{eqnarray}
\sigma _{pp\rightarrow p\Lambda K^+} = 0.024 {p_{max}\over m_K}
\; {\rm mb}\\
\sigma _{pp\rightarrow p\Sigma^0 K^+} \approx
\sigma _{pp\rightarrow p\Sigma^+ K^0} \approx
= 0.012 {p_{max}\over m_K} \; {\rm mb} ,
\end{eqnarray}
where $p_{max}$ is the maximum momentum of kaon given by
\begin{eqnarray}
p_{max}= {1\over 2\sqrt s} \sqrt {\left(s-(m_N+m_Y+m_K)^2)\right)
\left(s-(m_N+m_Y-m_K)^2\right)},
\end{eqnarray}
with $\sqrt s$ the total energy and $m_Y$ the mass of hyperon in
the final state.

Based on experimental data and isospin analysis, Randrup and Ko
also obtained the following isospin-averaged cross sections
\begin{eqnarray}
\sigma_{NN\rightarrow NY K} = 0.072 {p_{max}\over m_K} \; {\rm mb},
\end{eqnarray}
where $Y$ is either lambda or sigma hyperon.
Furthermore, they have analysed the experimental data for
processes with one pion in the final state, and by using 
the detailed-balance relation, they obtained the following
expressions for kaon production in $N\Delta$ and $\Delta\Delta$
collisions,
\begin{eqnarray}
\sigma_{N\Delta\rightarrow NY K} = 0.054 {p_{max}\over m_K}
\; {\rm mb},\\
\sigma_{\Delta\Delta\rightarrow NY K} = 0.036 {p_{max}\over m_K}
\; {\rm mb}.
\end{eqnarray}
Thus at the same center-of-mass energy, kaon production cross
section in the $N\Delta$ collision is 3/4 of that in the 
$NN$ collision, while that in the $\Delta\Delta$ collision 
is half of that in the $NN$ collision.

Later on, Sch\"urmann and Zwermann \cite{sch87} proposed another
parameterization of experimental data which treated more accurately
the threshold behavior by using a quartic $p_{max}$
dependence rather the linear dependence of Randrup and Ko,
\begin{eqnarray}
\sigma _{pp\rightarrow K^+X} = 0.8 p_{max}^4 \; {\rm mb}.
\end{eqnarray}
This parameterization was fitted to the inclusive
$K^+$ production cross section in proton-proton ($pp$) collisions. 
But it has often been identified with the lowest threshold 
process $NN\rightarrow N\Lambda K^+$.

Recently, some of the much needed experimental data for
kaon production in $pp$ collisions near threshold
have become available from the COSY-11 collaboration  
\cite{cosy,cosy11}. They have measured the cross section 
$\sigma _{pp\rightarrow p\Lambda K^+}$
at 2 MeV above the threshold, and obtained a cross section
of about 8 nb. Based on these as well as early experimental
data, Cassing {\it et al.} proposed  \cite{cass97a}
new parameterizations for exclusive kaon production
cross sections in $pp$ collisions,
\begin{eqnarray}
\sigma _{pp\rightarrow p\Lambda K^+} & =& 
0.732 \left(1-{s_0\over s}\right)^{1.8}\left({s_0\over s}\right)^{1.5}
\; {\rm mb},\\
\sigma _{pp\rightarrow p\Sigma^0 K^+} & =&
0.275 \left(1-{s_0\over s}\right)^{1.98}\left({s_0\over s}\right)
\; {\rm mb},\\
\sigma _{pp\rightarrow p\Sigma^+ K^0} & =&
0.338 \left(1-{s_0\over s}\right)^{2.25}\left({s_0\over s}\right)^{1.35}
\; {\rm mb},
\end{eqnarray}
where $\sqrt {s_0}$ are corresponding threshold.
The isospin factors and the scaling factor for
$N\Delta$ and $\Delta\Delta$ in these new parameterization are
the same as those in Randrup-Ko parameterization.

\begin{figure}
\begin{center}
\epsfig{file=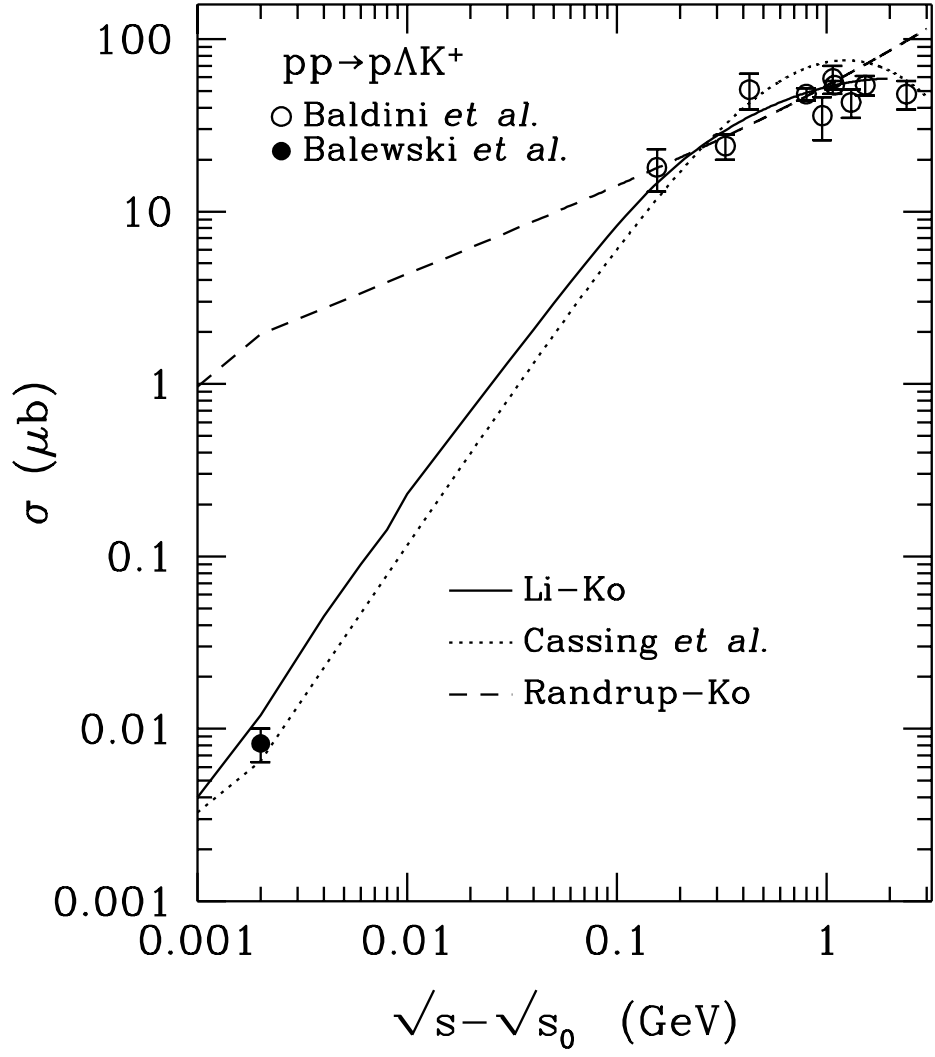,height=7.0in,width=7.0in}
\caption{Kaon production cross section $\sigma_{pp\rightarrow p\Lambda 
K^+}$. The solid curve is the results of boson-exchange
model of Ref. \protect\cite{liko95}. The dashed and dotted
curves are the parameterizations of Randrup and Ko \protect\cite{rk80}
and Cassing {\it et al.} \protect\cite{cass97a}, respectively.
The open circles are experimental data from Ref. 
\protect\cite{data}, while the solid circle is from Ref.
\protect\cite{cosy}.
\label{ppplk}}
\end{center}
\end{figure}

There have been also several theoretical calculations
of kaon production in baryon-baryon interactions, based
mostly on one-boson-exchange model 
\cite{wuko89,laget91,liko95,tsu97,likoc97}. 
We discuss here mainly the model of Ref. \cite{liko95,likoc97},
which includes one-pion and one-kaon exchanges. By adjusting
two cut-off parameters $\Lambda _\pi$ and $\Lambda _K$, very 
good agreement with experimental data on the kaon production
cross section in $NN$ interactions has been achieved.
In Fig. \ref{ppplk}, the results for $pp\rightarrow p\Lambda K^+$ are 
compared with experimental data and various parameterizations.
The open circles are early data from the compilation of
Baldini {\it et al,} \cite{data}, while the solid circle
is the experimental data available only very recently from
the COSY-11 collaboration that measured kaon production in proton-proton
interactions at 2 MeV above the threshold \cite{cosy}. 
It is seen that our model
provides a good description for both the old and new data.
The parameterization of Randrup and Ko \cite{rk80}
is shown in the figure by the dashed curve. It describes
the older experimental data (open circles) 
reasonably well. It however significantly
overestimates the newest data point at 2 MeV above the 
threshold (solid circle). The assumed linear dependence on
$p_{max}$ (the maximum momentum of kaon) is responsible
for this incorrect threshold behaviour. 
The dotted curve in the figure gives the 
recent parameterization of Cassing {\it et al.} \cite{cass97a},
who has specifically tried to fit the newest data from the COSY-11
collaboration.  

The extension of this model to kaon production in 
nucleon-delta ($N\Delta$) and delta-delta ($\Delta\Delta$) collisions
is straightforward, except for one complication arising from the
fact that in the $\Delta \pi N$ vertex energy momentum conservation 
allows the pion to go on-shell. A pole develops in the 
pion propagator at part of the kinematically allowed 
region of the phase space. As a result the cross section
becomes singular. This kind of singularity also appears 
in $N\Delta\rightarrow NN\eta$ \cite{peters}, 
$N\Delta \rightarrow NN\phi$ \cite{chung},
and $\pi\rho\rightarrow \pi\rho$ \cite{haglin95}, 
via pion exchange. The same kind of singularity occurs 
in the process $\mu^{+} \mu^{-} \rightarrow e \bar \nu W^{+}$, 
in which the $\mu$ can decay to $e \bar \nu \nu_{\mu}$ and
the exchanged neutrino goes on-shell in the diagram,
leading to a singular cross section \cite{muon}. 

In Ref. \cite{likoc97}, we applied the Peierls method \cite{peierls}, 
which takes care of the finite lifetime of the incoming $\Delta$ 
resonance, to regulate the singularity.  
The singularity is removed by the delta width
gained through the energy-momentum conservation relation.
This width is density independent and the resulting
cross section does not diverge as the density goes to zero.
The Peierls method has also been used in Ref. \cite{muon} 
to remove the singularity associated with muon collision.

\begin{figure}
\begin{center}
\epsfig{file=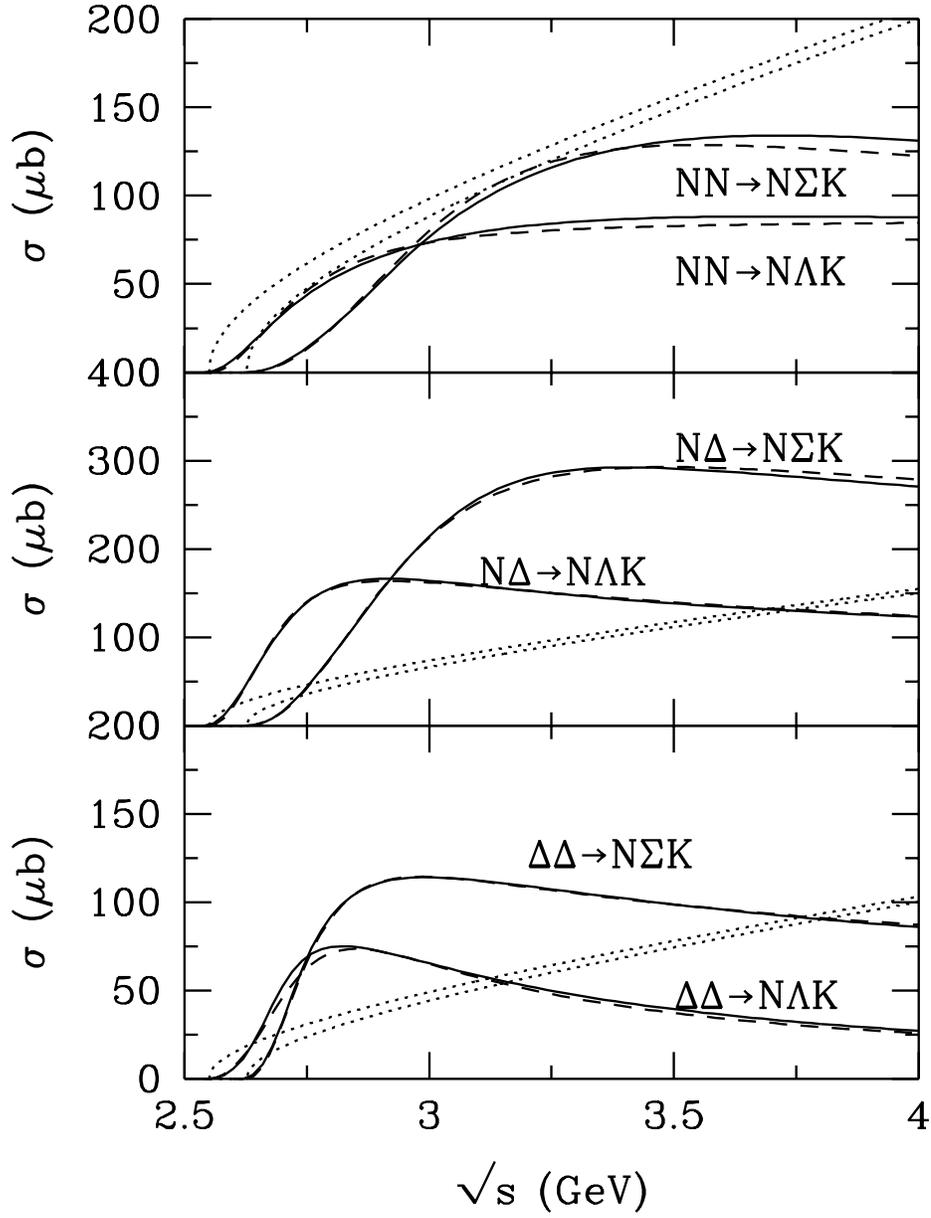,height=7.0in,width=7.0in}
\caption{Isospin-averaged kaon production cross sections
$\sigma_{BB\rightarrow NYK}$. The solid lines are the results
of boson-exchange model of Refs. \protect\cite{liko95,likoc97}.
The dotted lines are the parameterizations of Randrup and Ko
\protect\cite{rk80}. The dashed curves are the parameterizations
Eq. (\protect\ref{fit}).
\label{bbk}}
\end{center}
\end{figure}

The results for isospin-averaged kaon production cross sections
are shown in Fig. \ref{bbk}, together with those from the 
parameterization of Randrup and Ko \cite{rk80}, with the upper curves
represent the $\Lambda N K$ final state, and the lower curves
the $\Sigma N K$ final state. For the $NN$ interaction, the Randrup-Ko
parameterization is larger than our results, since in the former
only one-pion-exchange was considered which leads to a large 
isospin-average factor. For $N\Delta$ and $\Delta\Delta$ 
interactions, our results are generally larger than the 
Randrup-Ko parameterization, except near the threshold. 
To apply these cross sections in the transport model, we have 
parameterized our theoretical results in terms of the following 
expression
\begin{eqnarray}\label{fit}
\sigma _{BB\rightarrow NYK} = {a(\sqrt s-\sqrt {s_0})^2\over
b+(\sqrt s-\sqrt {s_0})^x} ~{\rm mb}.
\end{eqnarray}
The fitted parameters $a$, $b$ and $x$ for six channels are
listed in Table 1. These parameterizations are shown in Fig. 
\ref{bbk} by dashed curves, which are seen to reproduce
the solid curves quite accurately.

\vskip 0.5cm

{\bf Table 1} Fitted parameters in Eq. (\ref{fit}).

\vskip 0.5cm

\begin{center}
\begin{tabular}{ccccccc}
\hline
& $NN\rightarrow N\Lambda K$ & $NN\rightarrow N\Sigma K$ &
$N\Delta\rightarrow N\Lambda K$ & $N\Delta\rightarrow N\Sigma K$ &
$\Delta\Delta\rightarrow N\Lambda K$ & $\Delta\Delta\rightarrow N\Sigma K$\\
$a$ & 0.0865 & 0.1499 & 0.1397 & 0.3221 & 0.0361 & 0.0965 \\
$b$ & 0.0345 & 0.167  & 0.0152 & 0.107  & 0.0137 & 0.014 \\
$x$ & 2.0    & 2.4    & 2.3    & 2.3    & 2.9    & 2.3   \\
\hline
\end{tabular}
\end{center}
\vskip 0.5cm  

\begin{figure}
\begin{center}
\epsfig{file=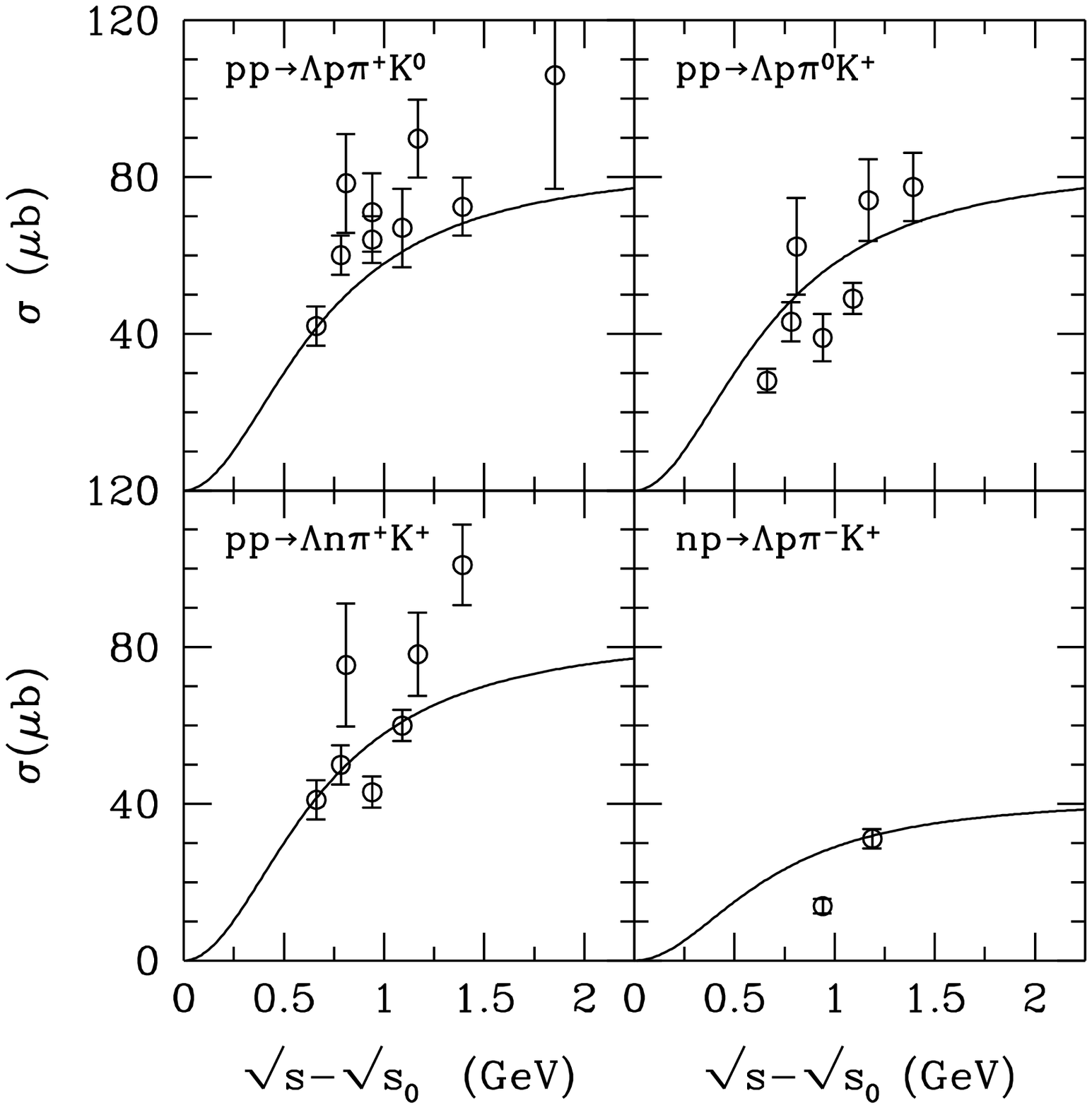,height=7.0in,width=7.0in}
\caption{Kaon production cross sections  $\sigma_{NN\rightarrow
N\Lambda \pi K}$. The open circles are the experimental
data from Ref. \protect\cite{data}, and the lines are 
parameterizations.
\label{nnlop}}
\end{center}
\end{figure}

In addition to processes with no pions in final states,
kaons can also be produced together with one or more
pions in the final states. We have analysed the available 
experimental data, and propose the following parameterizations
from processes with one and two pions in final states.
For $pp\rightarrow N\Lambda \pi K$, the three sets of 
experimental data can be fitted by the same expression
\begin{eqnarray}
\sigma _{pp\rightarrow \Lambda p\pi^+K^0}
\approx \sigma_{pp\rightarrow \Lambda p \pi^0K^+}
\approx \sigma_{pp\rightarrow \Lambda n \pi^+ K^+}
\approx {0.084(\sqrt s-\sqrt {s_0})^2\over 
0.45+(\sqrt s-\sqrt {s_0})^2} \; {\rm mb}.
\end{eqnarray}
We assume that other proton-proton ($pp$) and 
neutron-neutron ($nn$) channels have the same cross
section. The only set of data that is available for 
neutron-proton ($np$)
channels indicates a cross section which is about
half of the $pp$ one, and we assume all other $np$ channels
have the same cross section as this $np$ channel.
Under these assumption we get the isospin-averaged cross
section
\begin{eqnarray}
\sigma _{NN\rightarrow N\Lambda \pi K} 
\approx {0.21(\sqrt s-\sqrt {s_0})^2\over 
0.45+(\sqrt s-\sqrt {s_0})^2} \; {\rm mb}.
\end{eqnarray}
The comparison of these parameterizations
with experimental data is shown in Fig. \ref{nnlop}.

\begin{figure}
\begin{center}
\epsfig{file=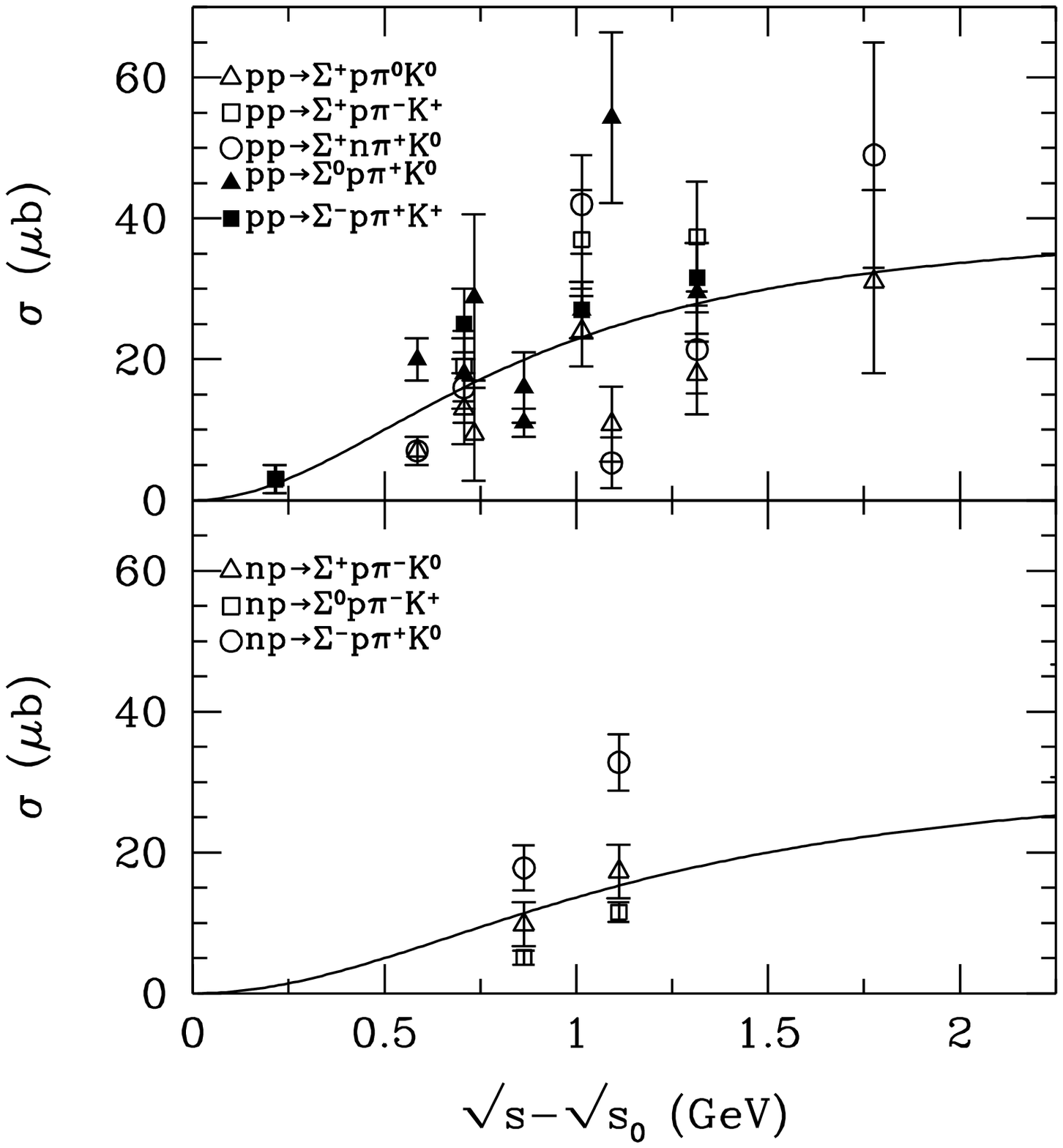,height=7.0in,width=7.0in}
\caption{Same as Fig. \protect\ref{nnlop}, for
$\sigma_{NN\rightarrow N\Sigma \pi K}$.   
\label{nnsop}}
\end{center}
\end{figure}

For $pp\rightarrow N\Sigma \pi K$, on the average, the five 
sets of available data can be fitted by the following expression
\begin{eqnarray}
{1\over 5}\left[\sigma _{pp\rightarrow p\Sigma^+\pi^0K^0}
+\sigma _{pp\rightarrow p\Sigma ^+\pi^-K^+}
+\cdot\cdot\cdot \right]
\approx {0.04(\sqrt s-\sqrt {s_0})^2\over 0.75 + (\sqrt s-\sqrt {s_0})^2}
\; {\rm mb}.
\end{eqnarray}
We assume that all the other $pp$ and $nn$ channels
have the similar cross sections. For $np$ channels
the three sets of available experimental data can be fitted, 
on the average, by the following expression,
\begin{eqnarray}
{1\over 3}\left[\sigma _{np\rightarrow p\Sigma^+\pi^-K^0}
+\sigma _{np\rightarrow p\Sigma ^0\pi^-K^+}
+\sigma_{np\rightarrow p\sigma^-\pi^+K^0}\right]
\approx {0.032(\sqrt s-\sqrt {s_0})^2\over 1.35 + (\sqrt s-\sqrt {s_0})^2}
\; {\rm mb}.
\end{eqnarray}
We assume that all the other $np$ channels have the similar
cross sections. We thus get the isospin-averaged cross section,
\begin{eqnarray}
\sigma _{NN\rightarrow N\Sigma \pi K} 
\approx {0.16(\sqrt s-\sqrt {s_0})^2\over 
0.75+(\sqrt s-\sqrt {s_0})^2} 
+ {0.16(\sqrt s-\sqrt {s_0})^2\over 
1.35+(\sqrt s-\sqrt {s_0})^2} \; {\rm mb}.
\end{eqnarray}
The comparisons of these parameterizations with
experimental data are shown in Fig. \ref{nnsop}.

\begin{figure}
\begin{center}
\epsfig{file=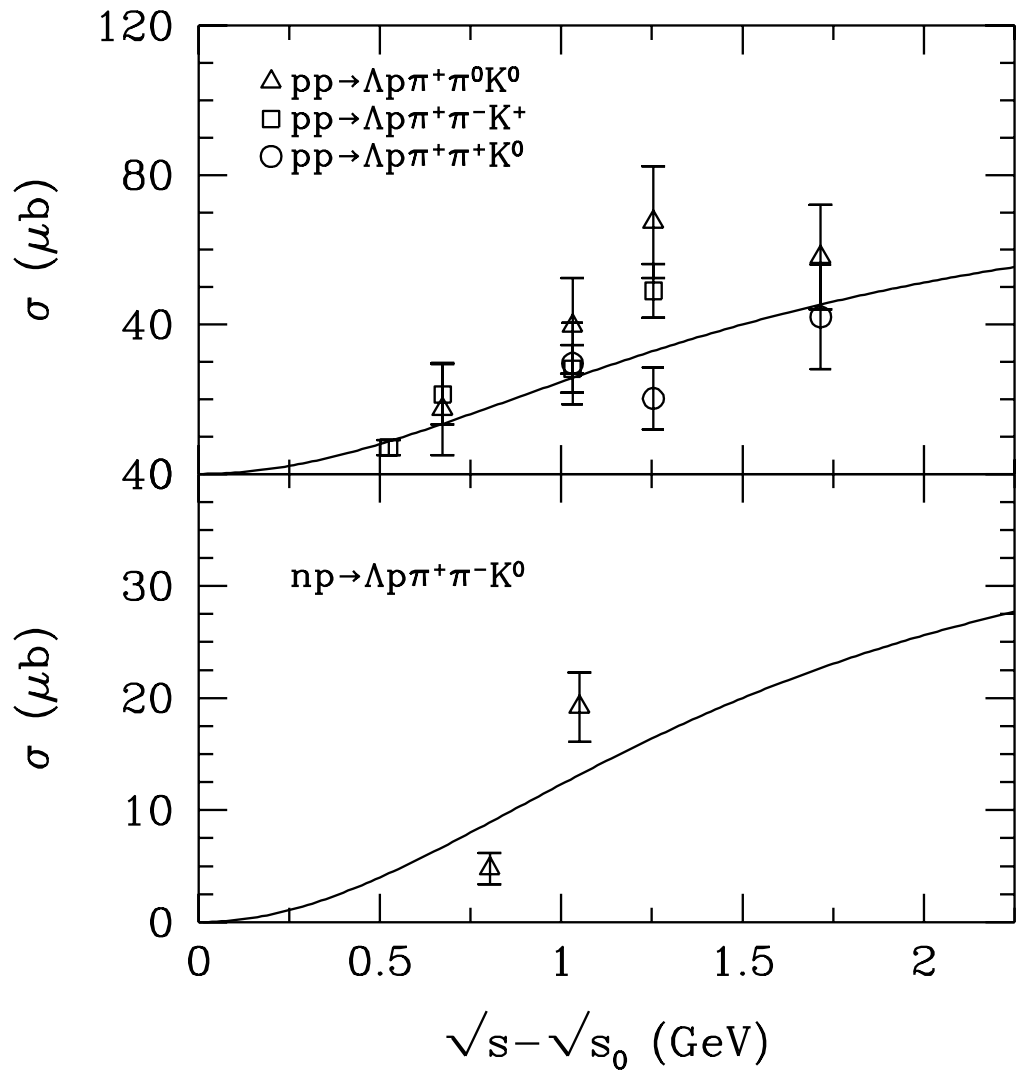,height=7.0in,width=7.0in}
\caption{Same as Fig. \protect\ref{nnlop}, for
$\sigma_{NN\rightarrow N\Lambda \pi\pi K}$.   
\label{nnltp}}
\end{center}
\end{figure}

Finally, we discuss cross sections with two pions in 
the final state. For $pp\rightarrow N\Lambda \pi\pi K$,
the available three sets of experimental data can be fitted,
on the average, by the following expression,
\begin{eqnarray}
{1\over 3}\left[\sigma _{pp\rightarrow p\Lambda \pi^+\pi^0K^0}
+\sigma _{pp\rightarrow p\Lambda \pi^+\pi^-K^+}
+\sigma _{pp\rightarrow p\Lambda \pi^+\pi^+K^0}\right]
\approx {0.08(\sqrt s-\sqrt {s_0})^2\over 2.25+(\sqrt s-\sqrt {s_0})^2}
\; {\rm mb}.
\end{eqnarray}
We assume that other $pp$ and $nn$ channels have the same cross
section. The only set of data that is available for $np$
channels indicates a cross section which is about
half of the $pp$ one, and we assume all other $np$ channels
have the same cross section as this $np$ channels.
Under these assumptions we get the isospin-averaged cross
section
\begin{eqnarray}
\sigma _{NN\rightarrow N\Lambda \pi \pi K} 
\approx {0.32(\sqrt s-\sqrt {s_0})^2\over 
2.25+(\sqrt s-\sqrt {s_0})^2} \; {\rm mb}.
\end{eqnarray}
The comparison of these parameterizations
with experimental data is shown in Fig. \ref{nnltp}.

\begin{figure}
\begin{center}
\epsfig{file=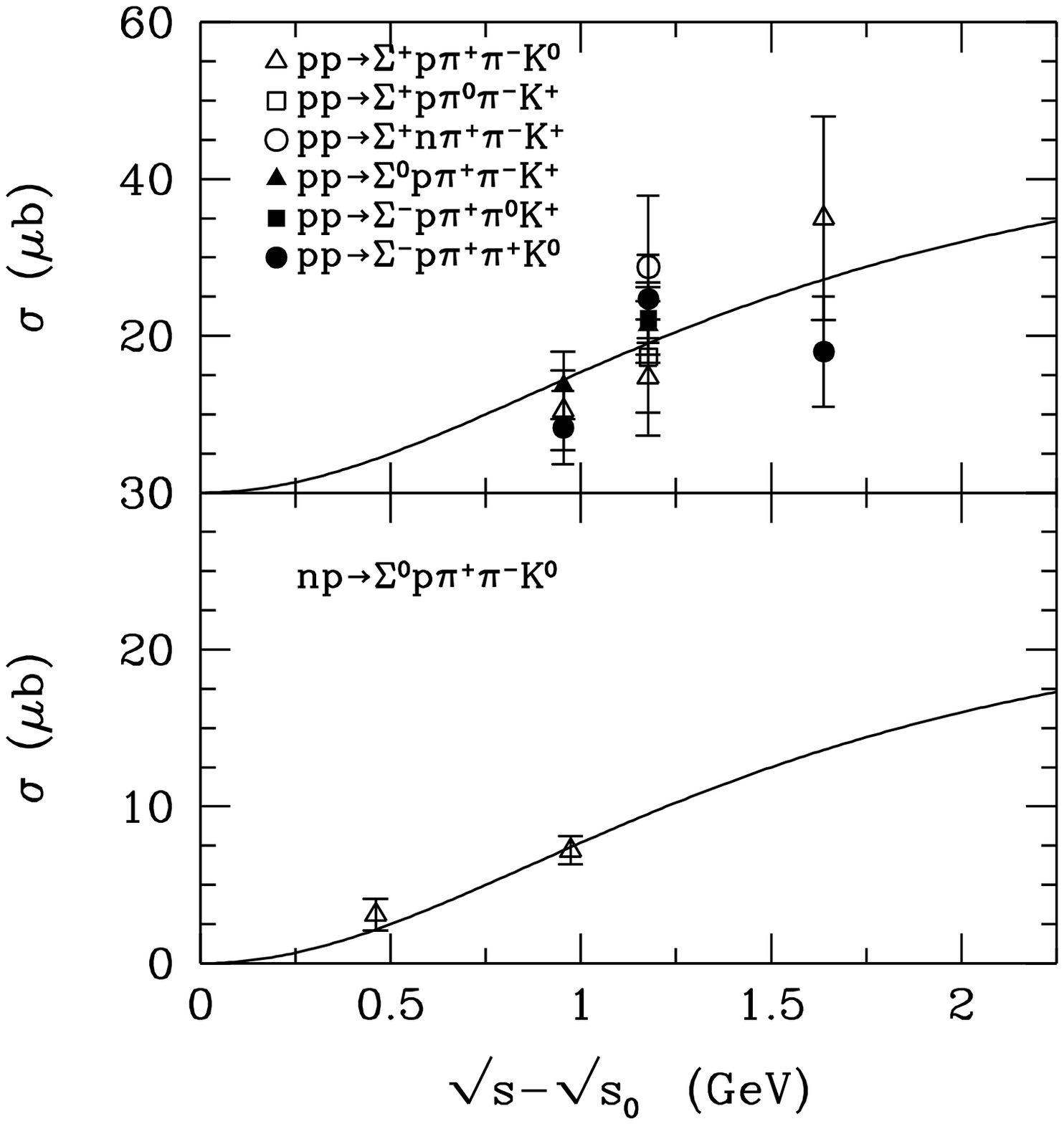,height=7.0in,width=7.0in}
\caption{Same as Fig. \protect\ref{nnlop}, for
$\sigma_{NN\rightarrow N\Sigma \pi\pi K}$.   
\label{nnstp}}
\end{center}
\end{figure}

A similar procedure is carried out for $NN\rightarrow N\Sigma\pi\pi K$.
The six sets of data for $pp$ channels can be fitted,
on the average, by the following expression,
\begin{eqnarray}
{1\over 6}\left[\sigma _{pp\rightarrow p\Sigma^+ \pi^+\pi^-K^0}
+\sigma _{pp\rightarrow p\Sigma^+ \pi^0\pi^-K^+}
+\cdot\cdot\cdot \right]
\approx {0.05(\sqrt s-\sqrt {s_0})^2\over 2.25+(\sqrt s-\sqrt {s_0})^2}
\; {\rm mb}.
\end{eqnarray}
Again, the $np$ channels have a cross section which is about 
half of that for $pp$ channels. Under the same assumptions
as before, the isospin-averaged cross section is found
to be
\begin{eqnarray}
\sigma _{NN\rightarrow N\Sigma \pi \pi K} 
\approx {0.55(\sqrt s-\sqrt {s_0})^2\over 
2.25+(\sqrt s-\sqrt {s_0})^2} \; {\rm mb}.
\end{eqnarray}
The comparisons of these parameterizations
with experimental data are shown in Fig. \ref{nnstp}.
                                                
\begin{figure}
\begin{center}
\epsfig{file=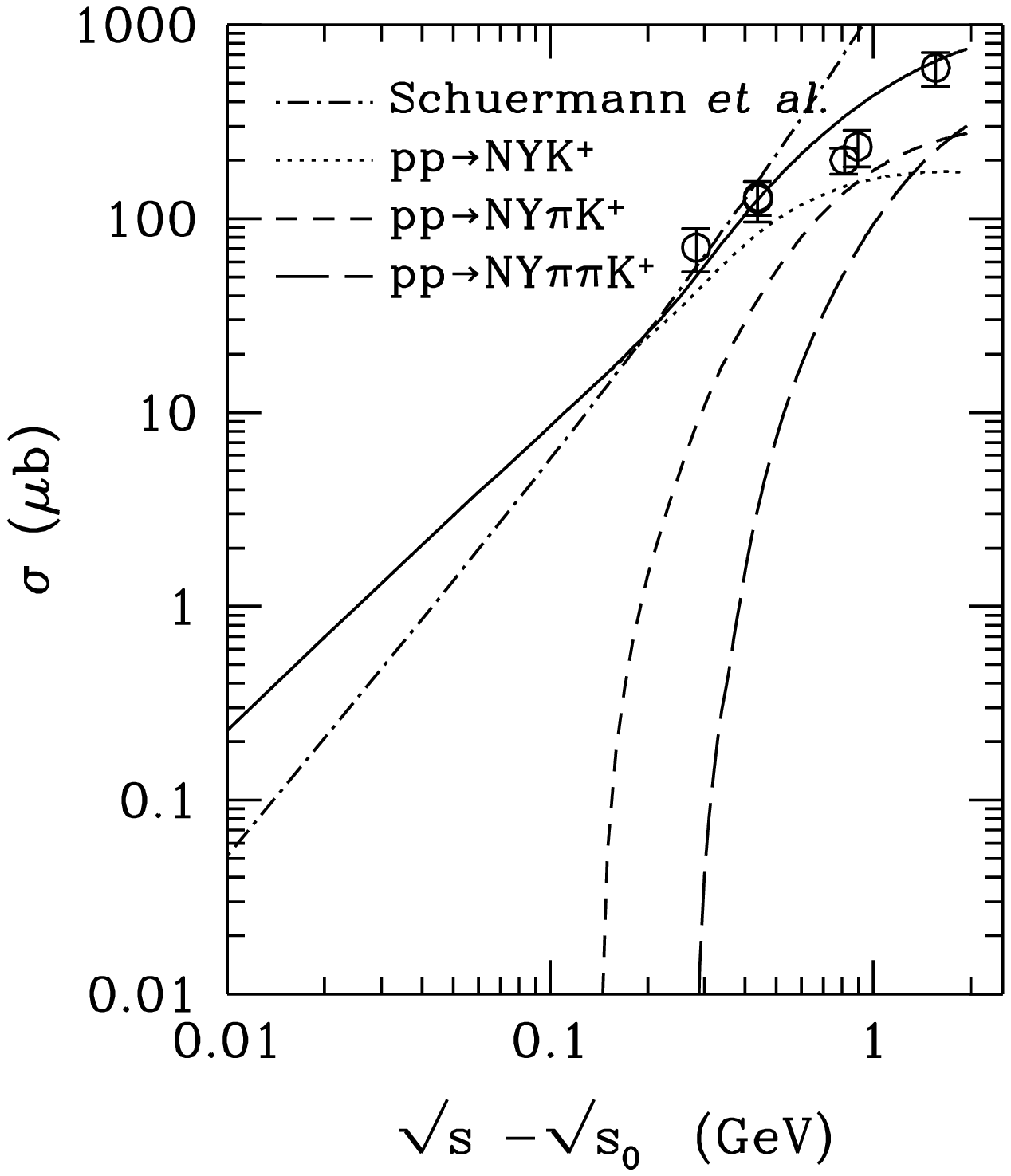,height=7.0in,width=7.0in}
\caption{Inclusive $K^+$ production cross section
in $pp$ collisions. The solid curve is the sum of
$pp\rightarrow NYK^+, ~NY\pi K^+,$ and $NY\pi\pi K^+$. 
\label{ppkp}}
\end{center}
\end{figure}

Combining our theoretical results for 
$pp\rightarrow NYK^+$ with our parameterizations
for channels with one and two pions in final state, 
we can calculate the inclusive $K^+$ production cross section
in $pp$ collision which is compared with experimental data 
in Fig. \ref{ppkp}. In the figure, the dotted, short-dashed, and 
long-dashed curves are for final states with zero, one and two 
pions, respectively, while the solid line
is the sum of the three contributions. It is seen that
our results are in good agreement with the data. 
In the figure we also show the Sch\"urmann-Zwermann 
parameterization which was originally proposed for
the inclusive $K^+$ production cross section in
$pp$ interaction.

\subsection{Elementary antikaon production cross section}

Similar to the kaon, at SIS energies, the antikaon can be produced
from $\pi N$, $\pi\Delta$, $NN$, $N\Delta$, and $\Delta\Delta$
interactions.  It can also be produced from the
strangeness-exchange processes such as $\pi Y\rightarrow {\bar K}N$.
Although the abundance of hyperons in heavy-ion collisions 
at SIS energies is small, these strangeness-exchange processes 
are quite important, because of a large cross section.

\subsubsection{antikaon production in pion-baryon collisions}

\begin{figure}
\begin{center}
\epsfig{file=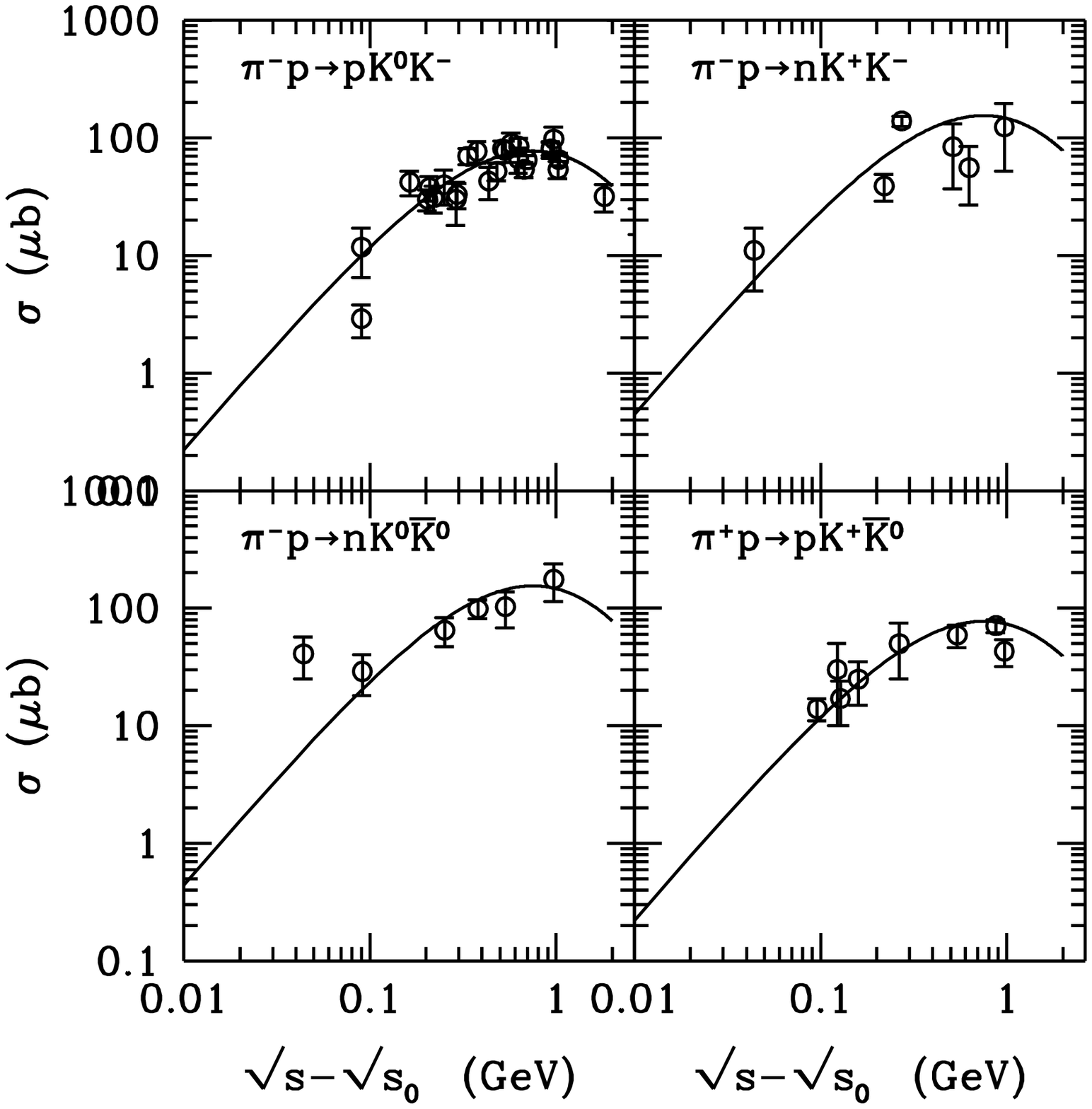,height=7.0in,width=7.0in}
\caption{Antikaon production cross sections 
$\sigma_{\pi N\rightarrow NK{\bar K}}$. The open circles
are experimental data from Ref. \protect\cite{data},
and the lines are the parameterizations of Sibirtsev
{\it et al.} \protect\cite{sib97}. 
\label{pinak}}
\end{center}
\end{figure}

The cross sections for $\pi N\rightarrow NK{\bar K}$ have
been analysed recently by Sibirtsev {\it et al.} \cite{sib97}
using a boson-exchange model. The authors found that the ratios
between experimental cross sections of different charge channels
can be well accounted for by this model. The isospin-averaged 
cross section from their model is
\begin{eqnarray}
\sigma _{\pi N\rightarrow NK{\bar K}}
=3\sigma _{\pi^-p\rightarrow pK^0K^-},
\end{eqnarray}
with the latter taken from the parameterization 
of the experimental data,
\begin{eqnarray}
\sigma _{\pi^-p\rightarrow pK^0K^-} = 
1.121 \left(1-{s_0\over s}\right)^{1.86} \left({s_0\over s}\right)^2
\; {\rm mb},
\end{eqnarray}
where $\sqrt {s_0}= m_N+2m_K$. The comparisons
of this parameterization, after including correct
isospin factor for different charge channels, 
with the experimental data are shown in Fig. \ref{pinak}.

\begin{figure}
\begin{center}
\epsfig{file=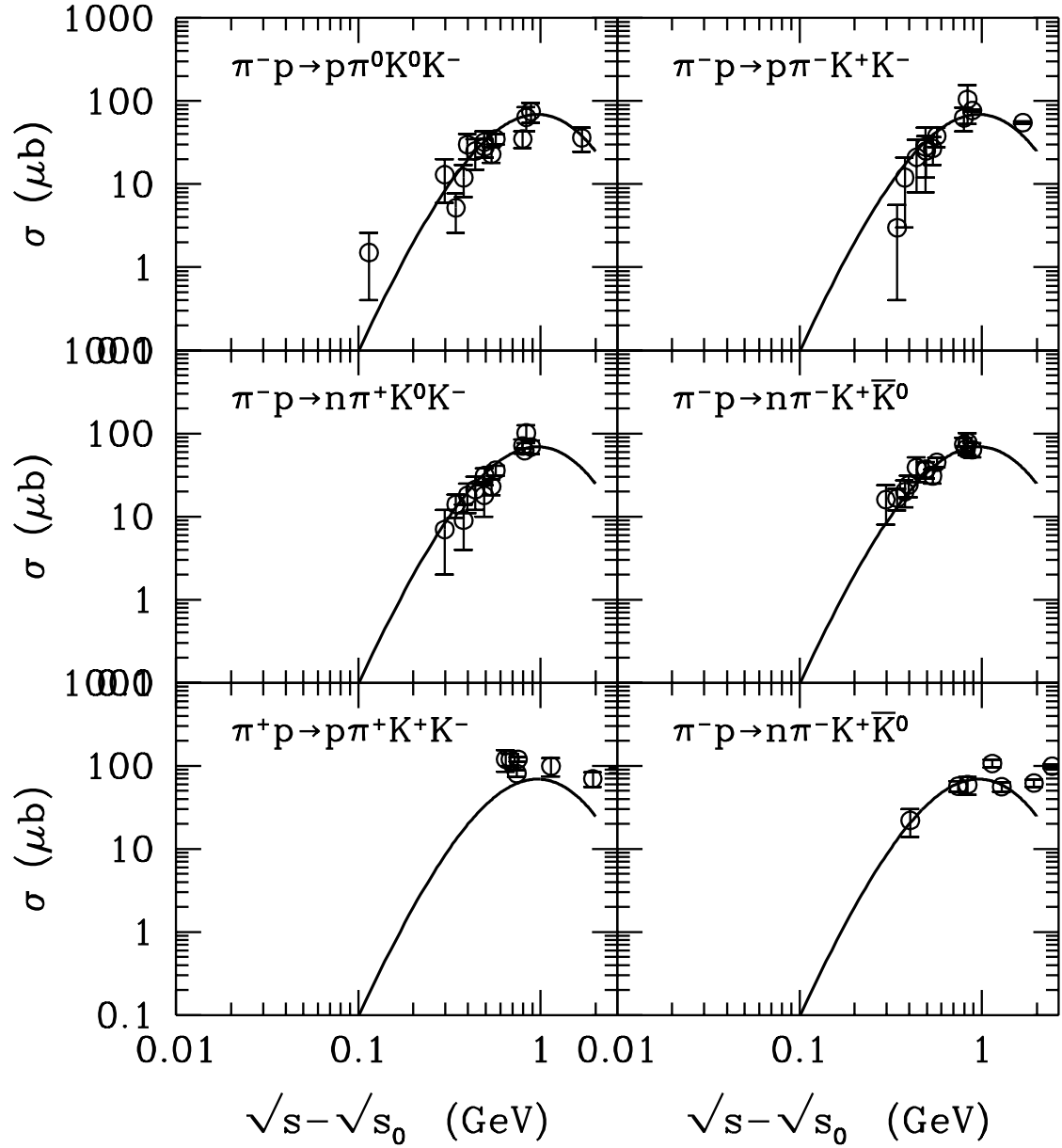,height=7.0in,width=7.0in}
\caption{Antikaon production cross section 
$\sigma_{\pi N\rightarrow N\pi K{\bar K}}$. The open circles
are experimental data from Ref. \protect\cite{data},
and the lines are the parameterizations.
\label{pinakop}}
\end{center}
\end{figure}

There are also processes with one pion in final states. The
available experimental data indicate charge independence
within a factor of two, and can be parameterized by
\begin{eqnarray}
\sigma _{\pi^- p\rightarrow p\pi^0K^0K^-}
\approx \sigma _{\pi^-p\rightarrow p\pi^-K^+K^-}
=\cdot\cdot\cdot 
\approx 85.0\left(1-{s_0\over s}\right)^{5.5}
\left({s_0\over s}\right)^{4.8} \; {\rm mb},
\end{eqnarray}
where $\sqrt {s_0}=m_N+m_\pi+2m_K$.
The comparisons of this parameterization with experimental
data are shown in Fig. \ref{pinakop}.
The isospin-averaged cross section is
\begin{eqnarray}
\sigma _{\pi N\rightarrow N\pi K{\bar K}}
=510.0\left(1-{s_0\over s}\right)^{5.5}
\left({s_0\over s}\right)^{4.8} \; {\rm mb} .
\end{eqnarray}
There has been no theoretical study on antikaon production
in pion-delta collisions. We assume that these cross sections,
after isospin averaging, are the same as those for $\pi N$
interactions at the same center-of-mass energy.

\subsubsection{antikaon production cross sections
in baryon-baryon collisions}

Again, the situation of experimental data for antikaon production
in $NN$ collisions is worse than that for $\pi N$ 
interactions. Zwermann and Sch\"urmann \cite{zwer84}
analysed the available experimental data and,
under the assumption that the cross section is charge 
independent, proposed the following parameterization for 
the isospin-averaged cross section,   
\begin{eqnarray}
\sigma _{NN\rightarrow NNK{\bar K}}
=0.05 p_{max}  \; {\rm mb},
\end{eqnarray} 
where $p_{max}$ is the maximum momentum of antikaon given by
\begin{eqnarray}
p_{max}={1\over 2\sqrt s}\sqrt {(s-4(m_N+m_K)^2)(s-4m_N^2)}.
\end{eqnarray}

\begin{figure}
\begin{center}
\epsfig{file=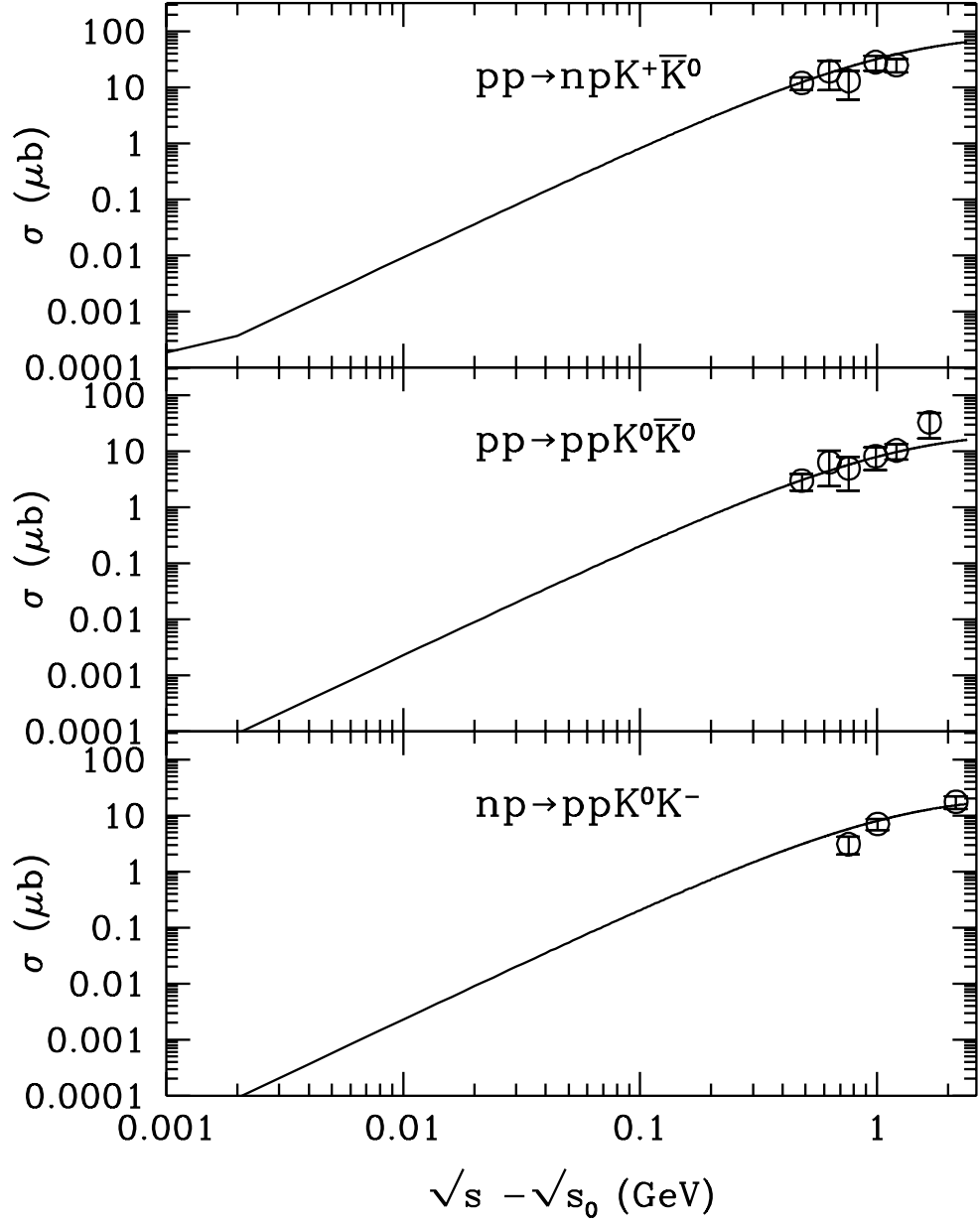,height=7.0in,width=7.0in}
\caption{Antikaon production cross section
$\sigma_{NN\rightarrow NNK\bar K}$. The open circles
are experimental data from Ref. \protect\cite{data},
and the curves are our parameterizations.
\label{nnak}}
\end{center}
\end{figure}

\begin{figure}
\begin{center}
\epsfig{file=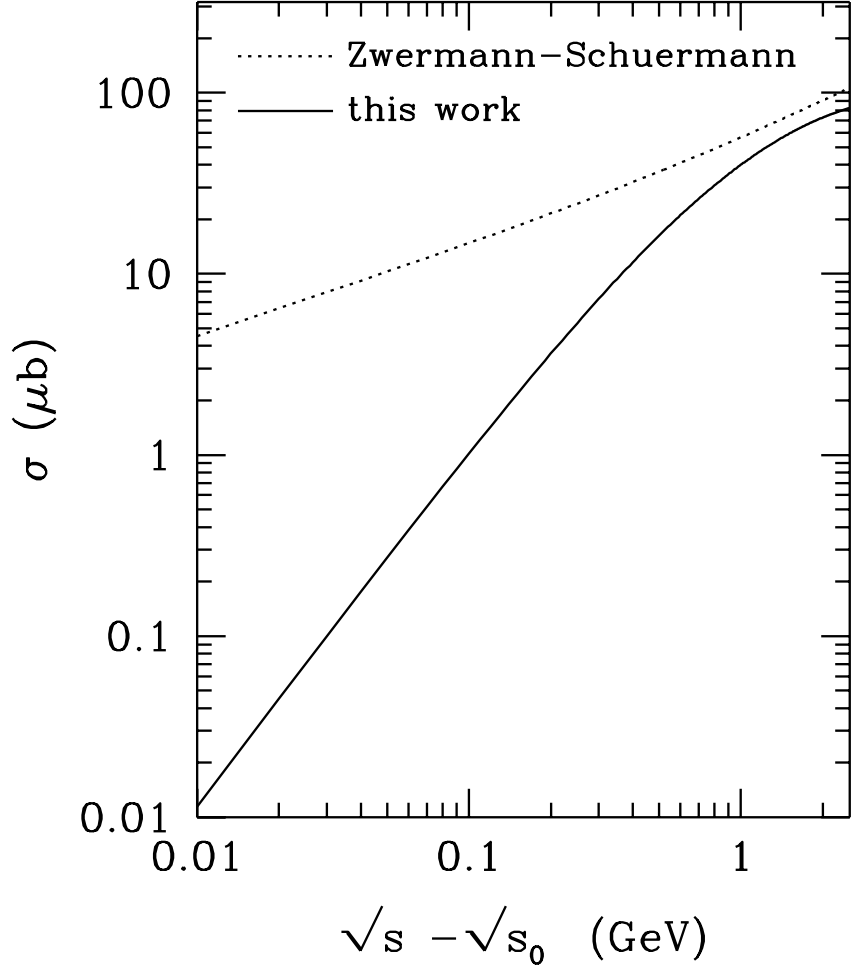,height=7.0in,width=7.0in}
\caption{Isospin-averaged antikaon production cross
section $\sigma_{NN\rightarrow NNK\bar K}$. 
The solid line is from this work, and the dotted line is
the parameterization of Zwermann and Sch\"urmann 
\protect\cite{zwer84}. 
\label{nnakiso}}
\end{center}
\end{figure}

Recently, Sibirtsev {\it et al.} \cite{sib97} analysed
this process in a boson-exchange model, using as
inputs the antikaon production cross sections from pion-nucleon
interactions outlined above.
Their results fit the experimental data for $pp\rightarrow
ppK^0{\bar K}^0$ quite well, but seem to underestimate
the data for $pp\rightarrow npK^+{\bar K}^0$.
We use in this work the following parameterization
for this process
\begin{eqnarray}
\sigma _{pp\rightarrow npK^+{\bar K}^0}
=0.19 \left(1-{s_0\over s}\right)^2\left({s_0\over s}\right)^{0.31}
\; {\rm mb}.
\end{eqnarray}
Experimental data indicate that
\begin{eqnarray}
\sigma _{pp\rightarrow ppK^0{\bar K}^0}
\approx \sigma _{np\rightarrow ppK^0K^-}
\approx {1\over 4} \sigma _{pp\rightarrow npK^+{\bar K}^0}.
\end{eqnarray}
The comparision of these parameterization with experimental
data are shown in Fig. \ref{nnak}.
Note that at 2 MeV above the threshold, our parameterization
gives a cross section of about 0.09 nb for $pp\rightarrow
ppK^0{\bar K}^0$. This is very close to the preliminary
data of about 0.1 nb from the COSY-11 collaboration for
$pp\rightarrow ppK^+K^-$ \cite{cosy11}. Both the $K^+$ and
$K^-$ data near the thresholds from 
the COSY-11 collaboration are indeed
very helpful in constraining the inputs of and reducing
the uncertainties in the transport models.
With some further assumptions about the relationship
between different charge channels, we
obtain the following isospin-averaged cross section
\begin{eqnarray}
\sigma _{NN\rightarrow NNK{\bar K}} = 
{5\over 4} \sigma _{pp\rightarrow npK^+{\bar K}^0}.
\end{eqnarray}
The comparision of this with the parameterization of
Zwermann and Sch\"urmann is shown in Fig. \ref{nnakiso}.
Apparently, the latter is much larger than ours near the
production threshold. Again the linear $p_{max}$ dependence
assumed in the Zwermann-Sch\"urmann parameterization is
responsible for the incorrect threshold behavior. It should 
be mentioned that the Zwermann-Sch\"urmann parameterization is for 
the isospin-averaged cross section $\sigma _{NN\rightarrow NNK{\bar K}}$. 
In Ref. \cite{sib97}, it was incorrectly compared with 
the inclusive $K^-$ production cross section in $pp$ collision.

\begin{figure}
\begin{center}
\epsfig{file=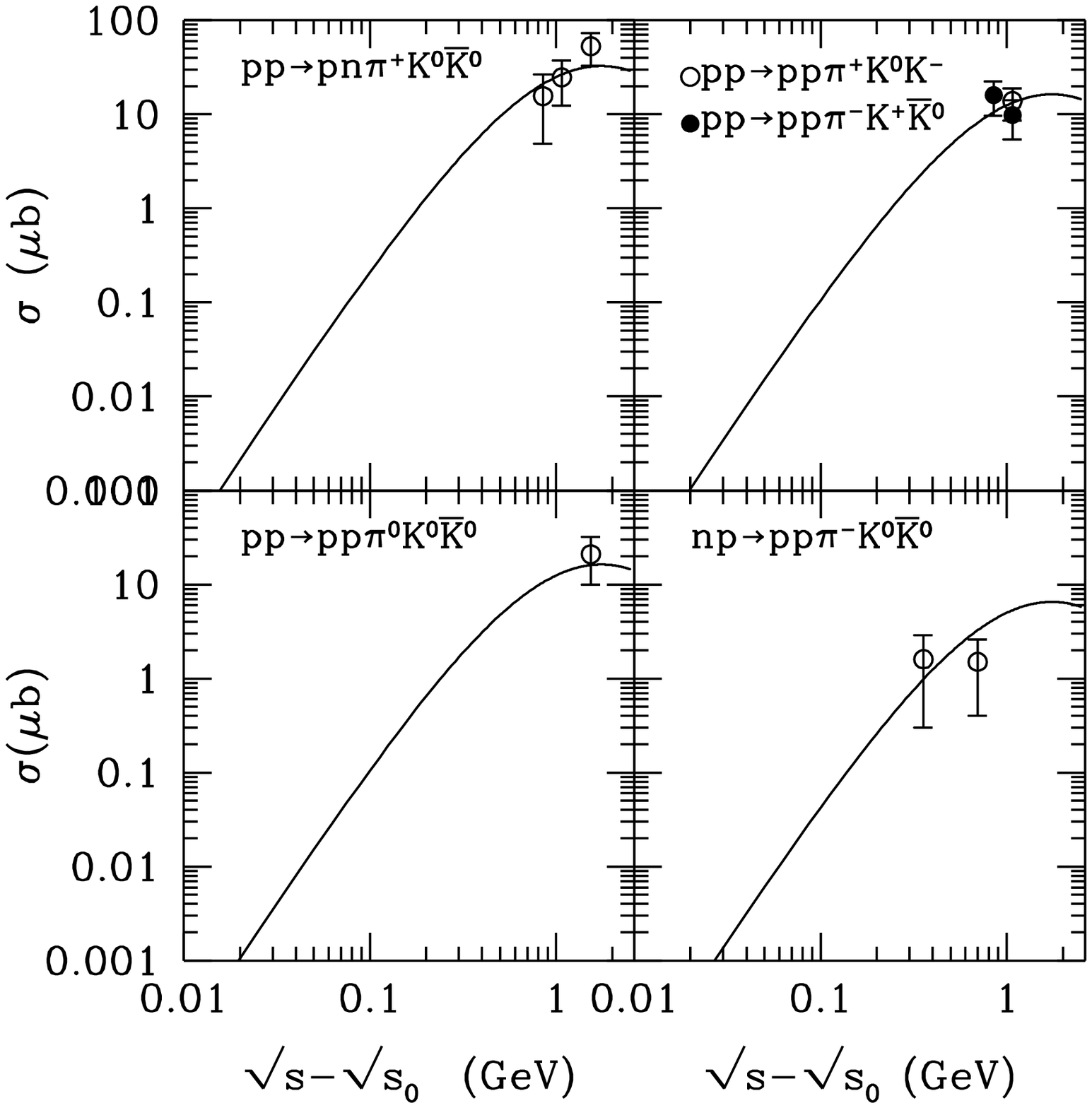,height=7.0in,width=7.0in}
\caption{Same as Fig. \protect\ref{nnak}, for
$\sigma_{NN\rightarrow NN\pi K\bar K}$.   
\label{nnakop}}
\end{center}
\end{figure}

For processes with one pion in the final state, experimental
data are quite scarce. We propose the following 
parameterization for $pp\rightarrow pn\pi^+K^0{\bar K}^0$,
\begin{eqnarray}
\sigma_{pp\rightarrow pn\pi^+K^0{\bar K}^0}
\approx 0.945 \left(1-{s_0\over s}\right)^3
\left({s_0\over s}\right)^2 \; {\rm mb}.
\end{eqnarray}  
The available data for other channels indicate that
\begin{eqnarray}
\sigma_{pp\rightarrow pp\pi^+K^0K^-}
\approx \sigma _{pp\rightarrow pp\pi^0K^0{\bar K}^0}
\approx {1\over 2}\sigma _{pp\rightarrow pn\pi^+K^0{\bar K}^0},\\
\sigma _{np\rightarrow pp\pi^-K^0{\bar K}^0}
\approx {1\over 5} \sigma _{pp\rightarrow pn\pi^+K^0{\bar K}^0}.
\end{eqnarray}
The comparisons of these parameterizations with experimental
data are shown in Fig. \ref{nnakop}.
With some further assumption about the relationship
between different charge channels, we obtain
the following isospin-averaged cross section
\begin{eqnarray}
\sigma _{NN\rightarrow NN\pi K{\bar K}} 
\approx 3.31 \left(1-{s_0\over s}\right)^3
\left({s_0\over s}\right)^2 \; {\rm mb}.
\end{eqnarray}  

\begin{figure}
\begin{center}
\epsfig{file=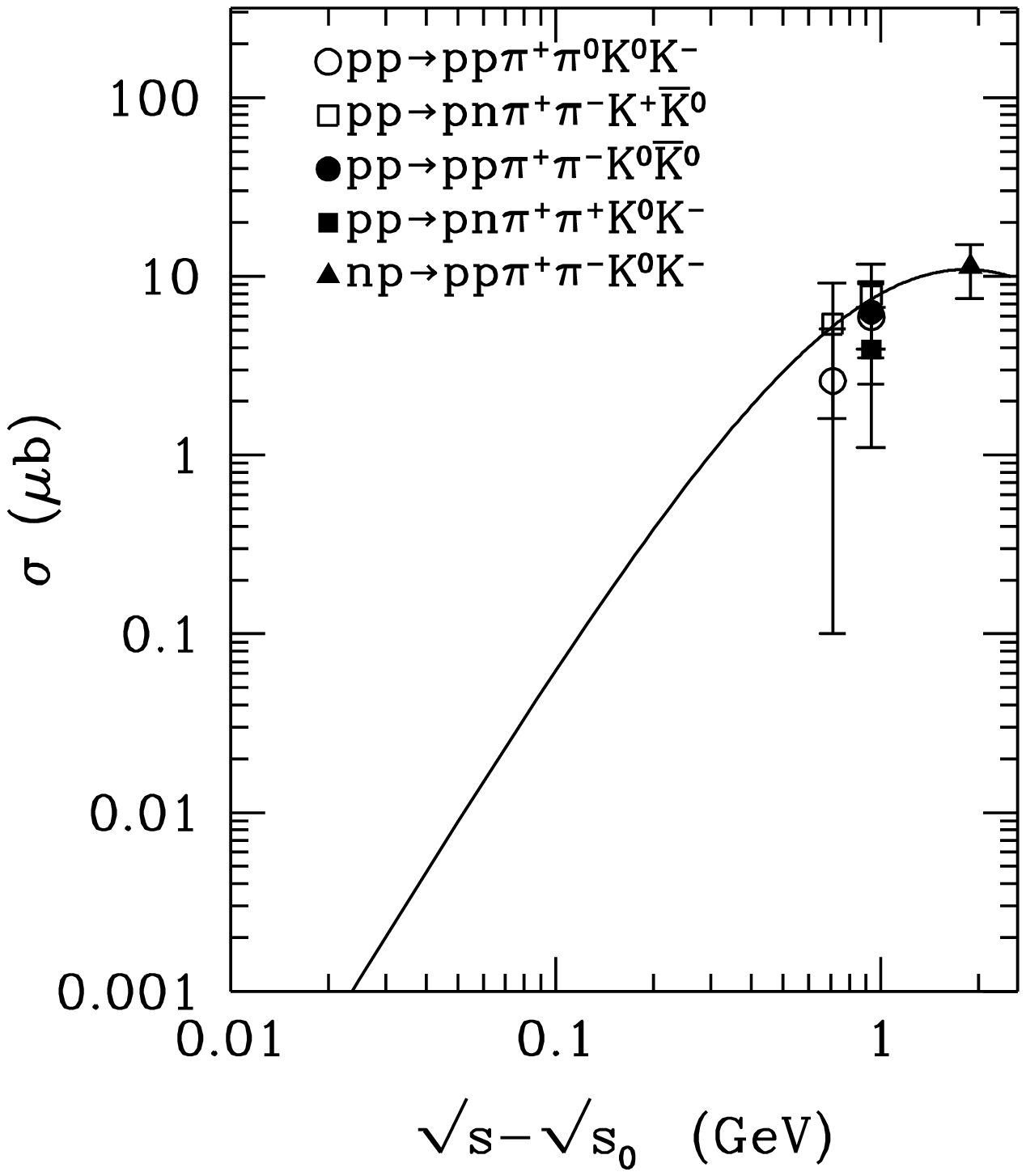,height=7.0in,width=7.0in}
\caption{Same as Fig. \protect\ref{nnak}, for
$\sigma_{NN\rightarrow NN\pi\pi K\bar K}$.
\label{nnaktp}}
\end{center}
\end{figure}

Finally, on the average, the available experimental data for
five channels with two pions in the final state can be fitted by
the following expression,
\begin{eqnarray}
{1\over 5}\left[\sigma _{pp\rightarrow pp\pi^+\pi^0K^0K^-}
+\sigma _{pp\rightarrow pn\pi^+\pi^-K^+K^0}
+\cdot\cdot\cdot \right]
\approx 0.316\left(1-{s_0\over s}\right)^3 \left({s_0\over s}\right)^2
\; {\rm mb}.
\end{eqnarray} 
The comparison of this parameterization with experimental data
is shown in Fig. \ref{nnaktp}. The isospin-averaged cross section,
assuming that all the other channels have the same cross section,
is given by
\begin{eqnarray}
\sigma _{NN\rightarrow NN\pi\pi K{\bar K}}
\approx 4.75 \left(1-{s_0\over s}\right)^3 \left({s_0\over s}\right)^2
\; {\rm mb}.
\end{eqnarray} 
 
\begin{figure}
\begin{center}
\epsfig{file=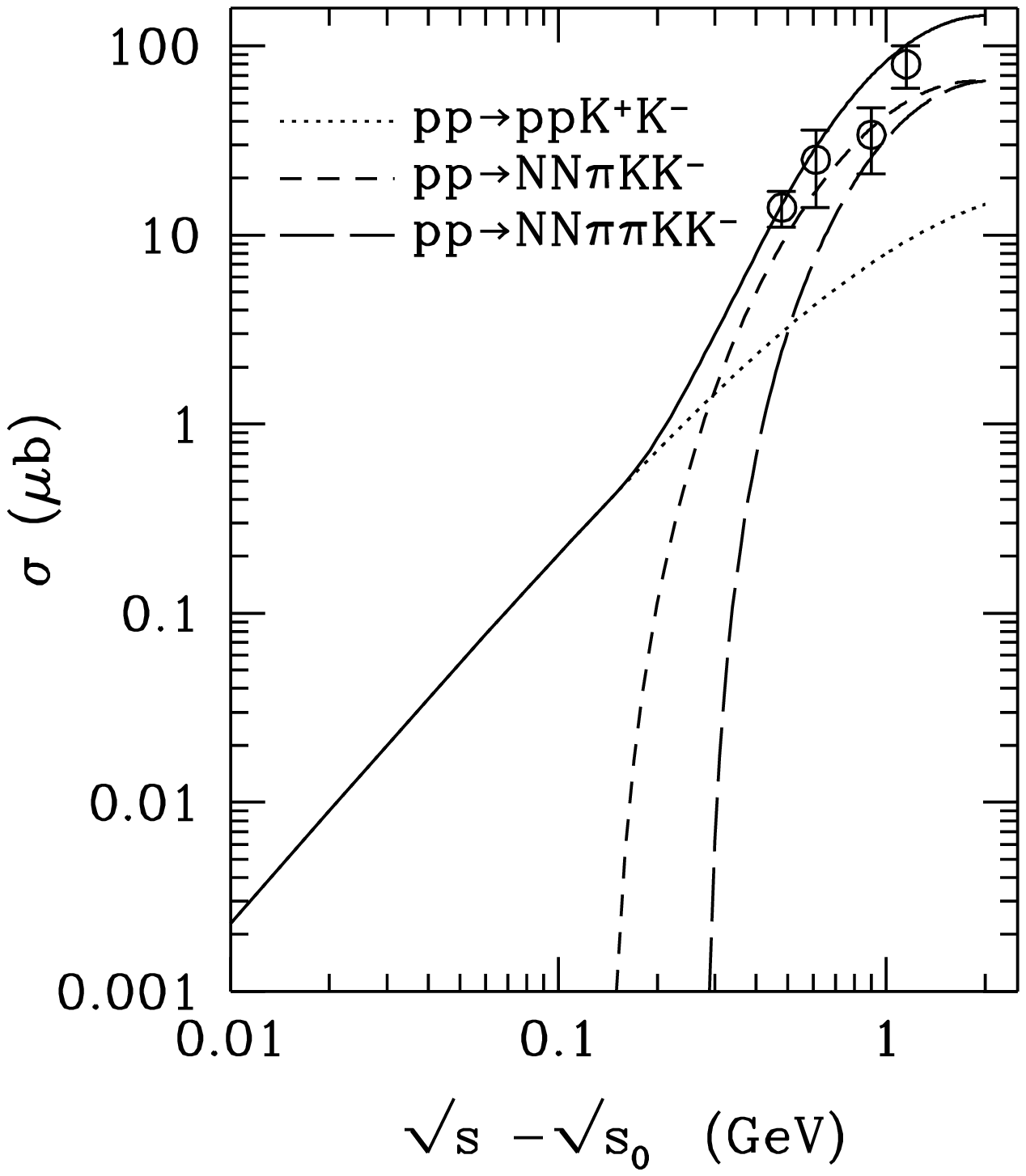,height=7.0in,width=7.0in}
\caption{Inclusive $K^-$ production cross section
in $pp$ collisions. The solid curve is the sum of
$pp\rightarrow ppK^+K^-, ~NN\pi KK^+,$  and
$NN\pi\pi KK^-$. 
\label{ppkm}}
\end{center}
\end{figure}

Summing up the cross sections with zero, one and two pions
in the final state, we obtain the inclusive $K^-$ production
cross section in $pp$ collision, which is compared with
experimental data in Fig. \ref{ppkm}. The dotted, short-dashed, and
long-dashed curves correspond to channels with zero, one and two
pions, respectively. The sum of these contribution, as shown 
in the figure by solid curve, agrees rather well with the
data.

\begin{figure}
\begin{center}
\epsfig{file=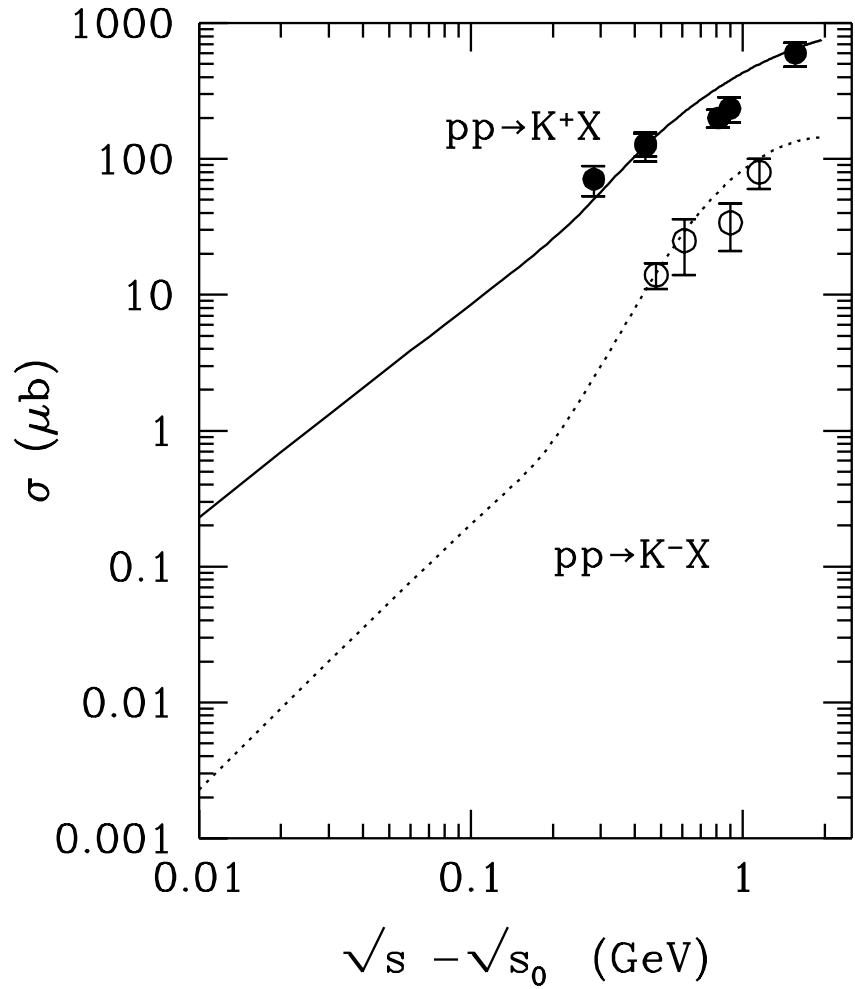,height=7.0in,width=7.0in}
\caption{Comparison of inclusive $K^+$ and $K^-$
production cross sections in $pp$ collisions.  
\label{ppkak}}
\end{center}
\end{figure}

It is also useful to compare inclusive $K^+$ and $K^-$
production cross sections in $pp$ collisions. This
is done in Fig. \ref{ppkak}. It is seen that at the
same $\sqrt s-\sqrt {s_0}$ ($\sqrt {s_0}=m_N+m_\Lambda+m_K$
for $K^+$ case, and $\sqrt {s_0}=2m_N+2m_K$ for $K^-$ case), 
the $K^+$ production cross section is one to two orders of 
magnitude larger than that of $K^-$ in $pp$ collisions.

\subsubsection{antikaon production in pion-hyperon
collisions}

As mentioned, in heavy-ion collisions, the antikaon can also
be produced from strange-exchange processes such as
$\pi \Lambda \rightarrow {\bar K}N$ and $\pi \Sigma 
\rightarrow {\bar K}N$. These cross sections are
not known empirically, but they can be obtained 
from the inverse processes, which have been studied
in great detail experimentally, by using the detailed
balance relation. At low energies, the experimental
data on ${\bar K}N\rightarrow \pi Y$ can be analysed
in terms of the K-matrix for three coupled channels 
${\bar K}N$, $\Lambda\pi$, and $\Sigma\pi$ \cite{dover82,martin70}. 
Based on this model, Ko obtained the cross sections
for $\pi \Lambda \rightarrow {\bar K}N$ and $\pi \Sigma 
\rightarrow {\bar K}N$ which can be as large as a few
mb \cite{ko83}. 

\begin{figure}
\begin{center}
\epsfig{file=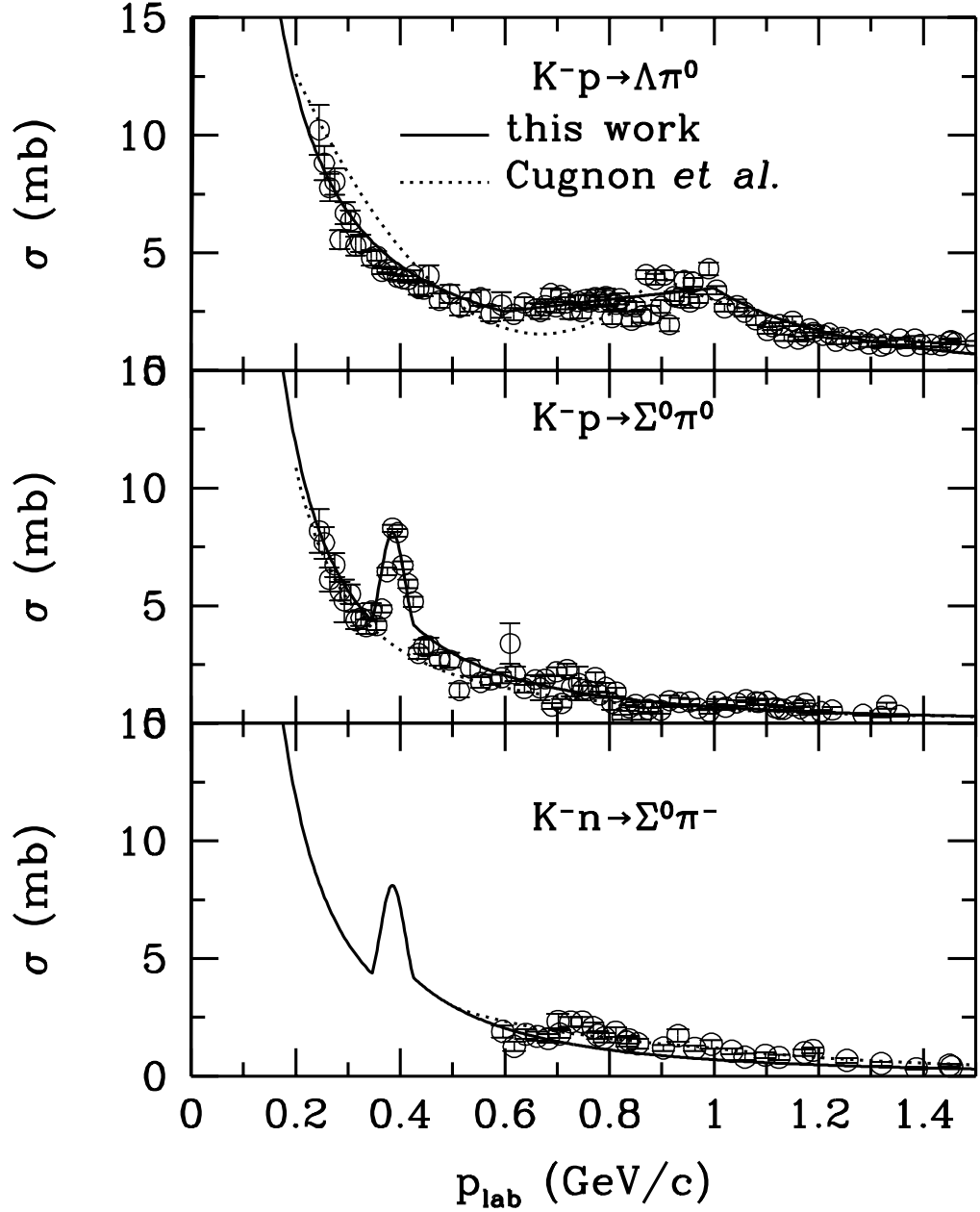,height=7.0in,width=7.0in}
\caption{Cross sections  $\sigma _{K^-N\rightarrow Y\pi}$.
The open circles are experimental data from 
Ref. \protect\cite{data}. The solid lines are the
parameterizations in this work, and the
dotted lines are the parameterizations of
Cugnon {\it et al.} \protect\cite{cugnon90}.
\label{aknypi}}
\end{center}
\end{figure}

\begin{figure}
\begin{center}
\epsfig{file=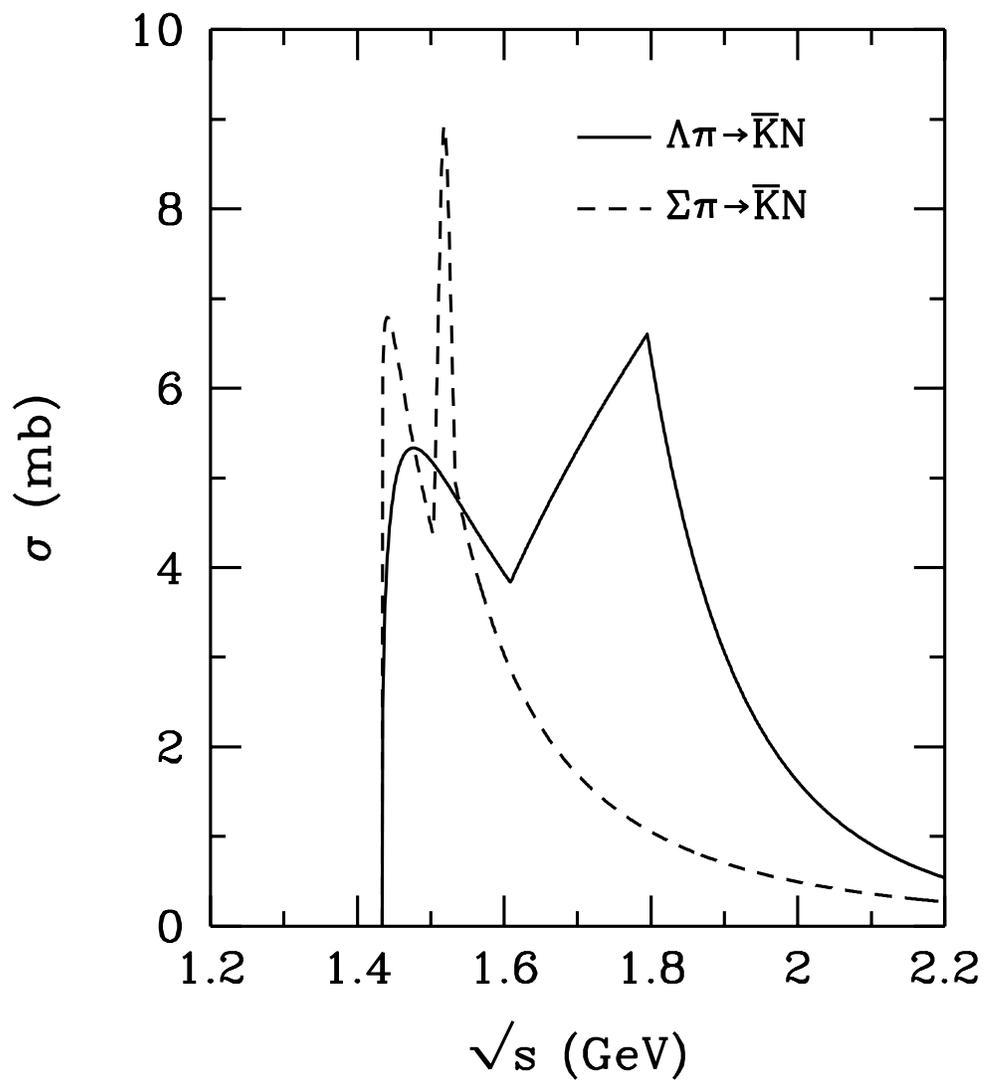,height=7.0in,width=7.0in}
\caption{Cross sections $\sigma _{\pi Y\rightarrow {\bar K}N}$.
\label{ypiakn}}
\end{center}
\end{figure}

In Ref. \cite{cugnon90}, Cugnon {\it et al.} proposed
the following parameterizations for the experimental data,
\begin{eqnarray}
\sigma _{K^- p\rightarrow \Lambda \pi^0} & = & \left\{
\begin{array}{lll}
50p^2-67p+24 \; {\rm mb} & {\rm if} & 
0.2<p<0.9 \; {\rm GeV}\\
3p^{-2.6} \; {\rm mb}   & {\rm if} &
0.9<p<10  \; {\rm GeV/c}\\
\end{array}\right.  ,
\end{eqnarray}
\begin{eqnarray}
\sigma _{K^-p\rightarrow \Sigma^0\pi^0} = 0.6p^{-1.8},
\end{eqnarray}
\begin{eqnarray}
\sigma _{K^- n\rightarrow \Sigma^0 \pi^-} & = & \left\{
\begin{array}{lll}
1.2p^{-1.3} \; {\rm mb} & {\rm if} & 
0.5<p<1 \; {\rm GeV/c}\\
1.2p^{-2.3} \; {\rm mb}   & {\rm if} &
1<p<6  \; {\rm GeV/c}\\
\end{array}\right.  ,
\end{eqnarray}
where $p$ is antikaon momentum in the laboratory frame.
These parameterization do not describe the resonance structure
in the data very well, as will be shown in Fig. \ref{aknypi}. 
We will use the following parameterizations instead,
\begin{eqnarray}
\sigma _{K^- p\rightarrow \Lambda \pi^0} & = & \left\{
\begin{array}{lll}
1.205 p^{-1.428} \; {\rm mb} & {\rm if} & 
p\le 0.6 \; {\rm GeV}\\
3.5p^{0.659} \; {\rm mb}   & {\rm if} &
0.6<p\le 1  \; {\rm GeV/c}\\
3.5p^{-3.97} \; {\rm mb}   & {\rm if} &
p>1  \; {\rm GeV/c}\\
\end{array}\right.  ,
\end{eqnarray}
\begin{eqnarray}
\sigma _{K^- p\rightarrow \Sigma^0 \pi^0} 
\approx  \sigma _{K^- n\rightarrow \Sigma^0 \pi^-} 
& = & \left\{
\begin{array}{lll}
0.624 p^{-1.83} \; {\rm mb} & {\rm if} & 
p\le 0.345 \; {\rm GeV}\\
0.0138/((p-0.385)^2+0.0017) \; {\rm mb}   & {\rm if} &
0.345<p\le 0.425  \; {\rm GeV/c}\\
0.7p^{-2.09} \; {\rm mb}   & {\rm if} &
p>0.425  \; {\rm GeV/c}\\
\end{array}\right.  .
\end{eqnarray}
The comparisons of these parameterizations with experimental
data are shown in Fig. \ref{aknypi}. The dotted lines give the
parameterizations of Cugnon {\it et al.} \cite{cugnon90}.
According to Ref. \cite{cugnon90}, the isospin-averaged cross
sections for the inverse processes are then
\begin{eqnarray}
\sigma _{\pi\Lambda \rightarrow {\bar K}N} =
2\left({p_K\over p_\pi}\right)^2 \sigma _{K^-p\rightarrow \Lambda \pi^0},
\end{eqnarray}
\begin{eqnarray}
\sigma _{\pi\Sigma\rightarrow {\bar K}N}
={2\over 3}\left({p_K\over p_\pi}\right)^2(\sigma _{K^-p\rightarrow
\Sigma^0\pi^0} + \sigma _{K^-n\rightarrow \Sigma ^0\pi^- }),
\end{eqnarray}
where $p_K$ and $p_\pi$ are, respectively, momenta of 
antikaon and pion in the center-of-mass frame.
These cross sections are shown in Fig. \ref{ypiakn}. At low
invariant energies, they are in agreement with those
obtained in Ref. \cite{ko83} based on the K-matrix method.
The first peak in these cross sections comes from the threshold
effect, while the second peak comes from the resonance structure
in the $\bar KN\rightarrow \pi Y$ cross sections.

\subsection{Kaon and antikaon final-state interactions}

Particles produced in elementary hadron-hadron 
interactions in heavy-ion collisions cannot 
escape the environment freely and be detected. Instead, they
are subjected to strong final-state interactions.
For the kaon, because of strangeness conservation,
its scattering with nucleons at low energies is
dominated by elastic and pion production processes,
which do not affect its final yield but changes its momentum 
spectra. Kaon-nucleon scattering was been studied in 
detail experimentally \cite{bland69}. Data
were nicely summarized in Ref. \cite{dover82}.

The experimental data for elastic kaon-nucleon scattering
can be parameterized by
\begin{eqnarray}
\sigma _{KN\rightarrow KN} & = & \left\{
\begin{array}{lll}
10.5+19.7(\sqrt s-\sqrt {s_0})^{1.58} \; {\rm mb} & {\rm if} &
\sqrt s-\sqrt {s_0} \le 0.2 \; {\rm GeV}\\
5.76(\sqrt s-\sqrt {s_0})/(0.056+(\sqrt s-\sqrt {s_0})^2) \;
{\rm mb} & {\rm if} & \sqrt s-\sqrt {s_0} > 0.2 \; {\rm GeV}
\end{array}\right.  ,
\end{eqnarray}
where $\sqrt {s_0} = m_N+m_K$.

\begin{figure}
\begin{center}
\epsfig{file=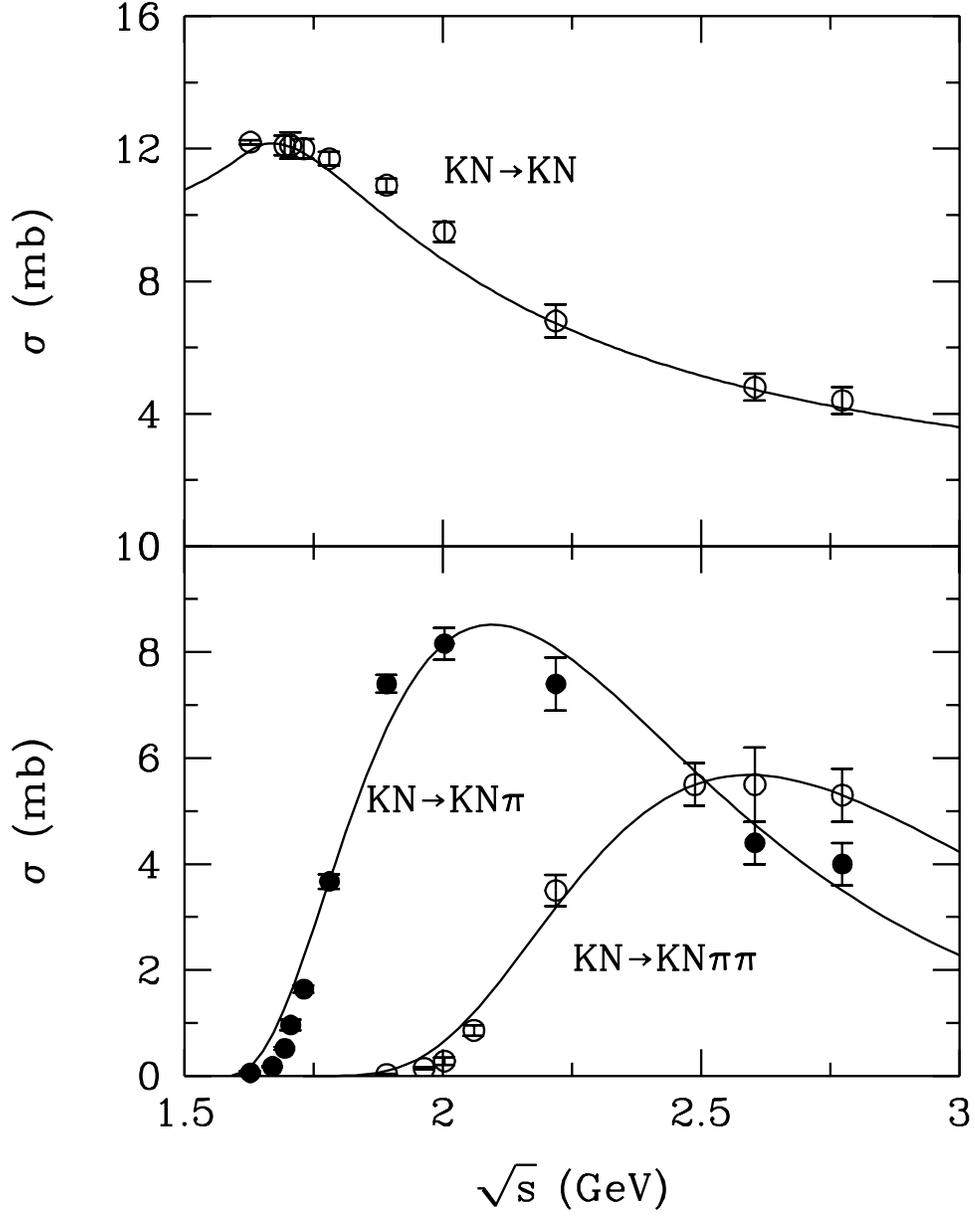,height=7.0in,width=7.0in}
\caption{
$KN$ elastic, one-pion production, and two-pion production
cross sections. The symbols are experimental data
from Ref. \protect\cite{bland69}, and the curves
are parameterizations.
\label{knxs}}
\end{center}
\end{figure}

Similarly, the one-pion ($KN\rightarrow KN\pi$) and two-pion 
production ($KN\rightarrow KN\pi\pi$) cross sections
can be parameterized as
\begin{eqnarray}
\sigma _{KN\rightarrow KN\pi} 
= 1.557 (1-{s_0\over s})^{3.28} ({s_0\over s})^{4.35} \; {\rm b},
\end{eqnarray}
with $\sqrt {s_0} = m_N+m_K+m_\pi$, and
\begin{eqnarray}
\sigma _{KN\rightarrow KN\pi\pi} 
= 12.87 (1-{s_0\over s})^{6.38} ({s_0\over s})^{4.9} \; {\rm b},
\end{eqnarray}
with $\sqrt {s_0} = m_N+m_K+2m_\pi$.
The comparison of these parameterizations with experimental
data is shown in Fig. \ref{knxs}.

\begin{figure}
\begin{center}
\epsfig{file=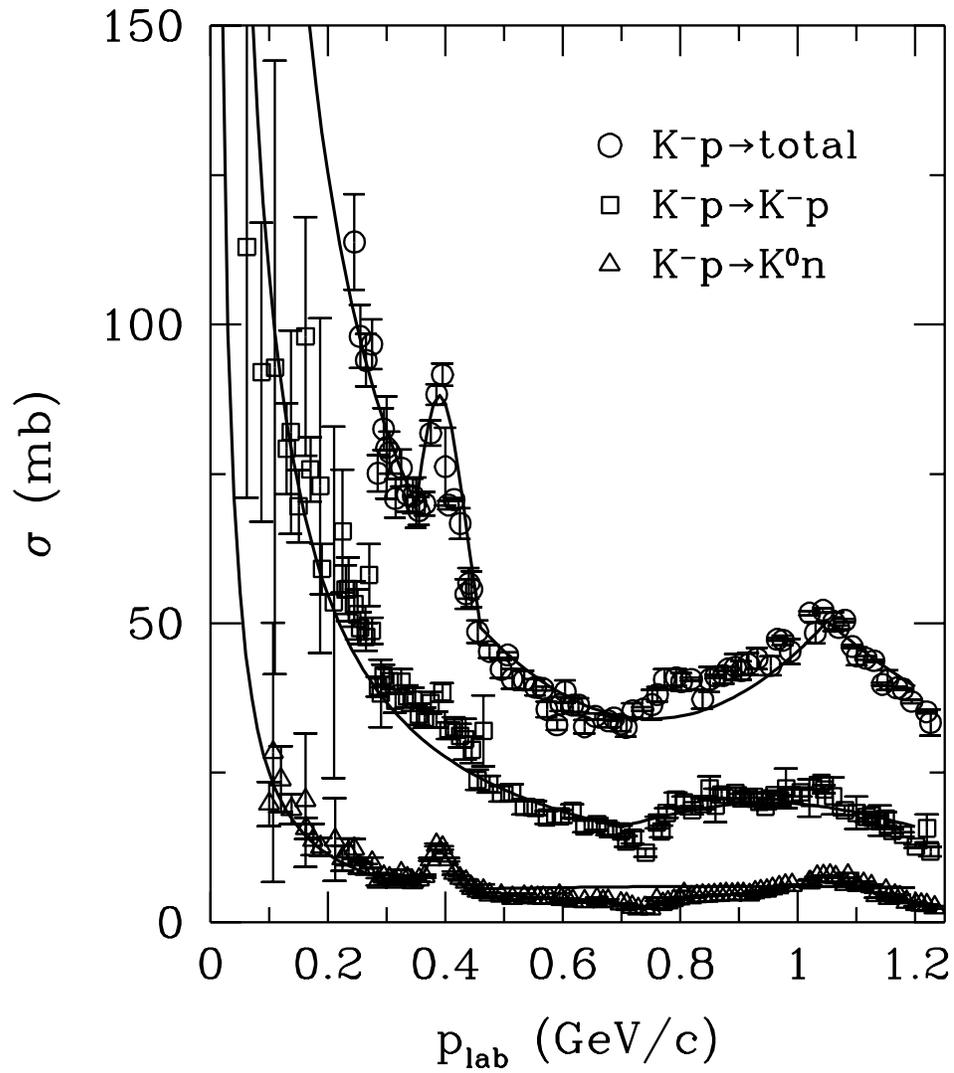,height=7.0in,width=7.0in}
\caption{$K^-p$ total, elastic, and charge-exchange
cross sections. The symbols are experimental data
from Ref. \protect\cite{data}, and the curves
are parameterizations.
\label{akp}}
\end{center}
\end{figure}

\begin{figure}
\begin{center}
\epsfig{file=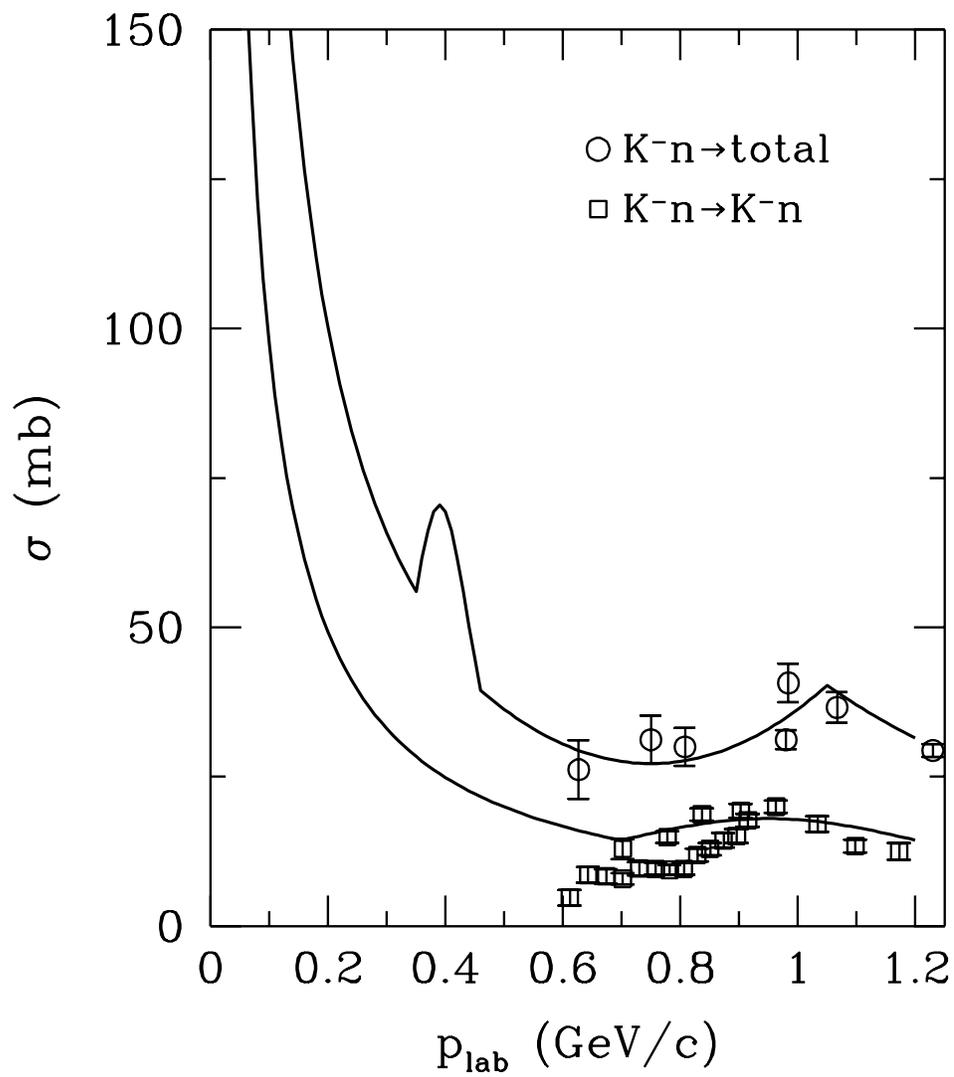,height=7.0in,width=7.0in}
\caption{Same as Fig. \ref{akp}, for $K^-n$ total and
elastic cross sections. 
\label{akn}}
\end{center}
\end{figure}

The final-state interaction for the antikaon is much 
stronger. As mentioned, antikaons can be destroyed in
the strangeness-exchange processes. They also undergo 
elastic scattering. We parameterize the experimental
data for $K^-p$ total, elastic, and charge-exchange
cross sections,
\begin{eqnarray}
\sigma _{K^- p\rightarrow total} & = & \left\{
\begin{array}{lll}
23.5 p^{-1.04} \; {\rm mb} & {\rm if} & 
p\le 0.35 \; {\rm GeV}\\
0.504/((p-0.39)^2+0.0056) \; {\rm mb}   & {\rm if} &
0.35<p\le 0.46  \; {\rm GeV/c}\\
181.9(p-0.75)^2+34.0\; {\rm mb}   & {\rm if} &
0.46<p<1.05  \; {\rm GeV/c}\\
55.2 p ^{-1.85}\; {\rm mb}   & {\rm if} &
p>1.05  \; {\rm GeV/c}\\
\end{array}\right.  ,
\end{eqnarray}
\begin{eqnarray}
\sigma _{K^- p\rightarrow K^-p} & = & \left\{
\begin{array}{lll}
11.2 p^{-0.986} \; {\rm mb} & {\rm if} & 
p\le 0.7 \; {\rm GeV}\\
5.0/((p-0.95)^2+0.25) \; {\rm mb}   & {\rm if} &
p> 0.7  \; {\rm GeV/c}\\
\end{array}\right.  ,
\end{eqnarray}
\begin{eqnarray}
\sigma _{K^- p\rightarrow {\bar K}^0n} & = & \left\{
\begin{array}{lll}
1.813 p^{-1.14} \; {\rm mb} & {\rm if} & 
p\le 0.35 \; {\rm GeV}\\
0.0192/((p-0.39)^2+0.0016) \; {\rm mb}   & {\rm if} &
0.35<p\le 0.43  \; {\rm GeV/c}\\
15.9/((p-0.9)^2+2.65)\; {\rm mb}   & {\rm if} &
p>0.43  \; {\rm GeV/c}\\
\end{array}\right.  .
\end{eqnarray}
where $p$ is the momentum of the antikaon in the laboratory
frame. The comparisions of these parameterizations with the
experimental data are shown in Fig. \ref{akp}. At low
momenta, our parameterizations for the total and elastic cross
sections are very similar to those proposed in Ref. \cite{dover82}.
For the $K^-n$ interaction, experimental data are available only
at relatively large momenta. These data indicate the
following approximate relations,
\begin{eqnarray}
\sigma _{K^-n\rightarrow total} \approx
0.8 \sigma _{K^-p\rightarrow total},
\end{eqnarray}
\begin{eqnarray}
\sigma _{K^-n\rightarrow K^-n} \approx
0.9 \sigma _{K^-p\rightarrow K^-p}.
\end{eqnarray}
The comparision of these relations with the experimental
data is shown in Fig. \ref{akn}.

We treat the charge-exchange process $K^-p\leftrightarrow 
{\bar K}^0n$ as `elastic', since we do not have
explicit isospin degrees of freedom in our transport model.
Under these approximations, the isospin-averaged ${\bar K}N$ 
elastic cross section is given by
\begin{eqnarray}
\sigma_{{\bar K}N\rightarrow {\bar K}N} 
\approx 0.5(1.9\sigma _{K^-p\rightarrow  K^-p} 
+\sigma _{K^-p\rightarrow {\bar K}^0n}),
\end{eqnarray}
and the antikaon absorption cross section is given by
\begin{eqnarray}
\sigma _{{\bar K}N\rightarrow absorption}
\approx 0.9\sigma _{K^-p\rightarrow total}-\sigma _{{\bar K}N
\rightarrow {\bar K}N}.
\end{eqnarray}
Both the elastic and absorption cross sections increase
rapidly with decreasing antikaon momenta. This will have
strong effects on the final $K^-$ momentum spectra in heavy-ion
collisions.
 
\subsection{Some Discussions}

{\it 1. Inclusive versus exclusive cross sections:}
As mentioned, kaon (and antikaon) can be produced in final
states with zero, one, and more pions. To calculate
their total production probability, inclusive cross
sections such as $pp\rightarrow K^+X$ and $pp\rightarrow
K^-X$ should be used. On the other hand, to treat their
final-state interactions and to calculate their spectra,
exclusive cross sections with specific final states,
such as $pp\rightarrow p\Lambda K^+$ and $pp\rightarrow p\Lambda \pi K^+$
are needed. For $NN$ collisions, we have shown that
by including exclusive processes with zero, one, and two pions
in the final states, the inclusive cross sections for
both $K^+$ and $K^-$ production with invariant energy
up to about 5 GeV can be saturated. For $\pi N$ interaction,
we have included exclusive processes with zero and one
pion in final states. They are expected to saturate the inclusive
cross sections with invariant energy up to about 3 GeV. 
Thus, as far as the heavy-ion collisions at SIS energies are concerned,
we have included sufficient exclusive processes 
for both the $NN$ and $\pi N$ interactions.

{\it 2. Total versus differential cross sections:}
So far we have discussed only total cross sections for 
kaon and antikaon production in a specific channel, which give 
their production probabilities. To calculate their spectra
and to treat their final-state interactions, we need also
to determine their momenta. We need thus also differential
cross sections, such as momentum spectra and angular 
distributions. There are far less experimental data on differential
cross sections than on total cross sections. Model calculations
trying to describe both the differential and total cross sections
are also far more involved \cite{laget91} than the one-boson-exchange
model of Ref. \cite{liko95} that attempted at total cross sections
only. 

\begin{figure}
\begin{center}
\epsfig{file=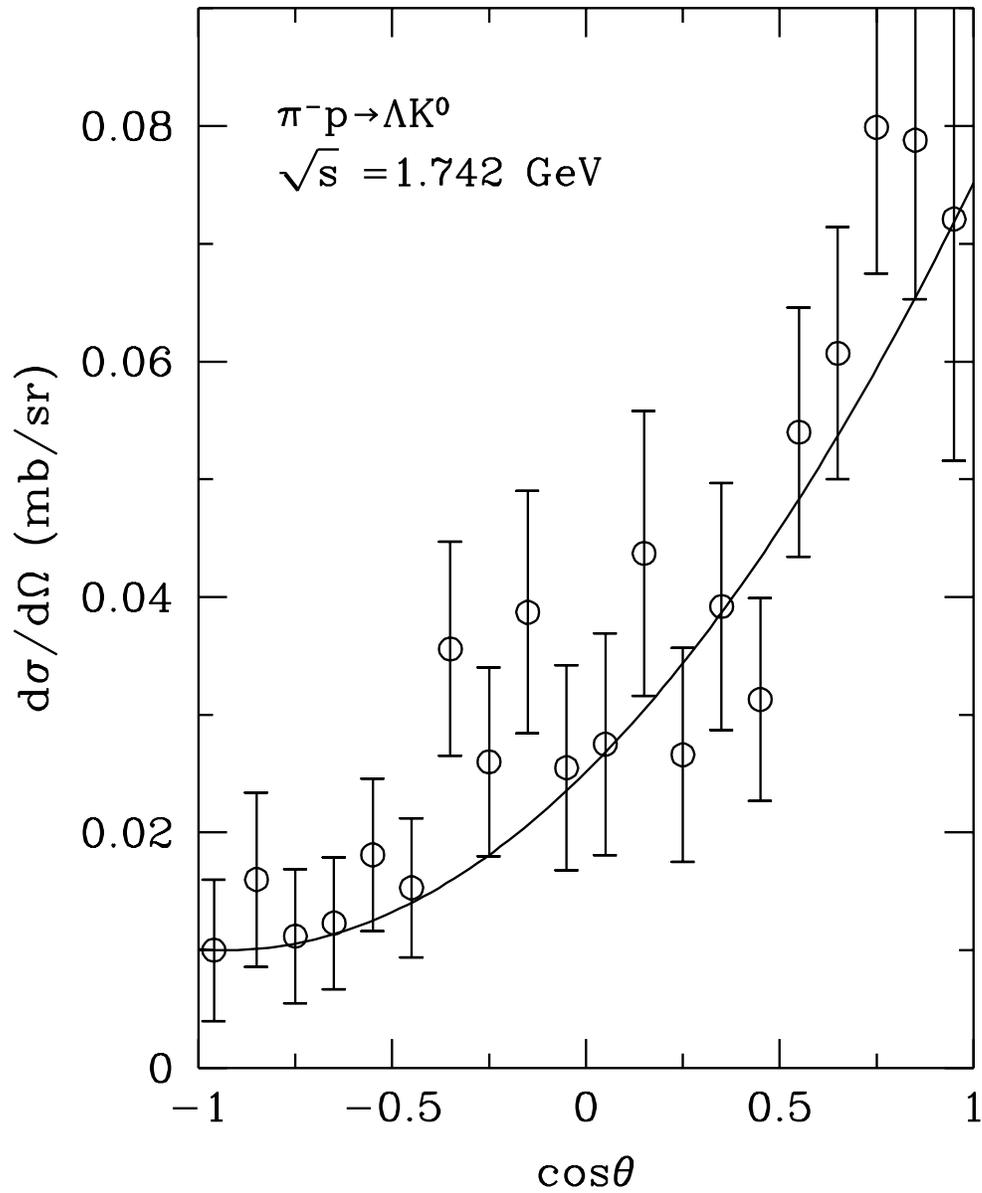,height=7.0in,width=7.0in}
\caption{Angular distribution in $\pi^- p \rightarrow
\Lambda K^0$ at $\sqrt s=1.742$ GeV. The experimental
data are taken from Ref. \protect\cite{kna75}, while the
solid curve is our parameterization.
\label{pindiff}}
\end{center}
\end{figure}

For two-body final states, such as $\pi N\rightarrow YK$, the
magnitude of the kaon momentum in the $\pi N$ center-of-mass
frame is fixed. We thus need only angular distribution
to determine the direction of the momentum. In all the transport
model calculations, this is assumed to be isotropic in the
$\pi N$ center-of-mass frame. We will also use this 
assumption in this work. It is expected that, because of
strong final-state interaction, the use of an anisotropic
angular distribution will not have significant effects on
the final kaon momentum spectra. 
To see more quantitatively the effects of angular distribution,
we will take $\pi N\rightarrow YK$ as an example, where 
limited experimental data are available \cite{kna75}. 
In Fig. \ref{pindiff} we show the angular distribution 
for $\pi^- p\rightarrow \Lambda K^0$ at $\sqrt s=1.742$ GeV. 
The open circles are experimental data, and the solid curve 
is our parameterization,
\begin{eqnarray}\label{pinang}
{d\sigma \over d\Omega }=0.01+0.0175 ({\rm cos}\theta +0.93)^2 ~
{\rm mb/sr}.
\end{eqnarray}
We neglect the beam energy dependence of the angular distribution.
The results for kaon final spectra using this angular distribution
will be compared with those using isotropic assumption.

\begin{figure}
\begin{center}
\epsfig{file=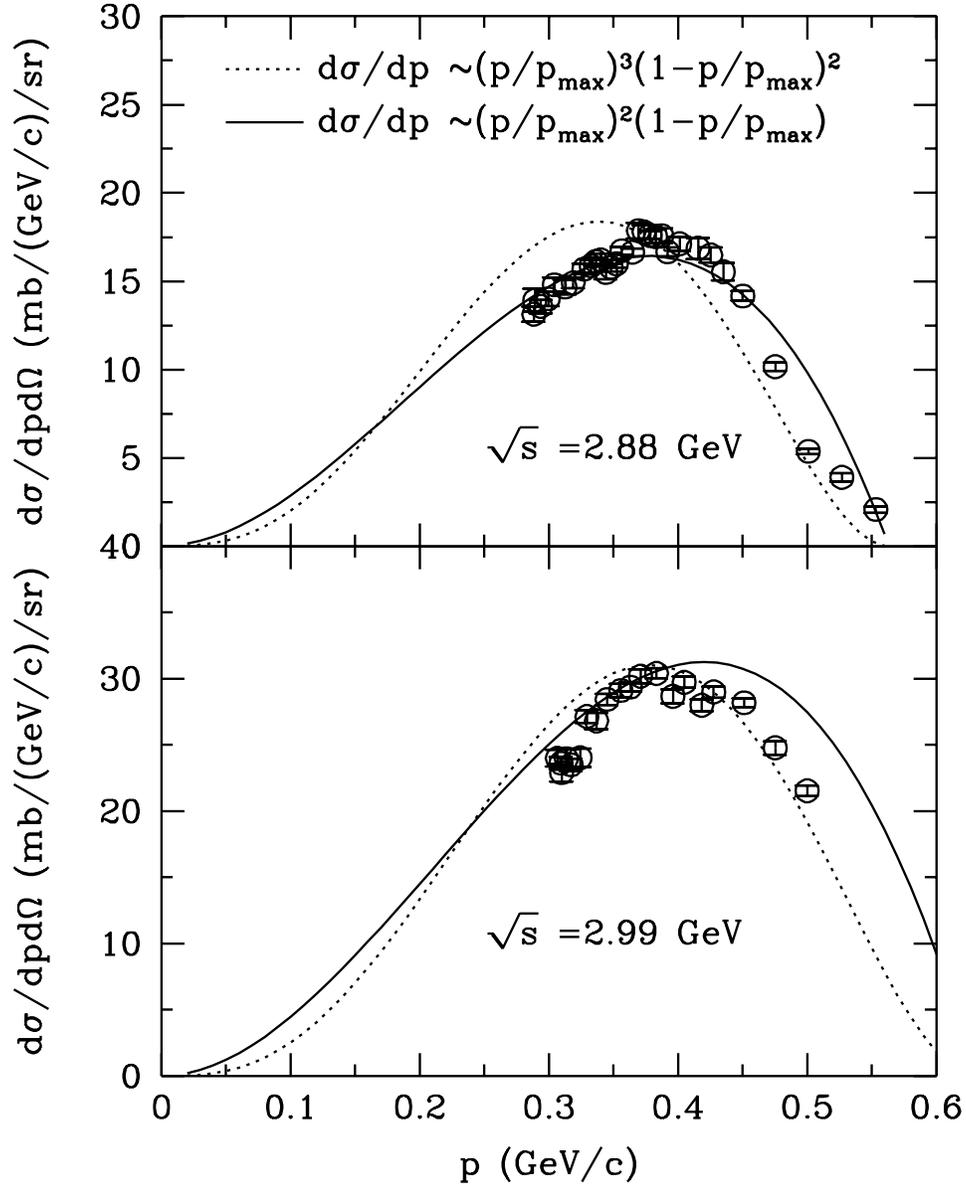,height=7.0in,width=7.0in}
\caption{Kaon momentum spectra in $pp$ collisions at 
$\sqrt s=2.88$ and 2.99 GeV. The experimental
data are taken from Ref. \protect\cite{hogan68}, while the
the curves are parameterizations.
\label{ppkdiff}}
\end{center}
\end{figure}

For three-body final states, such as $pp\rightarrow p\Lambda K^+$,
only the maximum momentum of the kaon is fixed by
energy-momentum conservation. Usually one first determine
the magnitude of the kaon momentum in the $pp$ center-of-mass
frame by using some sort of momentum spectra. 
Randrup and Ko proposed the following momentum spectra, which has
been frequently used in transport models,
\begin{eqnarray}\label{rkmom}
{d\sigma \over dp} \sim \left({p\over p_{max}}\right)^2
\left(1-{p\over p_{max}}\right),
\end{eqnarray}
where $p_{max}$ is the maximum momentum of the kaon in 
the $NN$ center-of-mass frame. This parameterization
is seen to describe the experimental data \cite{hogan68} 
quite well, as can be seen from Fig. \ref{ppkdiff}, which will also be
used in the present work. To see the effects
of elementary kaon momentum spectra on the final
kaon momentum distribution in heavy-ion collisions, we will also
use a somewhat different parameterization, which also provides
a reasonable description of the data,
\begin{eqnarray}\label{limom}
{d\sigma \over dp} \sim \left({p\over p_{max}}\right)^3
\left(1-{p\over p_{max}}\right)^2.
\end{eqnarray}
This parameterization is also compared with the data in
Fig. \ref{ppkdiff}.
  
{\it 3. In-medium versus free-space cross sections:} 
As cab seen from last subsections, we have parameterized 
all the elementary production cross sections in terms
of $\sqrt s$ and $\sqrt {s_0}$, where $\sqrt s$ is the
available energy and $\sqrt {s_0}$ is the threshold.
In nuclear medium, kaon and antikaon masses are modified,
so are their production thresholds. We will then use
$\sqrt {s_0^*}$, which are calculated with effective
masses, in evaluating the in-medium production cross sections.
This amounts to the change of threshold, or approximately,
to the change of available phase space. This kind of
treatment of medium effects is certainly incomplete. In 
nuclear medium, not only hadron masses, but also coupling
constants, cut-off parameters, and even the structure
of the cross sections might change. Nevertheless,
the change of the thresholds, or phase space, is the most apparent
and the simplest to implement. To study other medium effects,
one needs a complete theory for the elementary cross sections.

Since the $K^+$ mass increases and the $K^-$ mass decrease in
nuclear medium, the $K^+$ production cross sections are suppressed
while those of $K^-$ are enhanced in heavy-ion collisions, when
medium effects on their masses are included. This leads to an
increase of $K^-/K^+$ ratio with nuclear density, as illustrated
schematically in Ref. \cite{weise96}. This increase is 
partly responsible for the observed $K^-/K^+$ ratio by the
Kaos collaboration \cite{kaos}. We will come back to this point
later. 

In addition to the change in the production cross sections,
the medium effects on kaon and  antikaon also affect their 
momentum spectra, when they propagate in the mean-field potentials.
The Hamilton equations of motion for kaon and antikaon are very
similar to those for nucleons,
\begin{eqnarray}
{d{\bf r}\over dt} = {{\bf k}\over \omega _{K,{\bar K}} \mp b_k\rho _N},
\;\; {d{\bf k}\over dt} = - \nabla _x U_{K,{\bar K}},
\end{eqnarray}
where minus sign corresponds to kaon, and plus sign to antikaon.
It is clearly that the $K^+$ momentum increases and that of $K^-$
decreases when they propagate in their respective mean field potentials.
This is affect significantly their momentum spectra, and especially
the momentum spectra of their ratio. 

\subsection{Relativistic transport model}

Heavy-ion collisions involve very complicated nonequilibrium
dynamics. In order to extract useful information on,
e.g., nuclear equation of state and hadronic properties in
dense matter, it is necessary to use transport models.
At SIS energies, both mean-field and two-body scattering
play important roles in the dynamical evolution of
the system and need to be included. Over more than ten years,
Boltzmann-Uehling-Uhlenbeck (BUU) and similar models based on
either Skyrme-type or Walecka-type effective nucleon-nucleon
interactions have been developed \cite{cass90,koli96,bert88,aich91}. 
In this work we used the relativistic transport model
(RVUU) similar to that developed in Ref. \cite{ko87}
and used in Refs. \cite{fang94,likof94}. Instead of the usual linear
and non-linear $\sigma$-$\omega$ \cite{qhd}
as in Refs. \cite{ko87,fang94,likof94},
we base our model on an effective chiral Lagrangian
recently developed by Furnstahl, Tang, and Serot \cite{fst},
which will also be used in our studies of neutron 
star properties.

\begin{figure}
\begin{center}
\epsfig{file=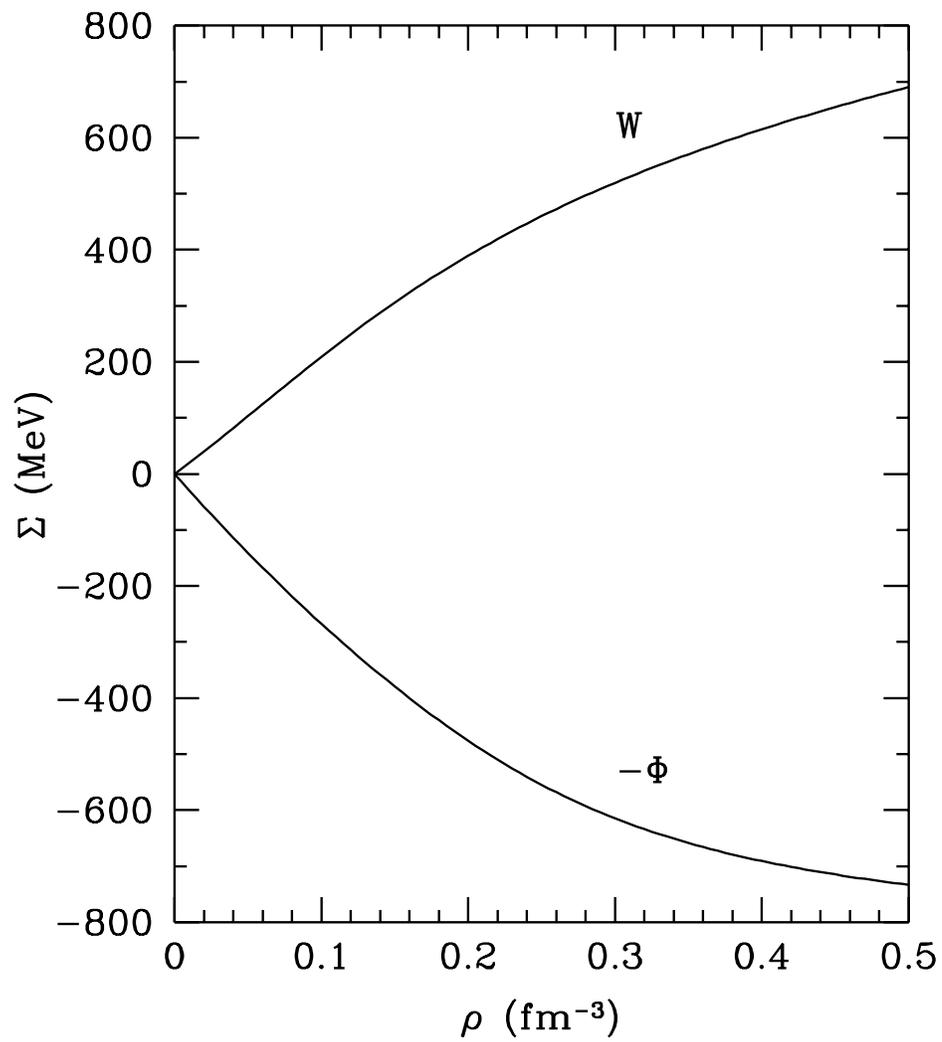,height=7.0in,width=7.0in}
\caption{Nucleon scalar and vector potentials from the
effective chiral Lagrangian of Furnstahl, Tang and 
Serot \protect\cite{fst}.
\label{spot}}
\end{center}
\end{figure}

The effective chiral Lagrangian of Furnstahl, Tang and 
Serot is derived using the dimensional analysis, naturalness 
arguments, and provide a very good description of nuclear 
matter and finite nuclei. In the mean-field approximation, the
energy density for the general case of asymmetric nuclear matter 
is given by
\begin{eqnarray}\label{fst}
\varepsilon & = & {2\over (2\pi )^3} \int _0^{K_{F_p}} 
d{\bf k} \sqrt {{\bf k}^2+m_N^{*2}} +{2\over (2\pi )^3} 
\int _0^{K_{F_n}} d{\bf k} \sqrt {{\bf k}^2+m_N^{*2}} \nonumber\\
 & + & W\rho +R {1\over 2}(\rho_p-\rho_n) -{1\over 2C_V^2}W^2
- {1\over 2C_\rho^2}R^2 + {1\over 2C_S^2}\Phi^2 \nonumber \\
 & +& {S^{\prime 2}\over 4C_S^2}d^2\left\{\left(1-{\Phi \over S^\prime}
\right)^{4/d}\left[{1\over d}{\rm ln}\left(1-{\Phi \over S^\prime}
\right) - {1\over 4}\right]+{1\over 4}\right\}
-{\xi\over 24}W^4 - {\eta \over 2C_V^2}{\Phi \over S^\prime}W^2. 
\end{eqnarray}
In the above the first line represents the `kinetic energy' of 
nucleons, with $K_{F_p}$ and $K_{F_n}$ being proton and neutron
Fermi momenta determined from their densities, $\rho_p$ and $\rho_n$.
The nucleon effective mass $m_N^*$ is related to its scalar
field $\Phi$ by $m_N^*=m_N-\Phi$. In the second line, $W$ (from 
$\omega$ meson exchange) and $R$ (from $\rho$ meson exchange) are the 
isospin-even and isospin-odd vector potentials,
respectively. The latter vanishes for symmetric nuclear matter.
The third line includes the self-interactions of
the scalar field (proportional to the anomalous dimension $d$), 
the vector field (proportional to $\xi$), and the coupling between
them (proportional to $\eta$). The parameters of this model
are adjusted to fit the properties of nuclear matter and
finite nuclei. We will use the parameter set T1 listed in table
1 of Ref. \cite{fst}. The nucleon scalar and vector potentials
for symmetric nuclear matter are shown in Fig. \ref{spot}
as a function of density. Note that because of self-interaction
of the vector field and the coupling between the scalar and
vector fields in this model, the vector potential tends
to saturate at high densities. In the usual linear Walecka model
and non-linear $\sigma$-$\omega$ models \cite{qhd}, the vector
potential increases linearly with density. In Ref. \cite{librown97}
we have shown that the inclusion of higher-order terms for
vector field is instructive in reproducing correctly both the nucleon
flow and dilepton data.

From the energy density, we
can derive a relativistic transport model for heavy-ion
collisions. At SIS energies, the colliding system consists
mainly of nucleons, delta resonances, and pions. While medium
effects on pions are neglected as in most transport models,
nucleons and delta resonances propagate in a common
mean-field potential according to Hamilton equation of
motion,
\begin{eqnarray}
{d{\bf x}\over dt} = {{\bf p}^*\over E^*}, \;\;\;
{d{\bf p}\over dt} = - \nabla _x (E^*+W),
\end{eqnarray}
where $E^*=\sqrt {{\bf p}^{*2} + m^{*2}}$.
In our transport model we have neglected the explicit
isospin degrees of freedom and thus the contributions
from $\rho$ meson exchange. 
In addition to propagations in their mean field potentials, 
we include typical two-body scattering processes such as 
$BB\leftrightarrow BB$, $NN\leftrightarrow
N\Delta$ and $\Delta\leftrightarrow N\pi$. 
Cugnon parameterizations and proper
detailed-balance prescriptions are used for 
describing these reactions \cite{bert88}. 

\subsection{Proton and pion observables}

\begin{figure}
\begin{center}
\epsfig{file=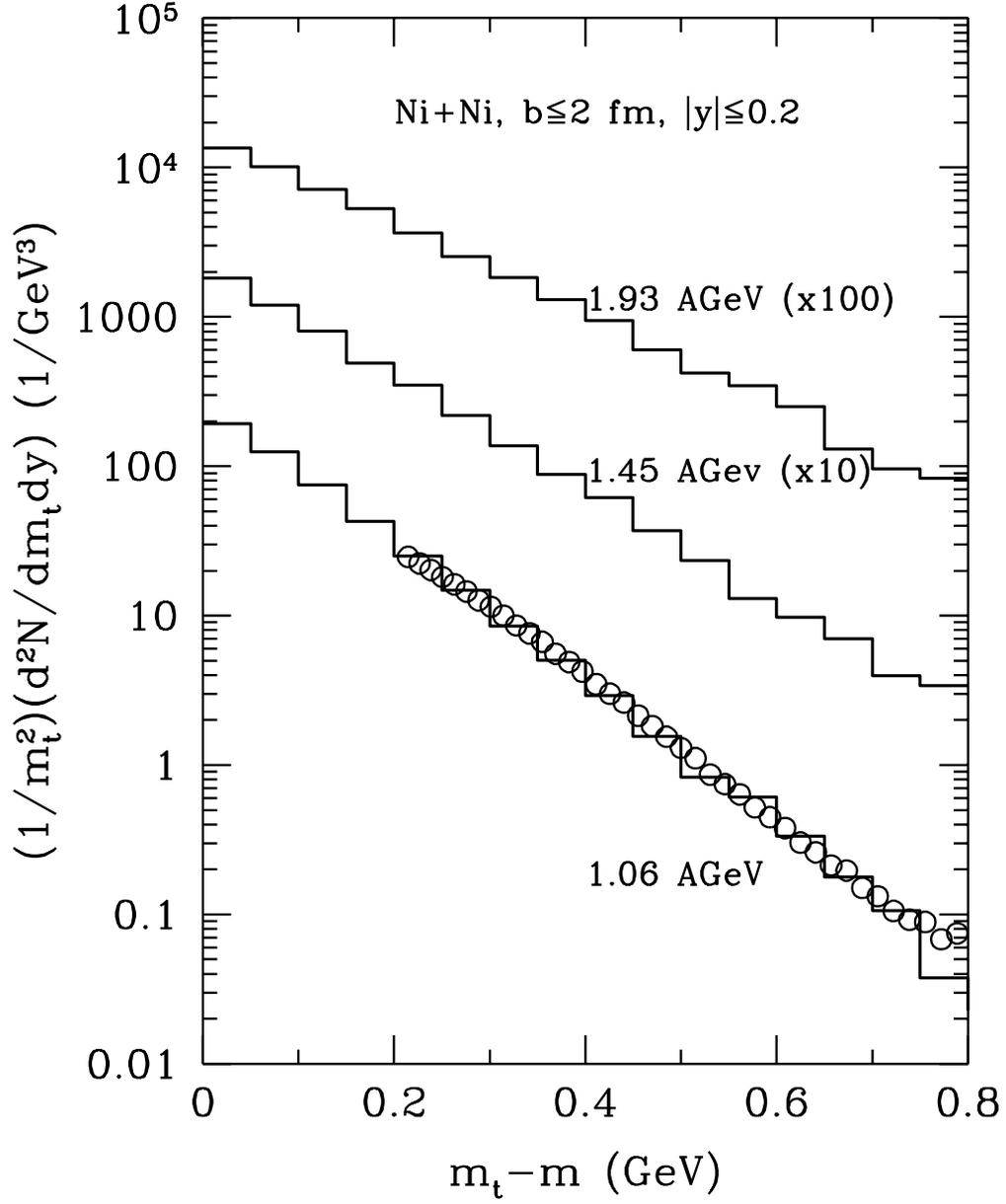,height=7.0in,width=7.0in}
\caption{Proton transverse mass spectra in central
Ni+Ni collisions. The open circles are experimental
data from Ref. \protect\cite{fopi}.
\label{protmtni}}
\end{center}
\end{figure}

To show that our transport model describes reasonably the
nucleon, delta and pion dynamics, we compare our predictions
for proton and pion observables with recent data from
the FOPI \cite{fopi} collaboration.
In Fig. \ref{protmtni} we show proton transverse mass spectra
in the mid-rapidity region in central Ni+Ni collisions 
at 1.06, 1.45, and 1.93 AGeV.
The impact parameter $b\le 2$ fm is chosen so that fair
comparison with FOPI data \cite{fopi}, which has a geometry cross
section of about 100 mb, can be made. It is seen that our
results for 1.06 AGeV agree very well with FOPI data \cite{fopi}.
Our spectra can be fitted by exponential function
exp$(-m_t/T)$. The slope parameter $T$ is found to be
about 95, 115, and 140 MeV at 1.06, 1.45 and 1.93 AGeV,
respectively. These are also in good agreement with what have been
extracted from FOPI data \cite{fopi}.

\begin{figure}
\begin{center}
\epsfig{file=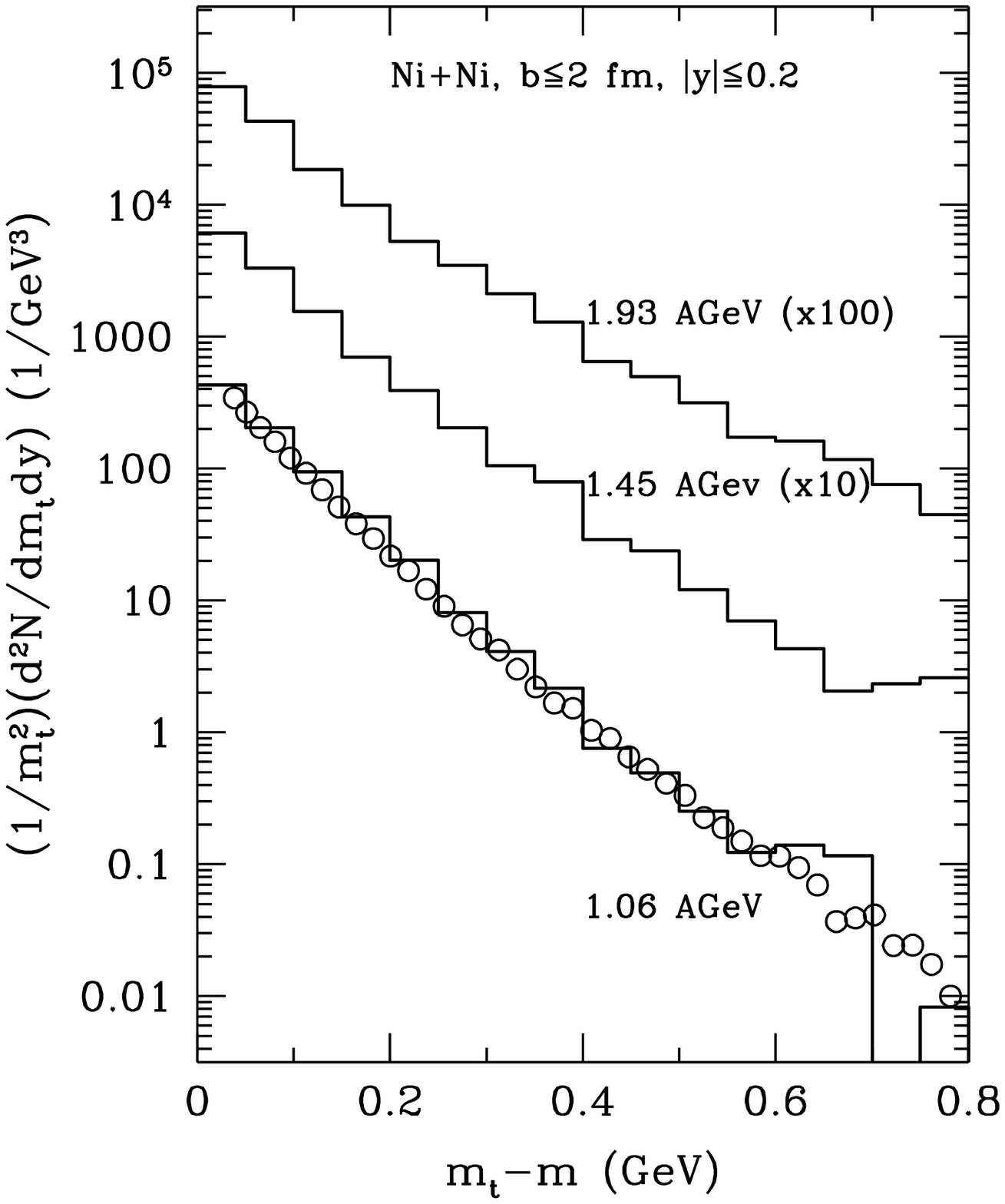,height=7.0in,width=7.0in}
\caption{Same as Fig. \protect\ref{protmtni}, for
$\pi ^-$.
\label{pionmtni}}
\end{center}
\end{figure}

In Fig. \ref{pionmtni} we show our results for $\pi^-$
transverse mass spectra in the same reactions as in Fig. \ref{protmtni}.
Again, our results are in very good agreement with FOPI data
for central Ni+Ni collisions at 1.06 AGeV. The slope parameters
of our pion spectra, fitted from $m_t-m$=0.1 to 0.5 GeV,
are about 70, 85, and, 95 MeV for 1.06, 1.45, and 1.93
AGeV, respectively. The difference between proton and pion slope
parameters can be explained by the collective transverse flow
which has stronger effects on the heavier protons than pions
\cite{peter}.

\begin{figure}
\begin{center}
\epsfig{file=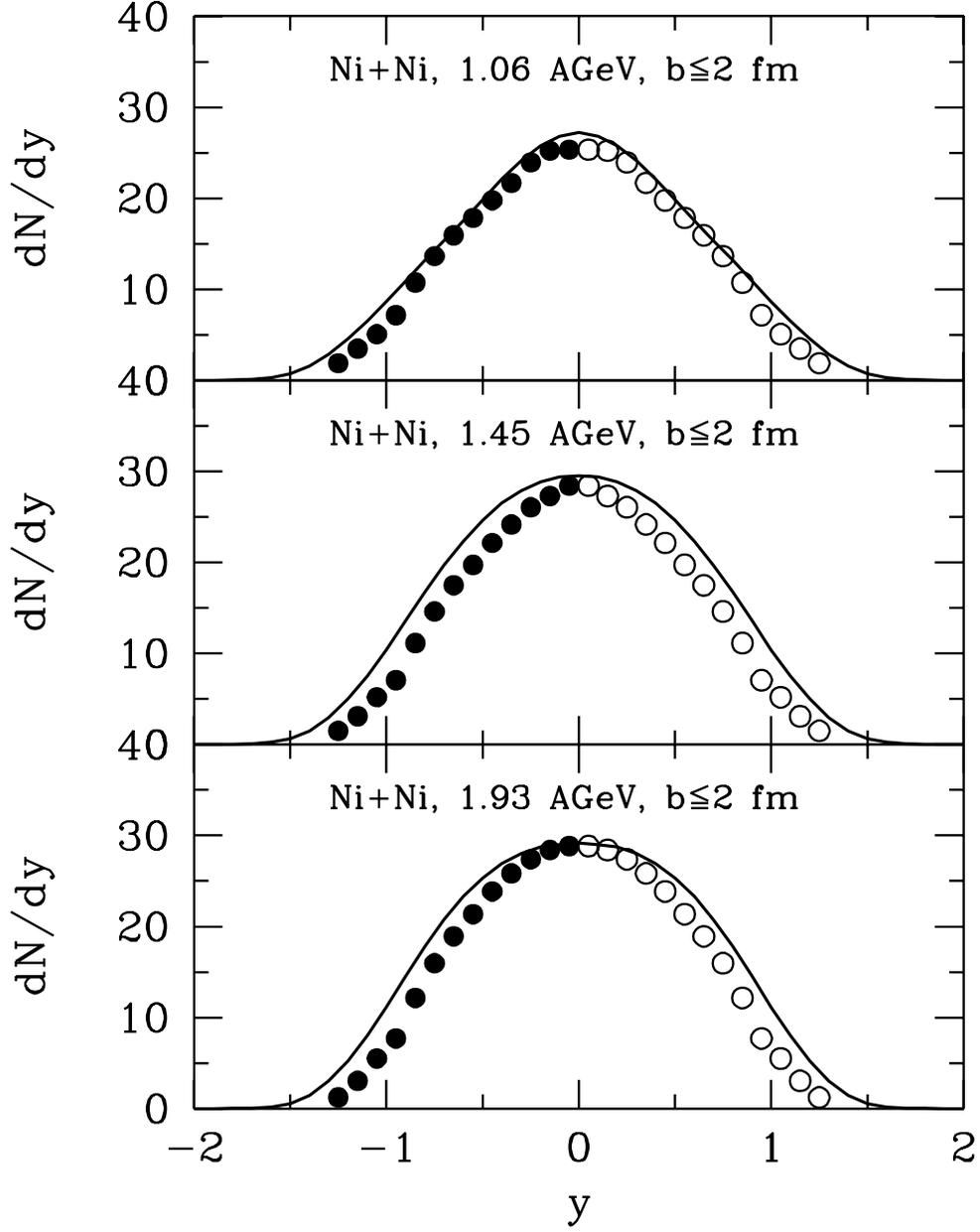,height=7.0in,width=7.0in}
\caption{Proton rapidity distribution in central
Ni+Ni collisions. The solid circles are experimental
data from Ref. \protect\cite{fopi}, the open circles
are reflected with respect to mid-rapidity.
\label{protdyni}}
\end{center}
\end{figure}

\begin{figure}
\begin{center}
\epsfig{file=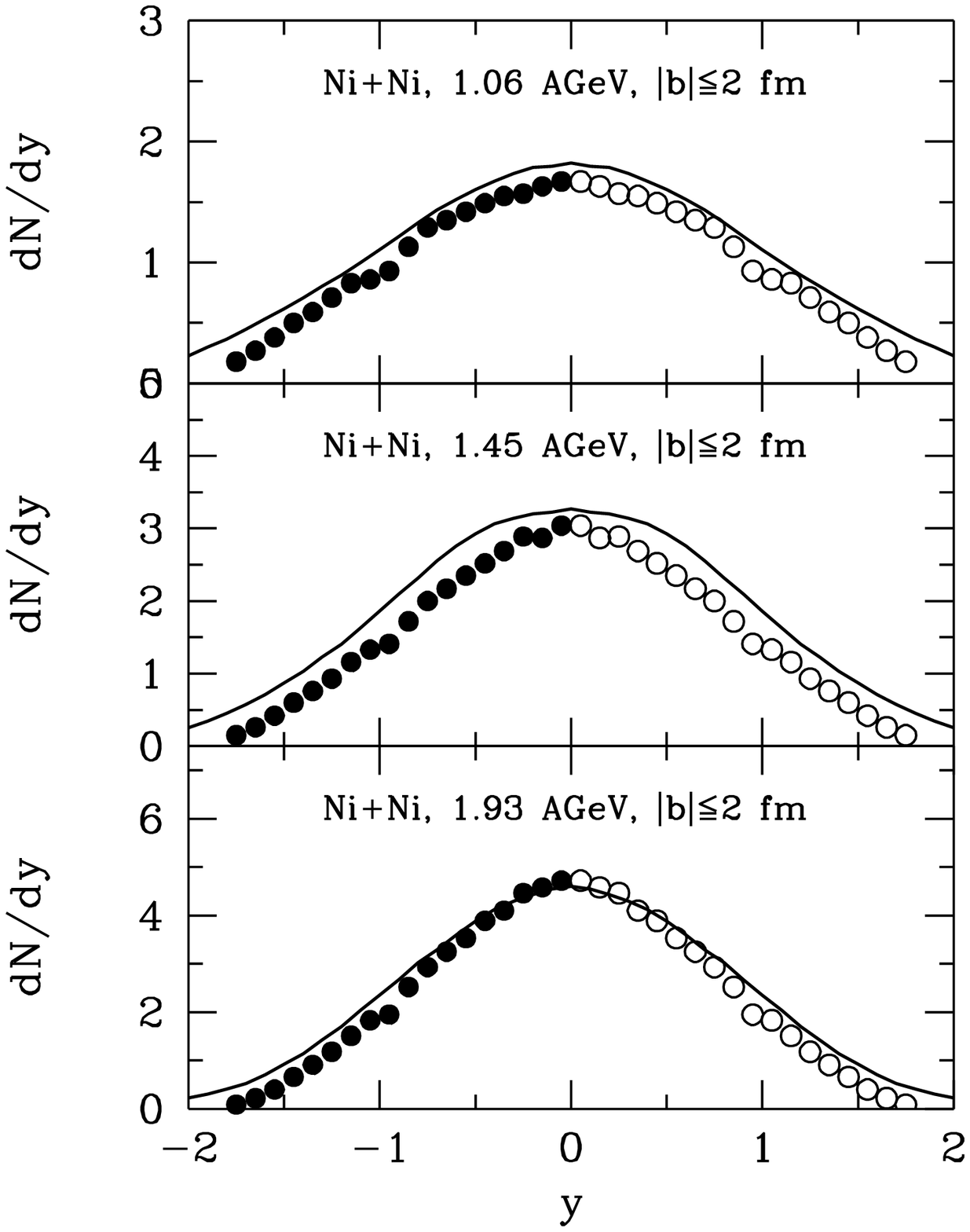,height=7.0in,width=7.0in}
\caption{Same as Fig. \protect\ref{protdyni}, for
$\pi ^-$.
\label{piondyni}}
\end{center}
\end{figure}

In Fig. \ref{protdyni} we compare our predictions
for the proton rapidity distribution in central Ni+Ni
collisions with FOPI data. Here the solid circles
are experimentally measured data, while the open 
circles are obtained from the measured data by
reflecting with respect to the mid-rapidity. As in 
the FOPI data, in this paper $y$ is defined as the
rapidity of a particle in the nucleus-nucleus
center-of-mass frame, normalized by the beam rapidity.
Overall very good agreement with the data have been achieved.
Similar comparisons for $\pi^-$ rapidity
distributions are shown in Fig. \ref{piondyni}. Again, the
agreement with the data is fairly good.
   
\subsection{Kaon and antikaon production in heavy-ion
collisions}

Since the proposal of Aichelin and Ko that subthreshold
$K^+$ yield may be sensitive to nuclear equation of
state at high densities \cite{aich85}, many works
have been done concerning the subthreshold kaon production
in heavy-ion collisions using various transport models
\cite{cass90,koli96,fang94,aich87,li92,lang92,huang93,maru94,aich94}.
There have been also several investigations on
subthreshold antikaon production 
\cite{cass97a,ko83,barz85,huang92,liko94}. In this paper, we
will concentrate on comparisons with recent data from 
the KaoS collaboration for Ni+Ni collisions \cite{kaos}
using the new sets of elementary cross sections as outlined in
subsections A-C. We consider two scenarios, namely, with and 
without kaon medium effects. As mentioned, we use $a_K\approx 0.22$
GeV$^2$fm$^3$ for $K^+$. For $K^-$, we adjust $a_{\bar K}$ such that we 
achieve a good fit to the experimental $K^-$ spectra.
We find $a_{\bar K}\approx 0.45$ GeV$^2$fm$^3$. With
this value and the nuclear scalar density from the model
of Furnstahl, Tang and Serot (the magnitude of the scalar 
attractive depends not only on the value of $a_{K,{\bar K}}$ 
but also on $\rho_S$, which in turn is model dependent 
\cite{sch94}), we find that the $K^-$ feels an attractive 
potential of about 110 MeV at normal nuclear matter
density, which is somewhat smaller than that extracted from
kaonic atoms \cite{gal94}. The $K^-$ production is not, however,
sensitive to the lower densities as those probed in kaonic atoms.
 
\subsubsection{Sensitivity to differential cross sections}

\begin{figure}
\begin{center}
\epsfig{file=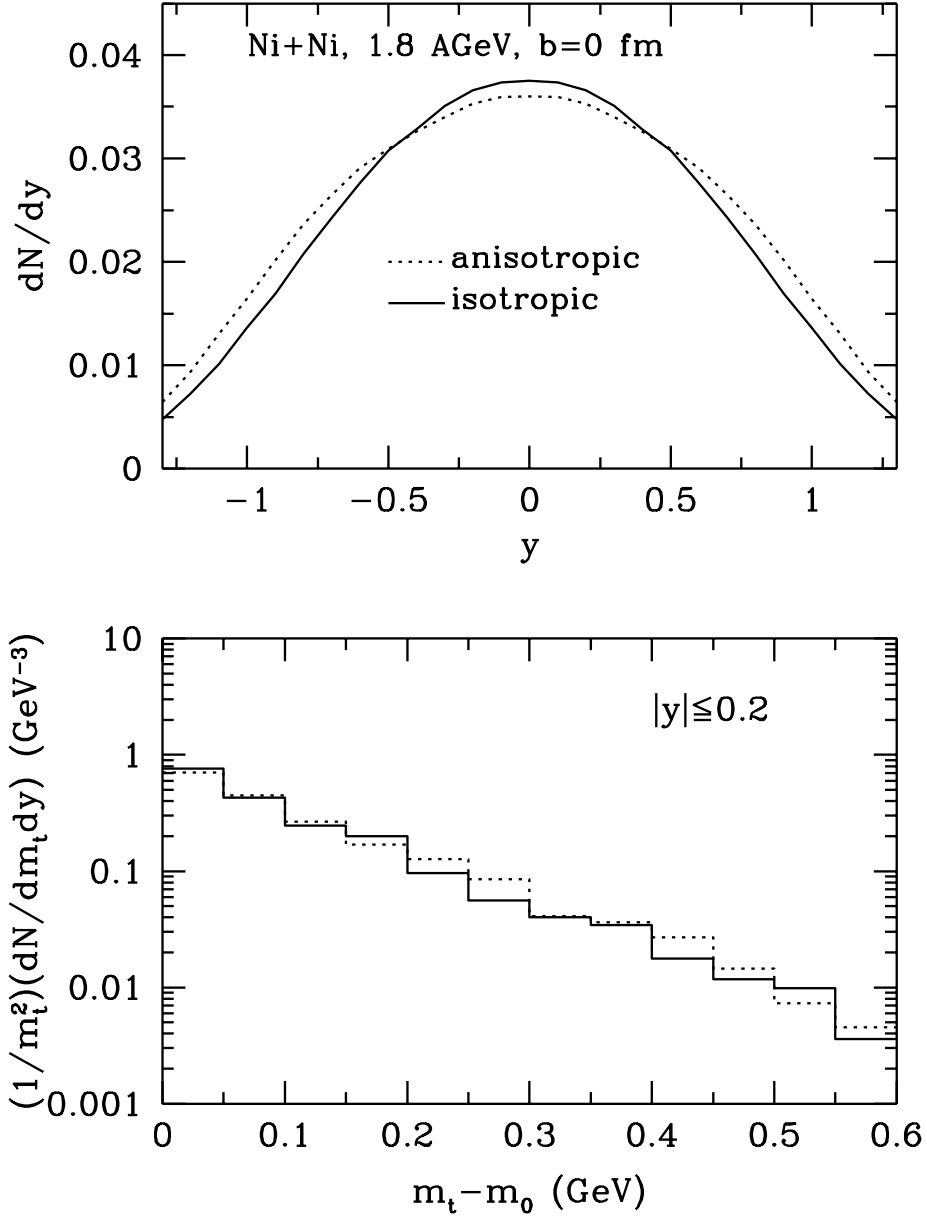,height=7.0in,width=7.0in}
\caption{$K^+$ tranverse mass spectra and rapidity 
distribution in Ni+Ni collisions at 1.8 AGeV and 0 fm. 
The solid and dotted lines are the results with isotropic 
and anisotropic angular distributions, respectively. 
\label{pinmtdydiff}}
\end{center}
\end{figure}

\begin{figure}
\begin{center}
\epsfig{file=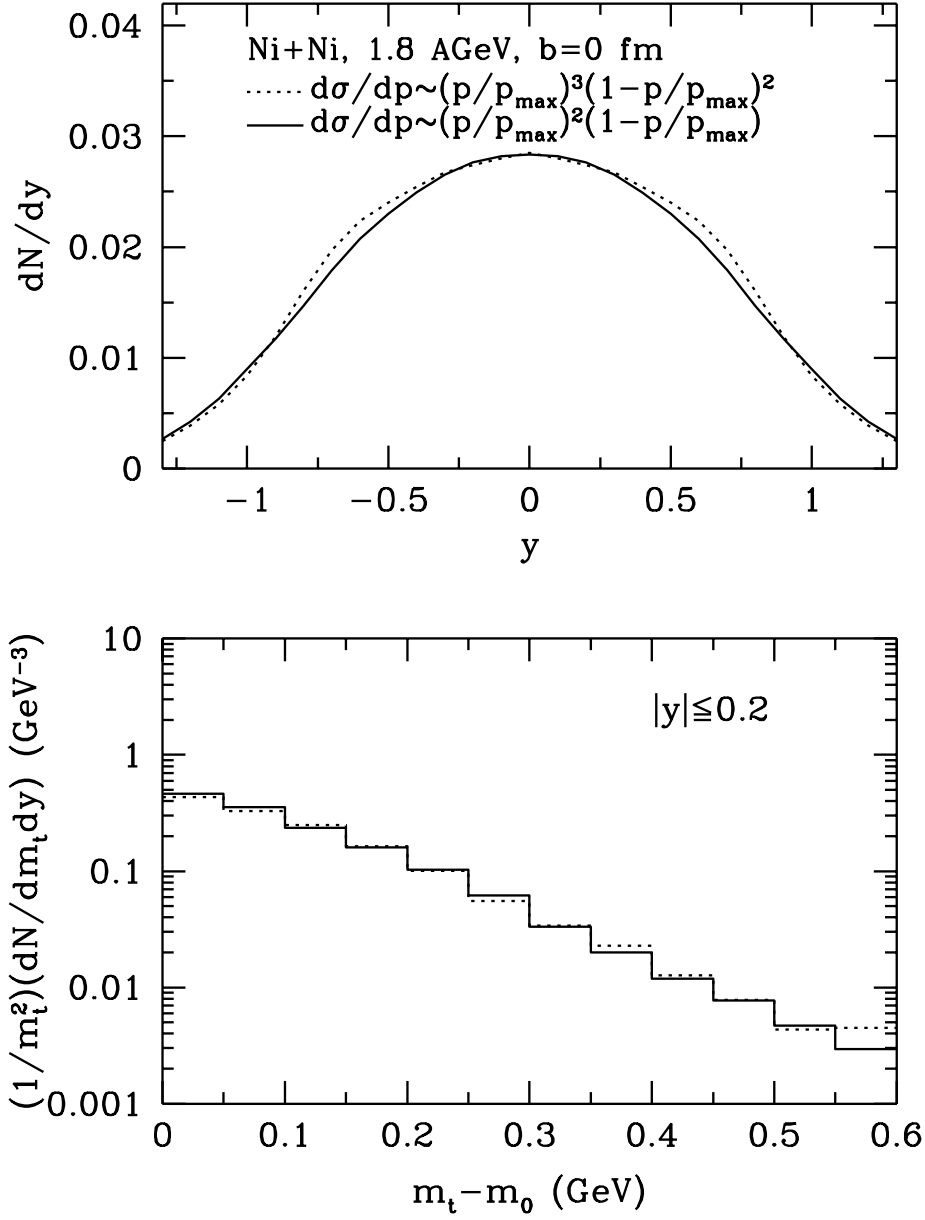,height=7.0in,width=7.0in}
\caption{$K^+$ tranverse mass spectra and rapidity 
distribution in Ni+Ni collisions at 1.8 AGeV and 0 fm. 
The solid and dotted lines are the results using Eq.
(\protect\ref{rkmom}) and Eq. (\protect\ref{limom}) 
as momentum spectra, respectively. 
\label{ppkmtdydiff}}
\end{center}
\end{figure}

Before presenting our main results and comparing with
experimental data, we discuss in this subsection the
sensitivity of final $K^+$ spectra to the differential
cross sections. We calculate $K^+$ transverse mass
spectra and rapidity distribution in Ni+Ni collisions
at 1.8 AGeV and b=0 fm. For the sensitivity to the
angular distribution we take $\pi N\rightarrow YK$ as
an example. In the first case, we assume that kaons are
emitted isotropically in the $\pi N$ center-of-mass
frame, while in the second case, we use the angular
distribution as given by Eq. \ref{pinang}.
The results are shown in Fig. \ref{pinmtdydiff}.
As can be seen, because of rescattering, the final
kaon spectra is not very sensitive to the elementary
angular distribution. For the sensitivity to the momentum
spectra, we take $NN\rightarrow NYK$ as an example. In
the first case we use the Randrup-Ko parameterization,
Eq. \ref{rkmom}, and in the second case we use Eq. \ref{limom}.
The results are shown in Fig. \ref{ppkmtdydiff}.
Again, as far as the elementary momentum spectra
reproduce the experimental data for {\it pp} reasonably,
the resulting $K^+$ spectra in heavy-ion collisions are
very similar.

\subsubsection{$K^+$ and $K^-$ kinetic energy spectra}

\begin{figure}
\begin{center}
\epsfig{file=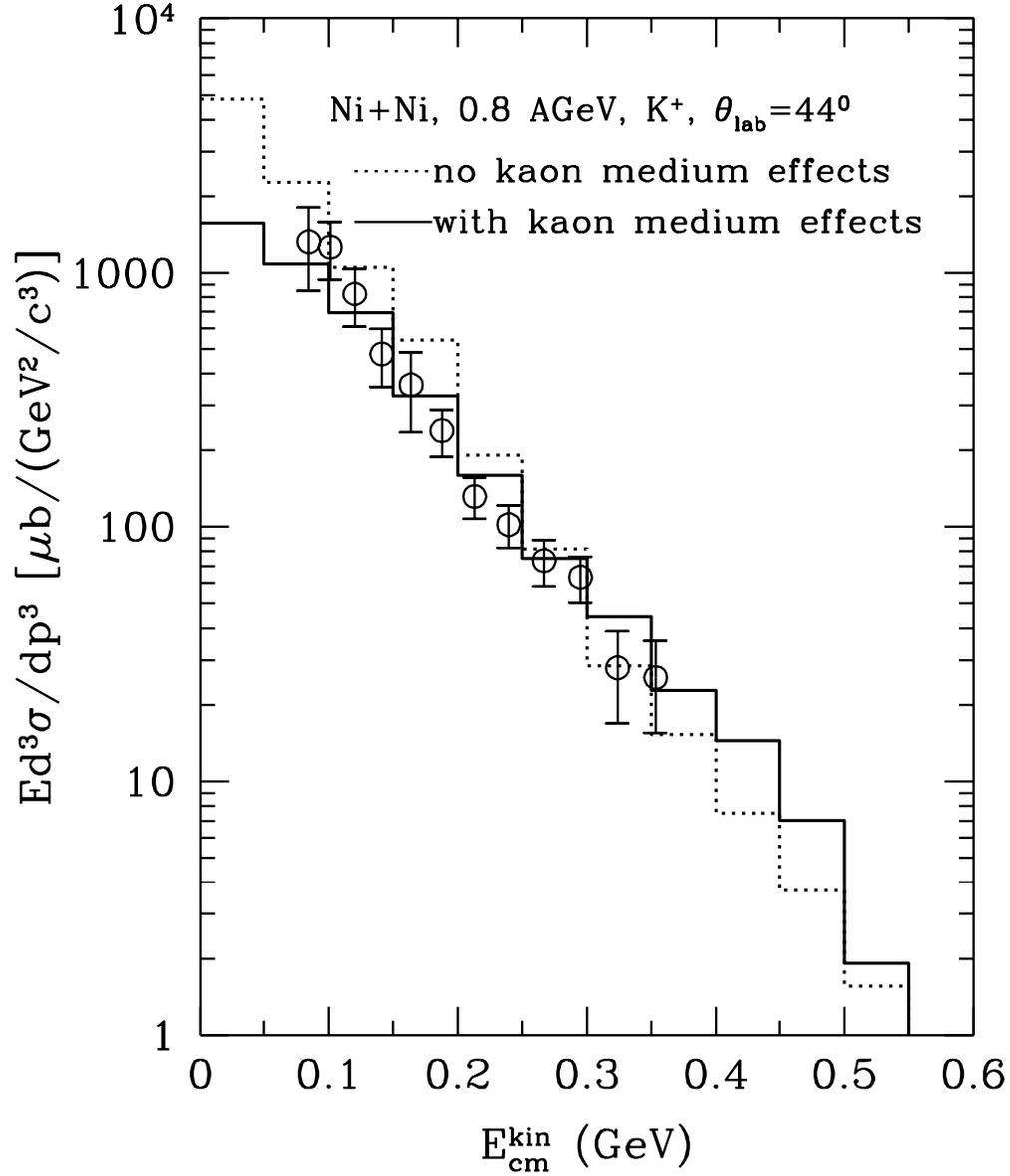,height=7.0in,width=7.0in}
\caption{$K^+$ kinetic energy spectra in Ni+Ni collisions
at 0.8 AGeV. The solid and dotted histograms are the
results with and without kaon medium effects,
respectively. The open circles are experimental
data from Ref. \protect\cite{kaos}.
\label{kaonni08}}
\end{center}
\end{figure}

\begin{figure}
\begin{center}
\epsfig{file=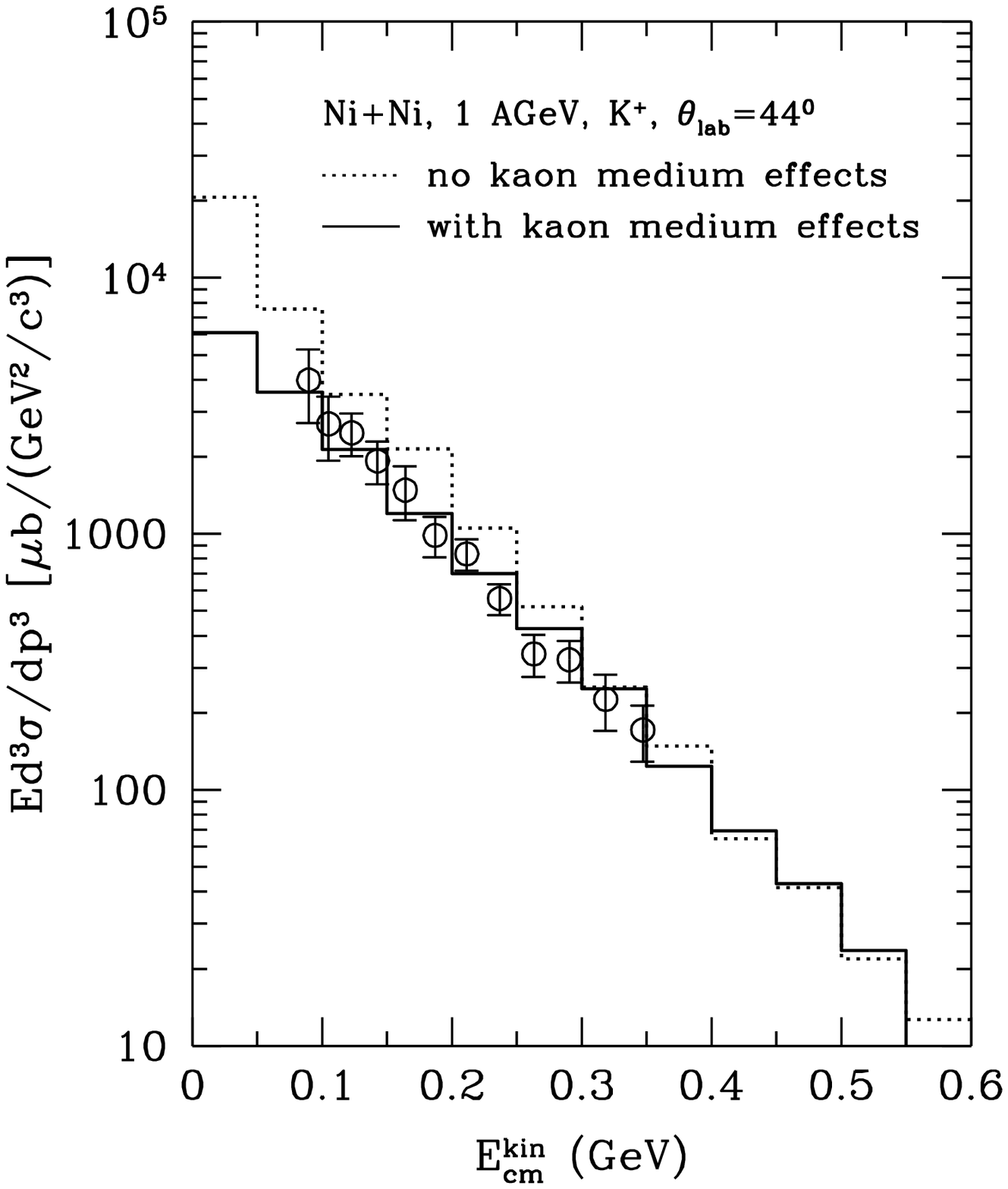,height=7.0in,width=7.0in}
\caption{Same as Fig. \protect\ref{kaonni08}, for
Ni+Ni at 1.0 AGeV.
\label{kaonni10}}
\end{center}
\end{figure}

\begin{figure}
\begin{center}
\epsfig{file=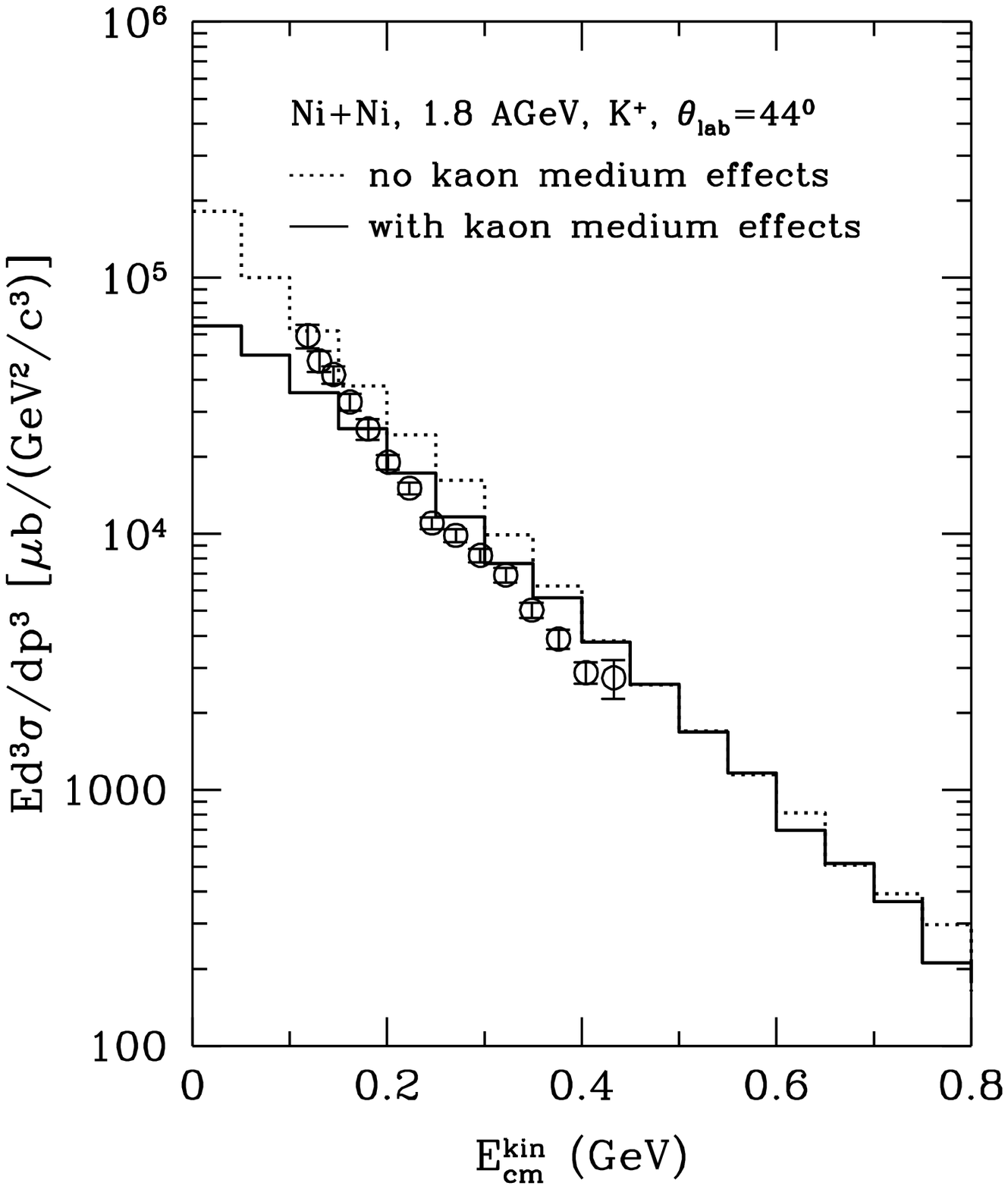,height=7.0in,width=7.0in}
\caption{Same as Fig. \protect\ref{kaonni08}, for
Ni+Ni at 1.8 AGeV.
\label{kaonni18}}
\end{center}
\end{figure}

The results for $K^+$ kinetic energy 
spectra in Ni+Ni collisions at 0.8, 1.0 and 1.8 AGeV are
shown in Figs. \ref{kaonni08}, \ref{kaonni10}, and
\ref{kaonni18}, respectively. The impact range is chosen to
be $b\le 8$ fm. The solid histogram gives
the results with kaon medium effects, while the dotted 
histogram is the results without kaon medium effects. 
The open circles are the experimental data from the KaoS 
collaboration {\cite{kaos}}. It is seen that 
the results with kaon medium effects are in good agreement 
with the data, while those without kaon medium effects slightly 
overestimate the data. We note that kaon feels a small 
repulsive potential; thus the inclusion of the kaon medium 
effects reduces the kaon yield. The slopes of the kaon spectra 
in the two cases also differ. With a repulsive potential, 
kaons are accelerated during the propagation, leading to a 
larger slope parameter as compared to the case without
kaon medium effects.

\begin{figure}
\begin{center}
\epsfig{file=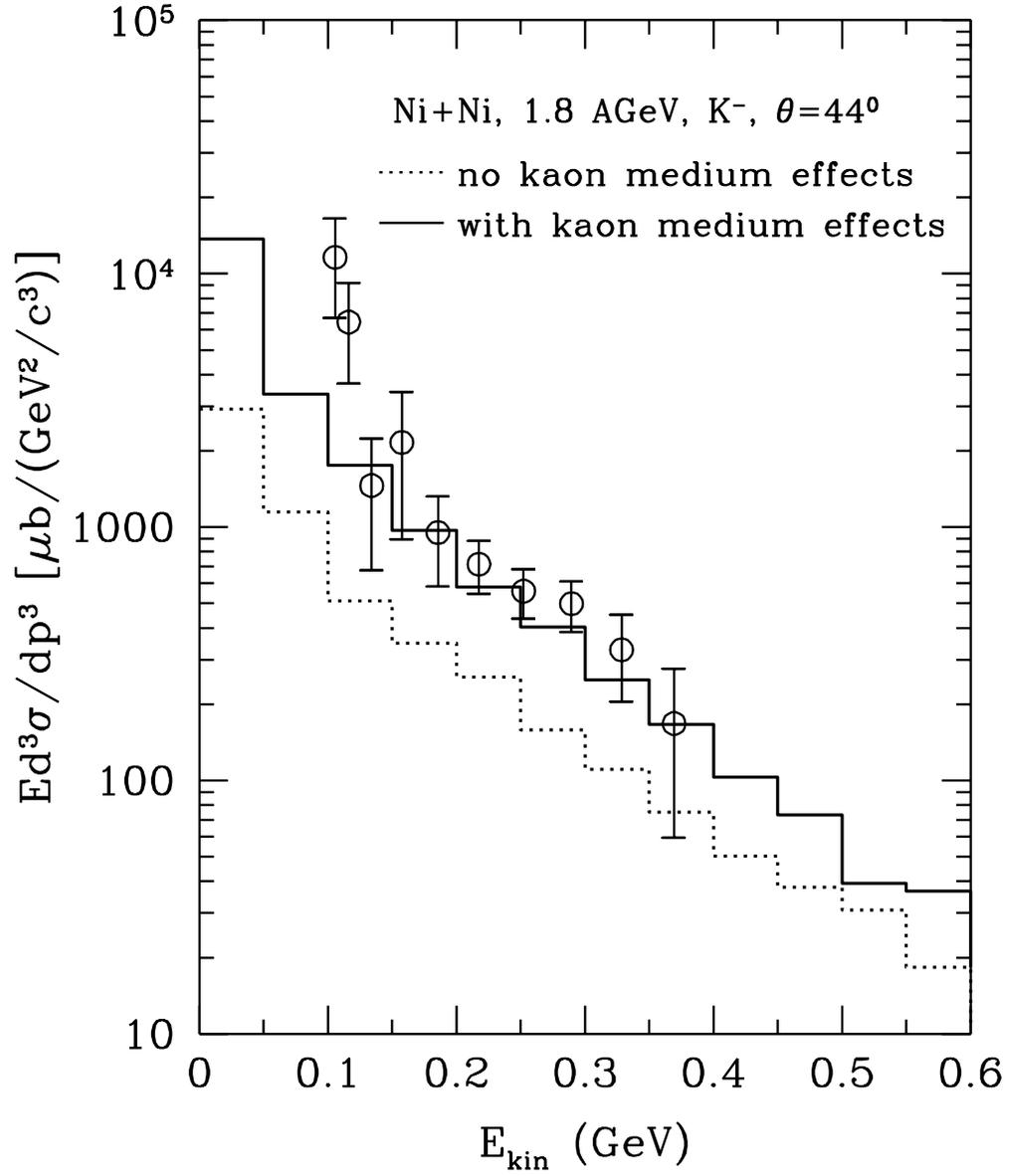,height=7.0in,width=7.0in}
\caption{Same as Fig. \protect\ref{kaonni08}, for
$K^-$ spectra in Ni+Ni at 1.8 AGeV.
\label{akni18}}
\end{center}
\end{figure}

The results for the $K^-$ kinetic energy spectra are shown 
in Fig. \ref{akni18} for Ni+Ni collision at 1.8 AGeV. 
It is seen that without medium effects, our
results are about a factor 3-4 below the experimental data.
With the inclusion of the medium effects which reduces the
antikaon production threshold, the $K^-$ yield increases by about a factor
of 3 and our results are in good agreement with the data. This
is similar to the findings of Cassing {\it et al.} {\cite{cass97a}}.
Also, with the inclusion of kaon medium effects, the slope parameter
of the $K^-$ spectra decreases, since the propagation of
antikaons in their attractive potential reduces their
momenta.

\begin{figure}
\begin{center}
\epsfig{file=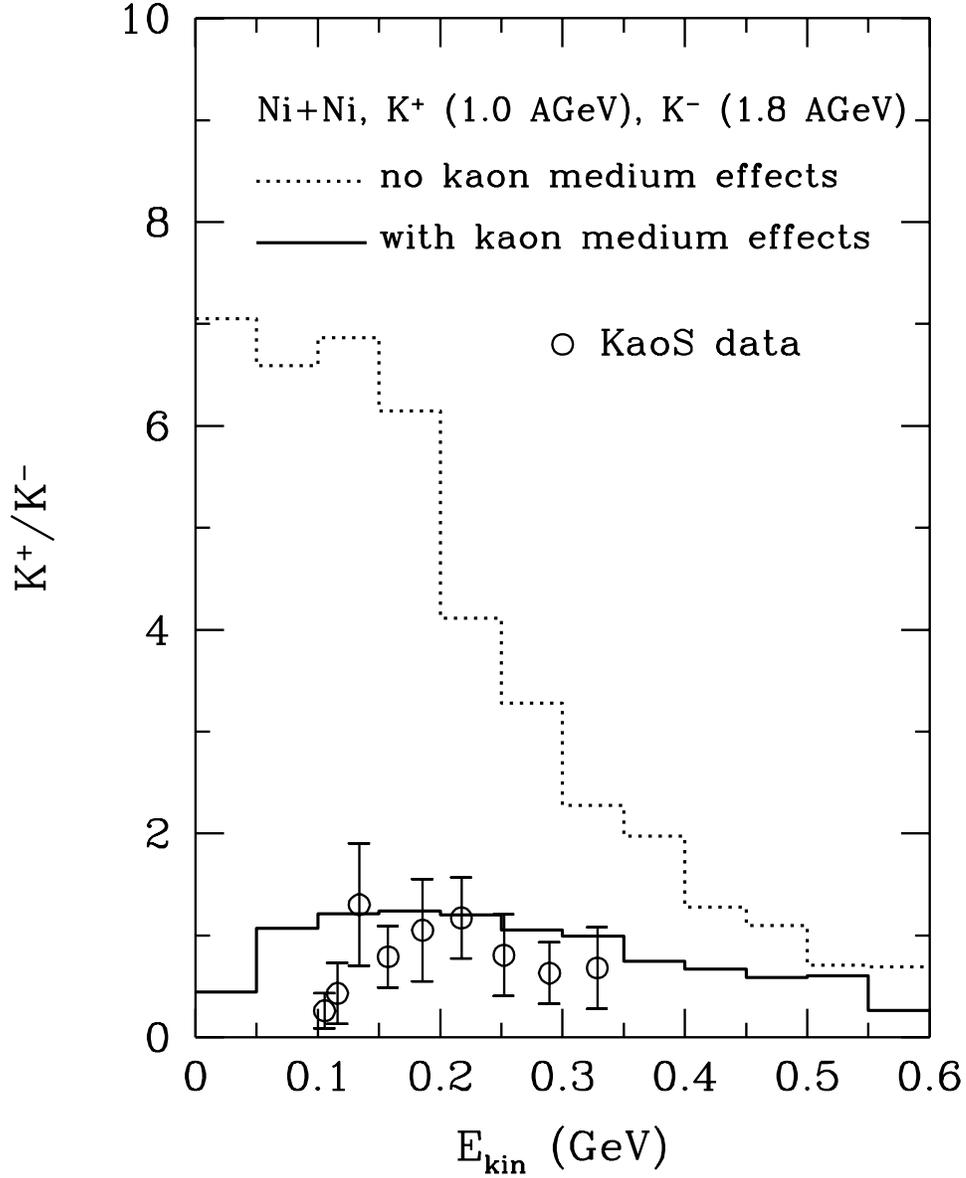,height=7.0in,width=7.0in}
\caption{kinetic energy spectra of $K^+/K^-$
in Ni+Ni collisions at 1.0 AGeV ($K^+$) and
1.8 AGeV ($K^-$). The solid and dotted histograms are the
results with and without kaon medium effects,
respectively. The circles are the experimental data
from the KaoS collaboration \protect\cite{kaos}.
\label{kakr}}
\end{center}
\end{figure}

As mentioned earlier, the KaoS observation that the $K^-$ yield at
1.8 AGeV is similar to $K^+$ yield at 1.0 AGeV is an
indication of the kaon medium effects. This can best be seen
in our results by looking at their ratio as a function
of kinetic energy. This is done in Fig. \ref{kakr}.
By doing this we can also see more clearly the effects 
of kaon and antikaon mean-field potentials on
the shapes of their momentum spectra.
It is seen that without kaon medium effects, the $K^+/K^-$
ratio decreases from about 7 at low kinetic energies to
about 1 at high kinetic energies. In central collisions
the ratio of their total yield is $K^+/K^-\approx 6$,
which is much greater than the KaoS observation of about 1.
Since the antikaon absorption cross section by nucleons
increases rapidly at low momentum, low-momentum
antikaons are more strongly absorbed by nucleons than 
high-momentum ones. This makes the $K^+/K^-$ ratio increase
with decreasing kinetic energies. When medium effects are
included, we find that the $K^+/K^-$ ratio is almost
a constant of about 1 in the entire kinetic energy region,
which is in good agreement with the experimental data from
the KaoS collaboration \cite{kaos}, shown in the figure
by open circles. The ratio between their total yields in the
central collisions is $K^+/K^-\approx 1.1$, very close to
the KaoS observation of about 1. As mentioned,
the shapes of $K^+$ and $K^-$ spectra change in 
opposite ways in the presence of their mean-field potential.
Kaons are `pushed' to high momenta by the repulsive potential,
while antikaons are `pulled' to low momenta. If there
were no antikaon absorption, or if the absorption
cross section were independent of antikaon momentum, 
the $K^+/K^-$ would increase with kinetic energy because
of propagation in the mean-field potential. A `flat' $K^+/K^-$
spectra is thus highly nontrivial, which must be result
from the combined effects of mean-field and energy-dependent
antikaon absorption. An accurate experimental measurement 
of this ratio can thus be very useful in determining whether
there are medium effects or not on kaons in nuclear matter. 

\subsubsection{$K^+$ and $K^-$ transverse mass spectra and
rapidity distributions}

\begin{figure}
\begin{center}
\epsfig{file=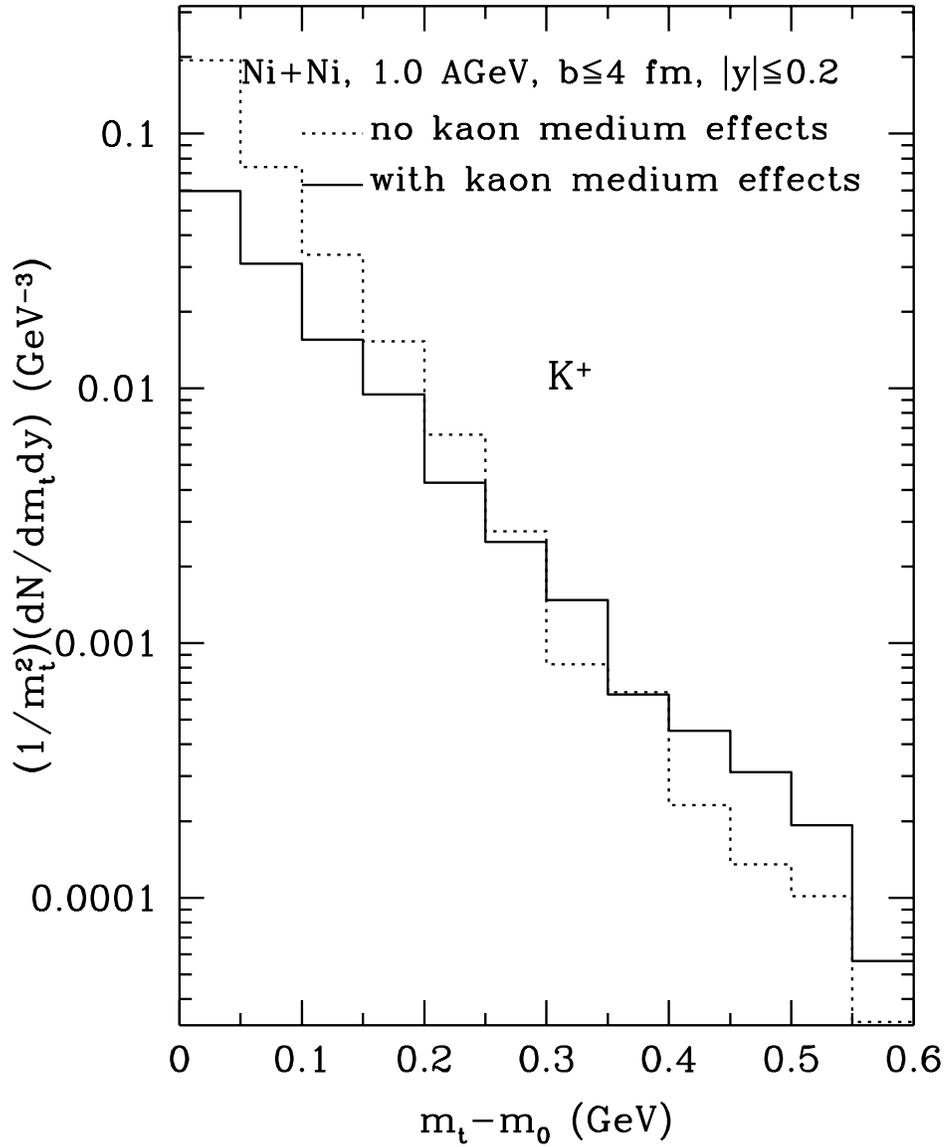,height=7.0in,width=7.0in}
\caption{$K^+$ transverse mass spectra in central Ni+Ni 
collisions at 1.0 AGeV. The solid and dotted histograms are the
results with and without kaon medium effects,
respectively. 
\label{kmt10}}
\end{center}
\end{figure}

\begin{figure}
\begin{center}
\epsfig{file=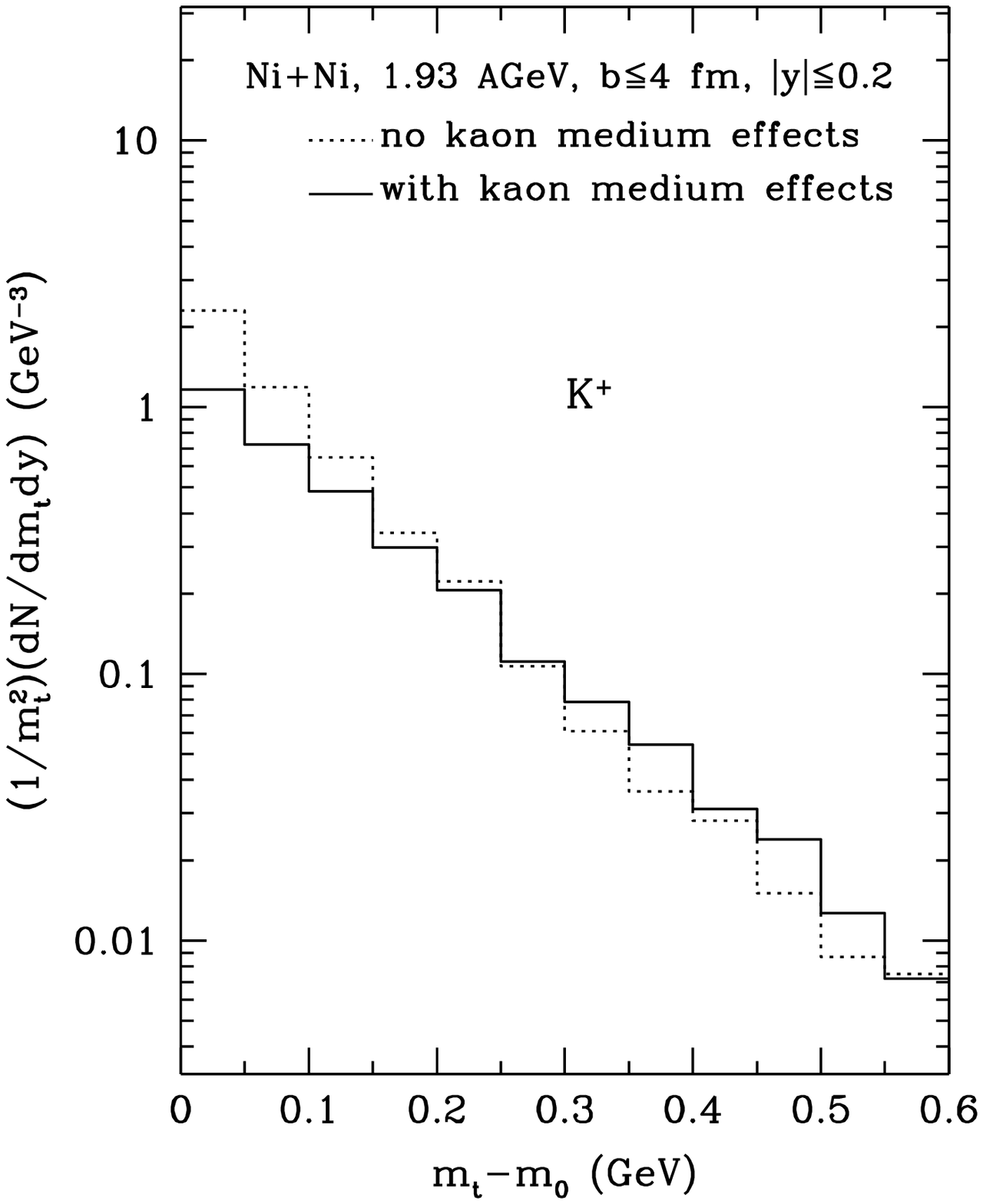,height=7.0in,width=7.0in}
\caption{Same as Fig. \protect\ref{kmt10},
for central Ni+Ni collisions at 1.93 AGeV. 
\label{kmt193}}
\end{center}
\end{figure}

\begin{figure}
\begin{center}
\epsfig{file=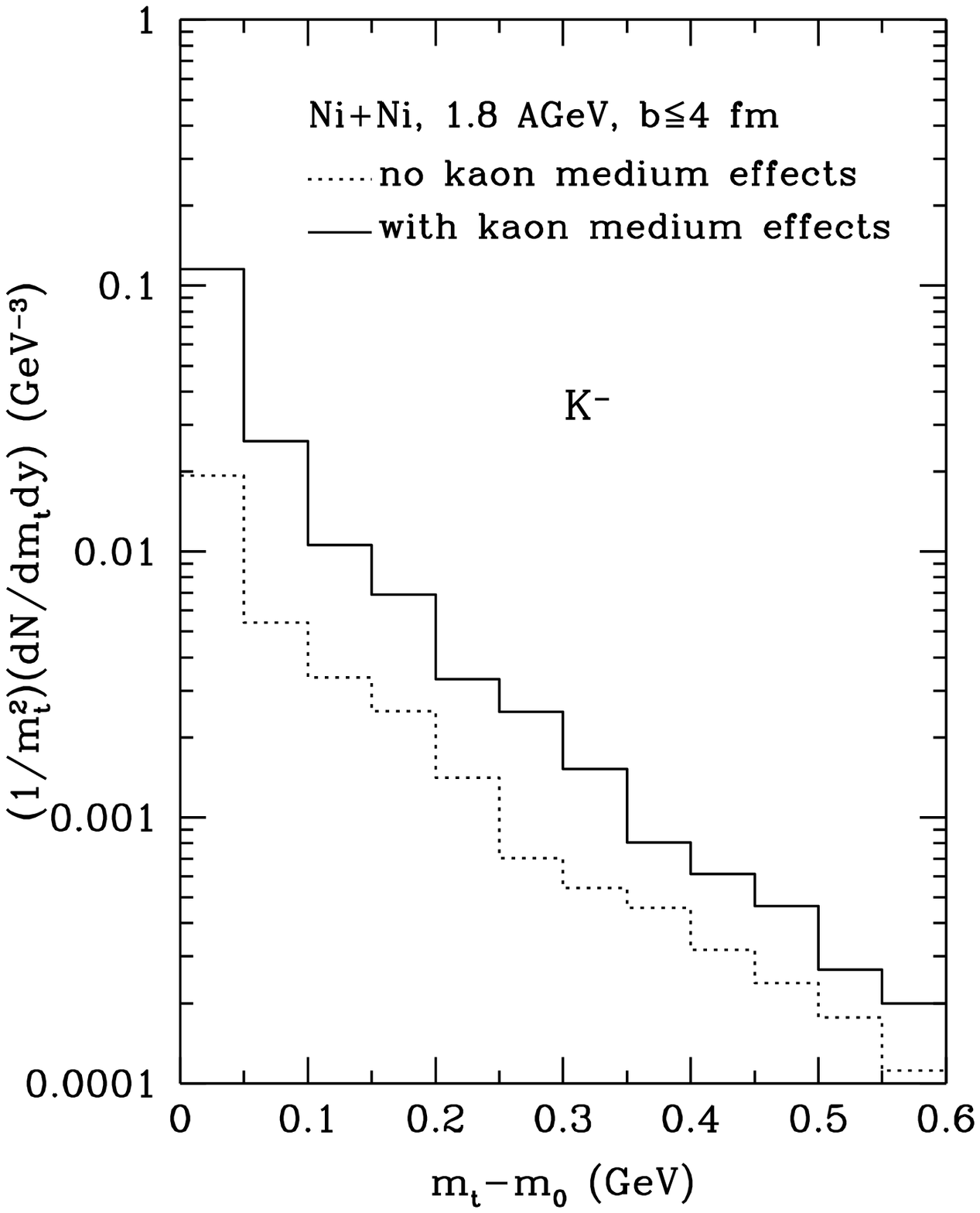,height=7.0in,width=7.0in}
\caption{Same as Fig. \protect\ref{kmt10},
for $K^-$ in central Ni+Ni collisions at 1.8 AGeV. 
\label{akmt18}}
\end{center}
\end{figure}

In Figs. \ref{kmt10} and \ref{kmt193} we show the $K^+$ transverse 
mass spectra at mid-rapidity in central Ni+Ni collisions at
1.0 and 1.93 AGeV, respectively. The impact parameter range
is chosen to be $b\le 4$ fm, comparable to the centrality selection
of the FOPI collaboration's measurement of kaon production in
Ni+Ni collisions. At 1.0 AGeV, the slope parameter of the
$K^+$ spectrum is about 60 MeV when kaon medium effects are
neglected. This increases to about 70 MeV when kaon
medium effects are included. At 1.93 AGeV, the slope parameters
are 95 and 110 MeV, respectively, without and with kaon medium
effects. Similarly, the $K^-$ transverse mass spectra in 
central Ni+Ni collisions at 1.8 AGeV are shown in Fig. \ref{akmt18}.
The difference between the scenarios with and without kaon medium
effects is shown to be most pronounced at low transverse mass.

\begin{figure}
\begin{center}
\epsfig{file=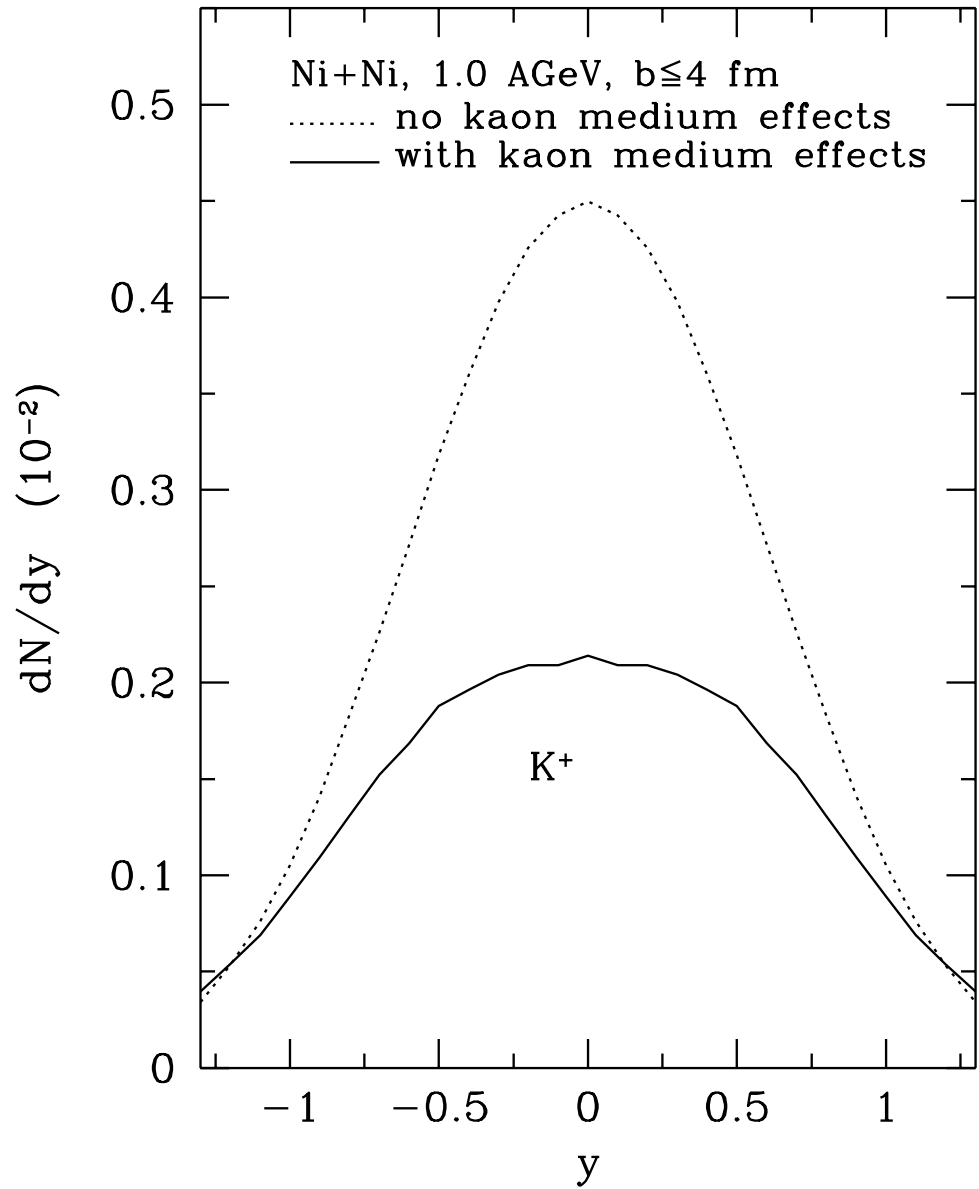,height=7.0in,width=7.0in}
\caption{$K^+$ rapidity distribution in central Ni+Ni 
collisions at 1.0 AGeV. The solid and dotted curves are the
results with and without kaon medium effects,
respectively. 
\label{ky10}}
\end{center}
\end{figure}

\begin{figure}
\begin{center}
\epsfig{file=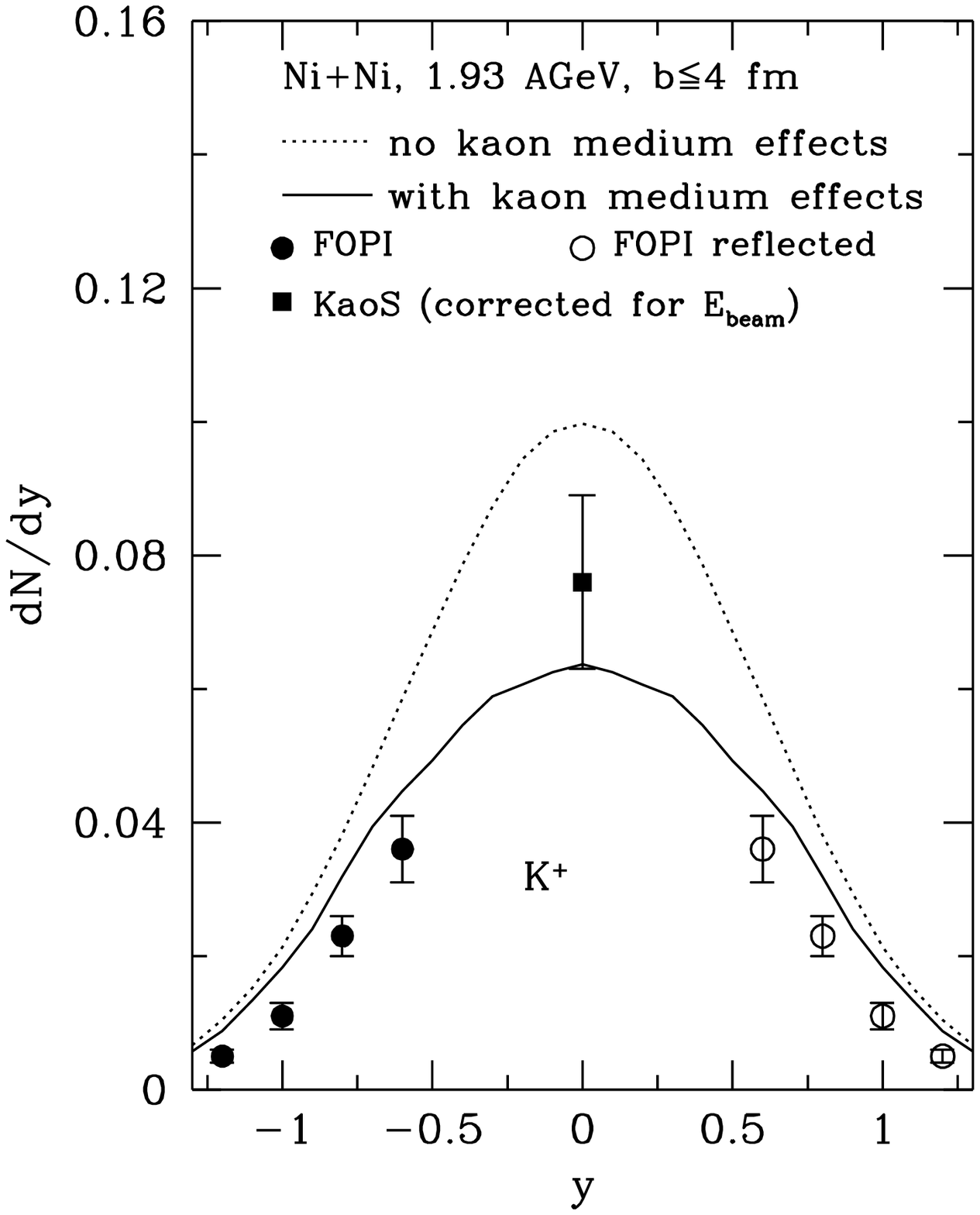,height=7.0in,width=7.0in}
\caption{Same as Fig. \protect\ref{ky10},
for central Ni+Ni collisions at 1.93 AGeV. 
\label{ky193}}
\end{center}
\end{figure}

In Fig. \ref{ky10}, we show the $K^+$ rapidity distribution 
in central Ni+Ni collisions at 1.0 AGeV. The solid and dashed
curves are obtained with and without kaon medium effects. The
most significant difference between the two cases is seen around
the mid-rapidities. The full width at half maximum (FWMH) is
about 1.4 in the case without kaon medium effects. This increases
to about 1.8 when kaon medium effects are included. The results
for the $K^+$ rapidity spectra in Ni+Ni collisions at 1.93 AGeV
are shown in Fig. \ref{ky193}. Here our results are compared with
experimental data from the FOPI and KaoS collaborations \cite{best97}. 
The FOPI measured data are shown in the figure by solid circles,
while the open circles are obtained by reflecting with
respect to mid-rapidity. The KaoS data point, shown 
by solid square, is obtained by multiplying the original KaoS
data measured at 1.8 AGeV by a factor of 1.45 as in Ref. \cite{best97}
to take account the beam energy dependence.
The agreement with the data is better when kaon medium
effects are included. In this case, while the total $K^+$ yield
agrees with the data, our rapidity distribution is somewhat
broader than what the data indicate.

\begin{figure}
\begin{center}
\epsfig{file=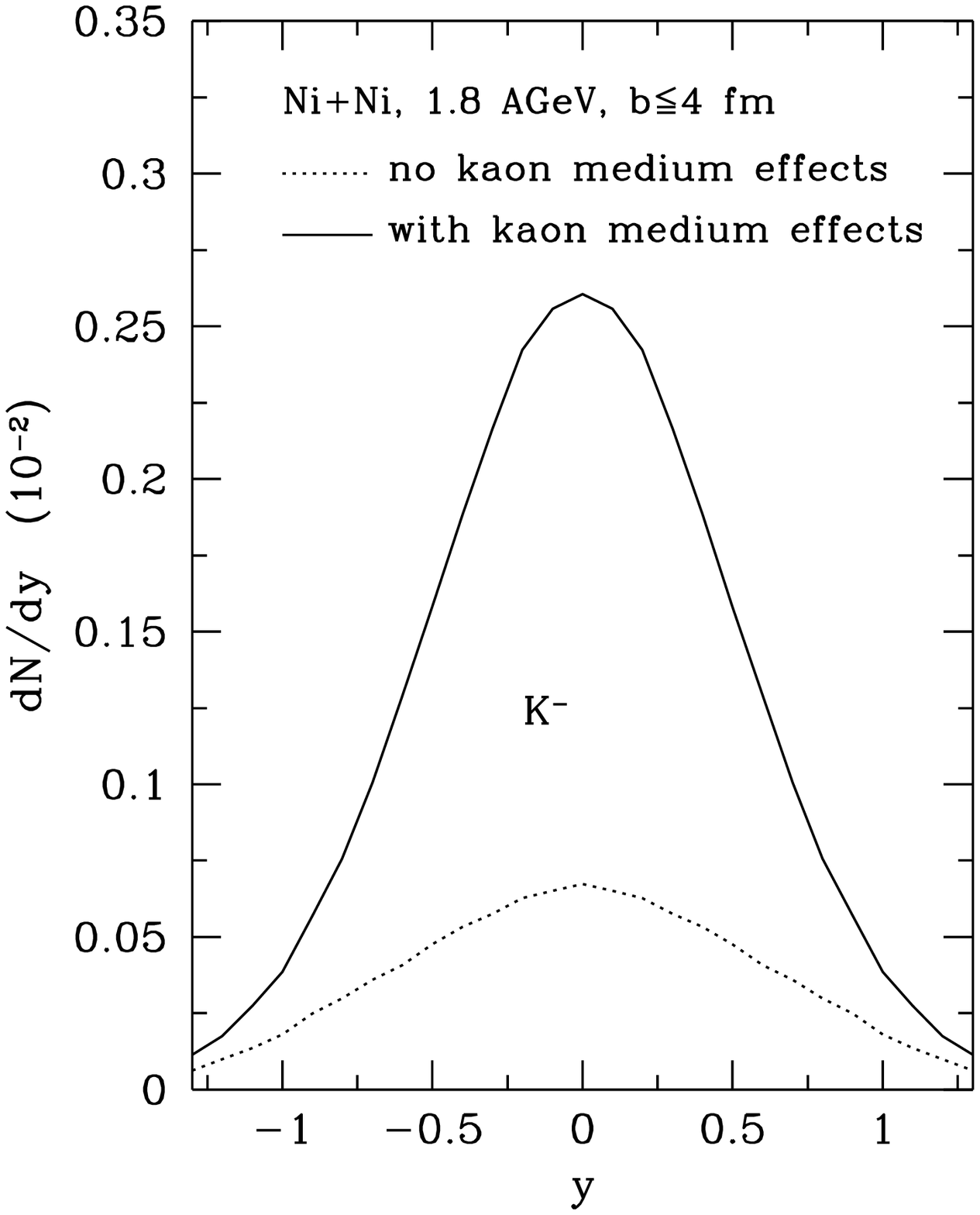,height=7.0in,width=7.0in}
\caption{Same as Fig. \protect\ref{ky10},
for $K^-$ in central Ni+Ni collisions at 1.8 AGeV. 
\label{aky18}}
\end{center}
\end{figure}

\begin{figure}
\begin{center}
\epsfig{file=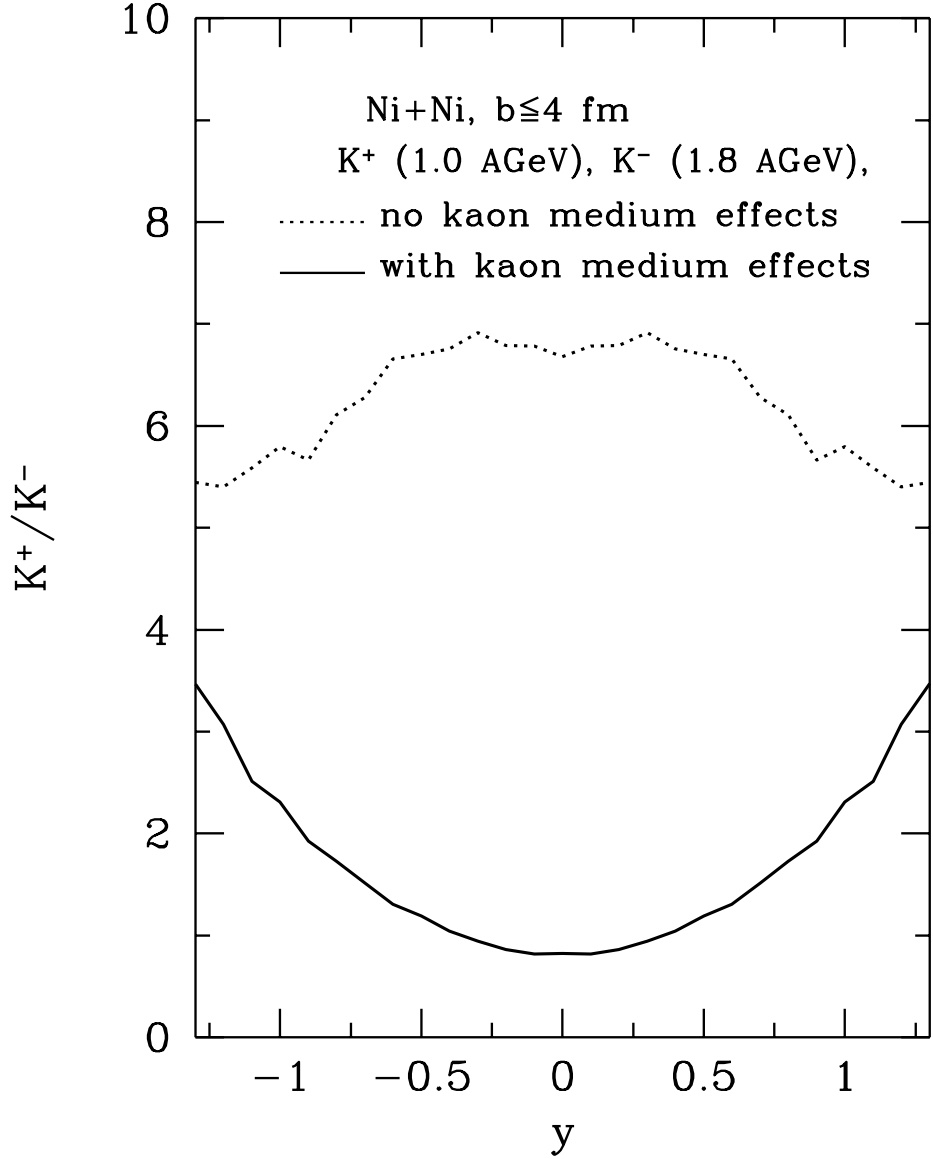,height=7.0in,width=7.0in}
\caption{Rapidity distribution of $K^+/K^-$
in Ni+Ni collisions at 1.0 ($K^+$) and 1.8 ($K^-$)
AGeV. The solid and dotted curves are the results with and without 
kaon medium effects,
respectively. 
\label{kaky}}
\end{center}
\end{figure}

We show in Fig. \ref{aky18} the $K^-$ rapidity distribution in
central Ni+Ni collisions at 1.8 AGeV. The difference between the
scenarios with and without kaon medium effects is most
visible near mid-rapidity. In Fig. \ref{kaky} we show the
ratio of the $K^+/K^-$ as a function of normalized rapity $y$.
As in Fig. \ref{kakr}, the $K^+$ is for beam energy of 1.0 AGeV,
and the $K^-$ is for beam energy of 1.8 AGeV. Without kaon 
medium effects, we find that the ratio {\it decreases} from
about 6.5 around the mid-rapidity to about 5 around the target and
projectile rapidities. When kaon medium effects are included,
the trend is just the opposite: the ratio {\it increases} 
from about 0.9 around mid-rapidity to about 2 at target and
projectile rapidities. Experimental information on the
rapidity dependence of the $K^+/K^-$ ratio should be
very useful in distinguishing the two scenarios. 
 
\subsubsection{$K^+$ excitation functions}

\begin{figure}
\begin{center}
\epsfig{file=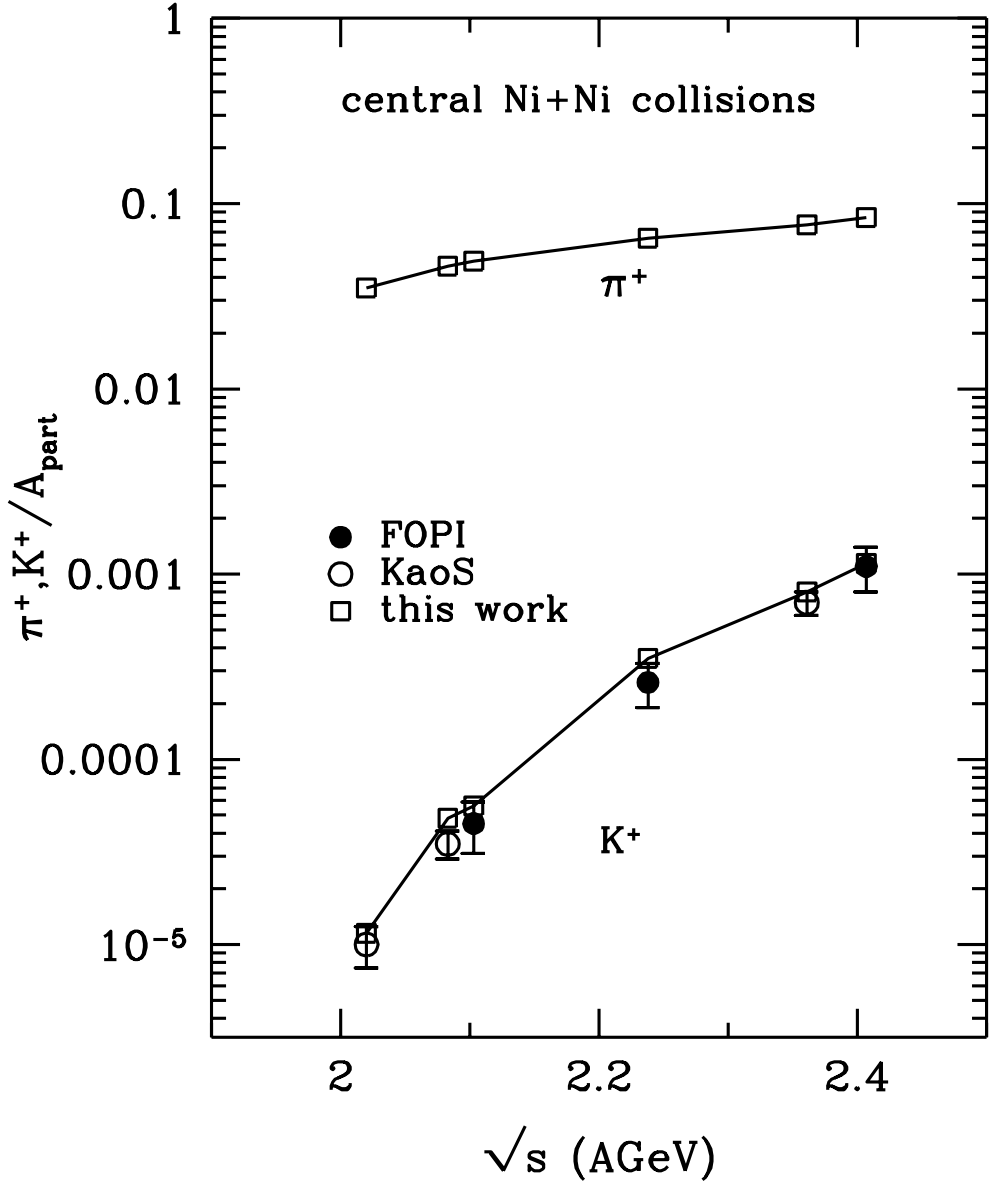,height=7.0in,width=7.0in}
\caption{Excitation function of $K^+/A_{part}$ and
$\pi^+/A_{part}$ in central Ni+Ni collisions.
The experimental data are from Ref. \protect\cite{best97}
\label{excini}}
\end{center}
\end{figure}

Finally, we show in Fig. \ref{excini} the beam energy dependence of 
$K^+$ yield in central Ni+Ni collisions. What is shown in the figure
is the ratio between $K^+$ yield and participant nucleon number
$A_{part}$. Our results, shown by open squares connected by solid
lines, are obtained with kaon medium effects included. The data
from the FOPI and KaoS collaborations are shown by
solid and open circles, respectively \cite{best97,best96}. For comparison,
we also plot in the figure the beam energy dependence of
$\pi^+/A_{part}$ ratio. From 0.8 AGeV to 1.93 AGeV, the
pion yield increases by about a factor of two and half,
indicating an almost linear dependence on $E_{beam}$.
This is in agreement with experimental observations.
The $K^+$ yield, on the other hand, increases by about 
two orders of magnitude. Its beam energy dependence
is about $E_{beam}^{5.3}$. The difference between 
the beam energy dependence of pion and kaon yield
reflects the fact that at these energies kaon 
production is `subthreshold' or 'near-threshold', while
pion production is well above the threshold.

\section{kaon condensation in neutron stars}

As mentioned, the medium modification of kaon properties
affects not only kaon observables in heavy-ion collisions,
it also bears important consequences in the structure and
evolution of compact objects in astrophysics, especially
the maximum mass of neutron stars. There have been many 
studies on this subject 
\cite{muto92,lee94,thor94,knor95,brown92,thor97}.
In this paper, we emphasize the constraints on 
kaon in-medium properties from heavy-ion data as discussed 
in the last section. We shall concentrate on 
cold, neutrino-free neutron stars which are in chemical 
equilibrium under $\beta$ decay processes.
We will consider two scenarios: one excludes kaons and the other 
includes them. In the first case, we need to find the ground
state for a system of nucleons, electrons and muons in
chemical equilibrium, while in the second scenario,
we try to find the ground state of a system of nucleons, 
electrons, muons and kaons in chemical equilibrium.
The ground state energy density and pressure are then used
in the Tolman-Oppenheimer-Volkov (TOV) equation to 
study neutron star properties \cite{thor94}.

In the absence of kaons, the chemical equilibrium conditions
require that the chemical potentials should satisfy
\begin{eqnarray}
\mu = \mu_n-\mu_p = \mu _e = \mu_\mu ,
\end{eqnarray}
and in the presence of kaons, 
\begin{eqnarray}
\mu = \mu_n-\mu_p = \mu _e = \mu_\mu =\mu _K.
\end{eqnarray}
In both cases $\mu$ is the overall charge chemical potential.
The local charge neutrality can be imposed by minimizing
the thermodynamical potential
\begin{eqnarray}
\Omega = \varepsilon _N + \varepsilon _K
+\varepsilon _L - \mu (\rho _p -\rho _K - \rho _e -\rho _\mu ).
\end{eqnarray} 
For the energy density of nucleons we use the results
of Furnstahl, Tang, and Serot, as given in Eq. (\ref{fst}).
The energy density of leptons is given by
\begin{eqnarray}
\varepsilon _L= \varepsilon _e + \eta (\mu -m_\mu )\varepsilon _\mu ,
\end{eqnarray}
where $\eta$ is the Heaviside function
($\eta (x) = 1$ if $x>0$ and $\eta (x)= 0$ if $x<0$). 
The electron and muon energy densities are, respectively, 
\begin{eqnarray}
\varepsilon _e= {\mu ^4\over 4\pi^2},
\end{eqnarray} 
and 
\begin{eqnarray}
\varepsilon _\mu= {m_\mu ^4\over 8\pi^2}\left[t(1+2t^2)\sqrt {1+t^2}
-{\rm ln}(t+\sqrt {1+t^2})\right],
\end{eqnarray} 
where $t={K_{F_\mu}\over m_\mu}$, and $K_{F_\mu}^2 = \mu ^2- m_\mu^2$.
The energy density of kaons, according to Ref. \cite{thor94}, is
\begin{eqnarray}
\varepsilon _K & = & -f^2{\mu^2\over 2}{\rm sin}^2\theta 
+2m^2_Kf^2{\rm sin}^2{\theta \over 2} + \mu x \rho \nonumber\\
 & - & \mu (1+x)\rho {\rm sin}^2{\theta \over 2} -
2f^2 a_{\bar K} \rho {\rm sin}^2{\theta \over 2},
\end{eqnarray}
where $f$ is the pion decay constant, $x=\rho_p /\rho$ is proton fraction,
and the chiral angle $\theta$ is defined in terms of kaon amplitude
$v_K$, 
\begin{eqnarray}
\theta = {\sqrt 2 v_K\over f}.
\end{eqnarray}

The energy density $\varepsilon$ of the ground state of the 
system is obtained by extremizing $\Omega$ with respect to 
$x$, $\mu$, and $\theta$.
The pressure of the system is then obtained from the energy
density 
\begin{eqnarray}
P= \rho \left({\partial (\varepsilon /\rho )\over \partial \rho}
\right).
\end{eqnarray}
The energy density $\varepsilon $ and pressure $P$
are then used in the TOV equation to obtain the properties
of neutron stars,
\begin{eqnarray}
{dM\over dr} & = & 4\pi r^2 \varepsilon \nonumber\\
{dP\over dr} & = & - {G[\varepsilon +P][1+4\pi r^3P]\over r(r-2GM)}.
\end{eqnarray}

\begin{figure}
\begin{center}
\epsfig{file=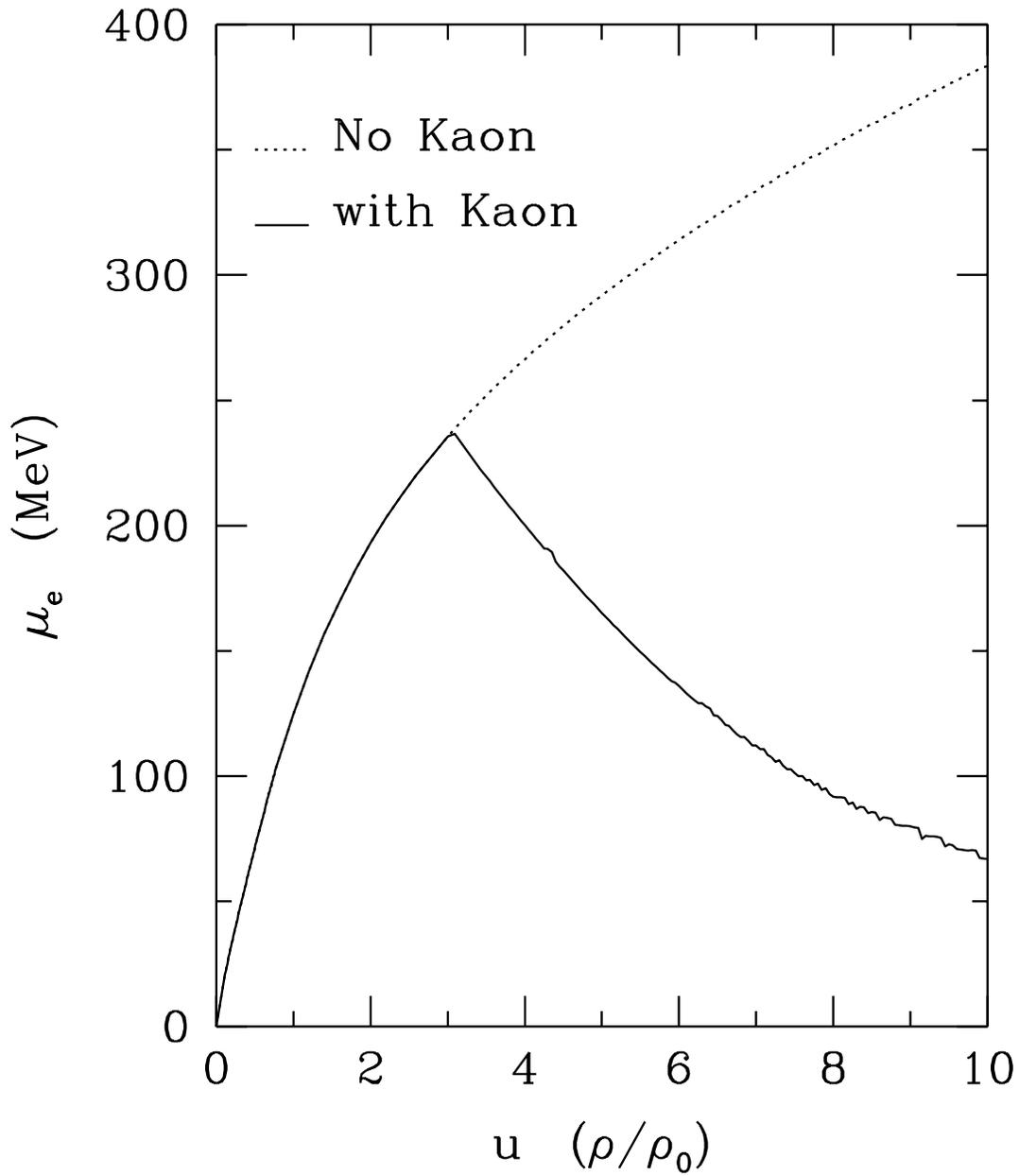,height=7.0in,width=7.0in}
\caption{Electron chemical potential as a function of
nucleon density. The solid and dotted curves
are obtained with and without kaons.
\label{mu}}
\end{center}
\end{figure}

In Fig. \ref{mu} we show the electron chemical potential as
a function of nucleon density. In the case without kaons, the
electron chemical potential continues to increase with 
nucleon density. When kaons are included, the electron chemical
potential first increases up to about three times normal nuclear
matter density $\rho_0$. Afterwards it starts to decrease because
of the onset of the kaon condensation. Thus based on the 
kaon in-medium properties as constrained by heavy-ion data and
nuclear equation of state from the model of Furnstahl, Tang
and Serot, we find that the critical density for kaon condensation
to be about 3$\rho_0$.

\begin{figure}
\begin{center}
\epsfig{file=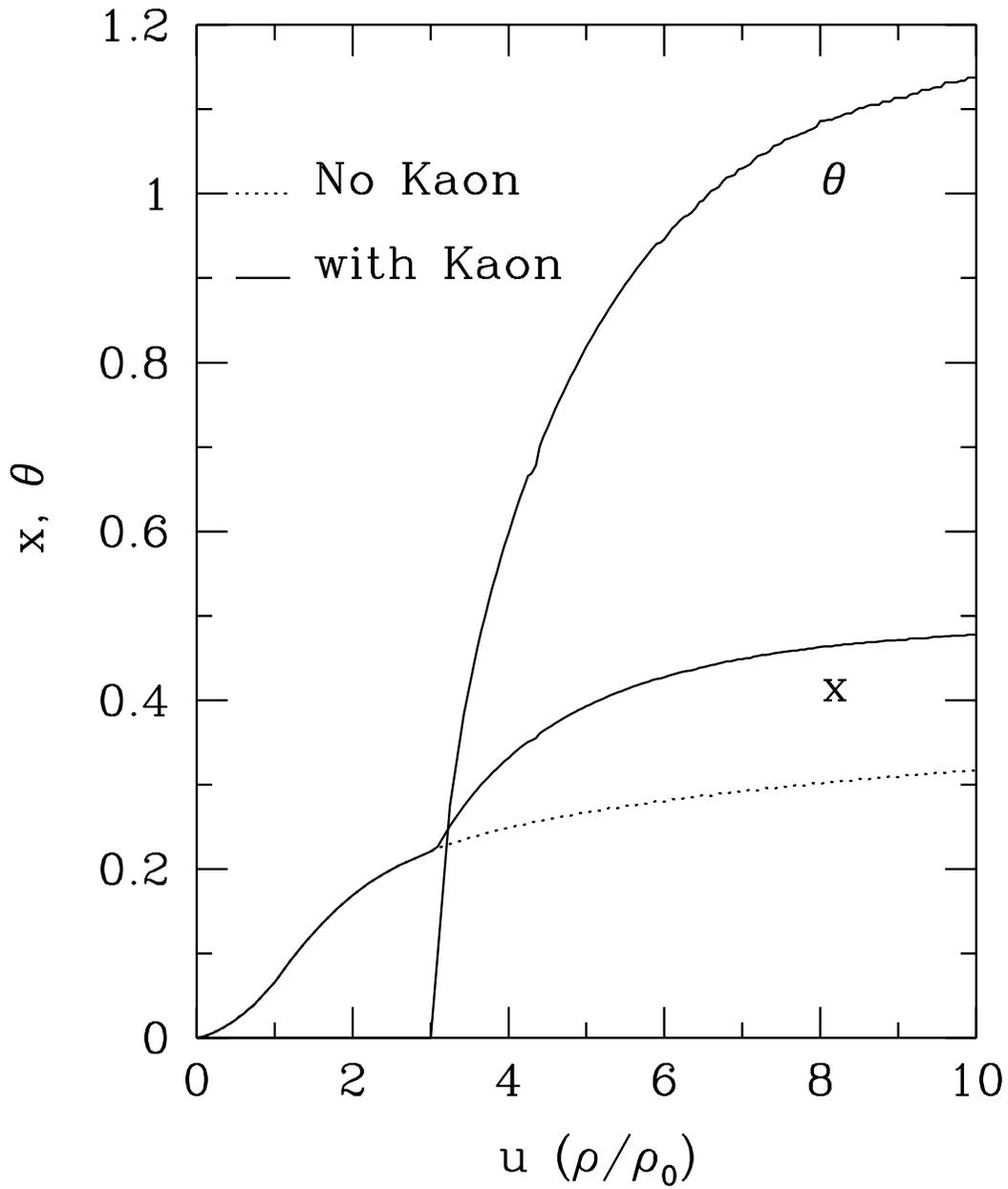,height=7.0in,width=7.0in}
\caption{Proton fraction and chiral rotation angle as a function of
nucleon density. The solid and dotted curves
are obtained with and without kaons.
\label{fraction}}
\end{center}
\end{figure}

In Fig. \ref{fraction}, we show the proton fraction $x$ and
rotation angle $\theta$ as a function of nucleon density.
The solid and dotted curves are obtained with and without
kaons. The proton fraction increases rapidly to become
almost as large as the neutron fraction, after kaon condensation 
sets in. This reduces the asymmetry parameter, and
hence the symmetry energy, of the system. The rotation
angle, or the condensate amplitude, rises rapidly near
the threshold, and then slowly increases to about 60$^0$.

\begin{figure}
\begin{center}
\epsfig{file=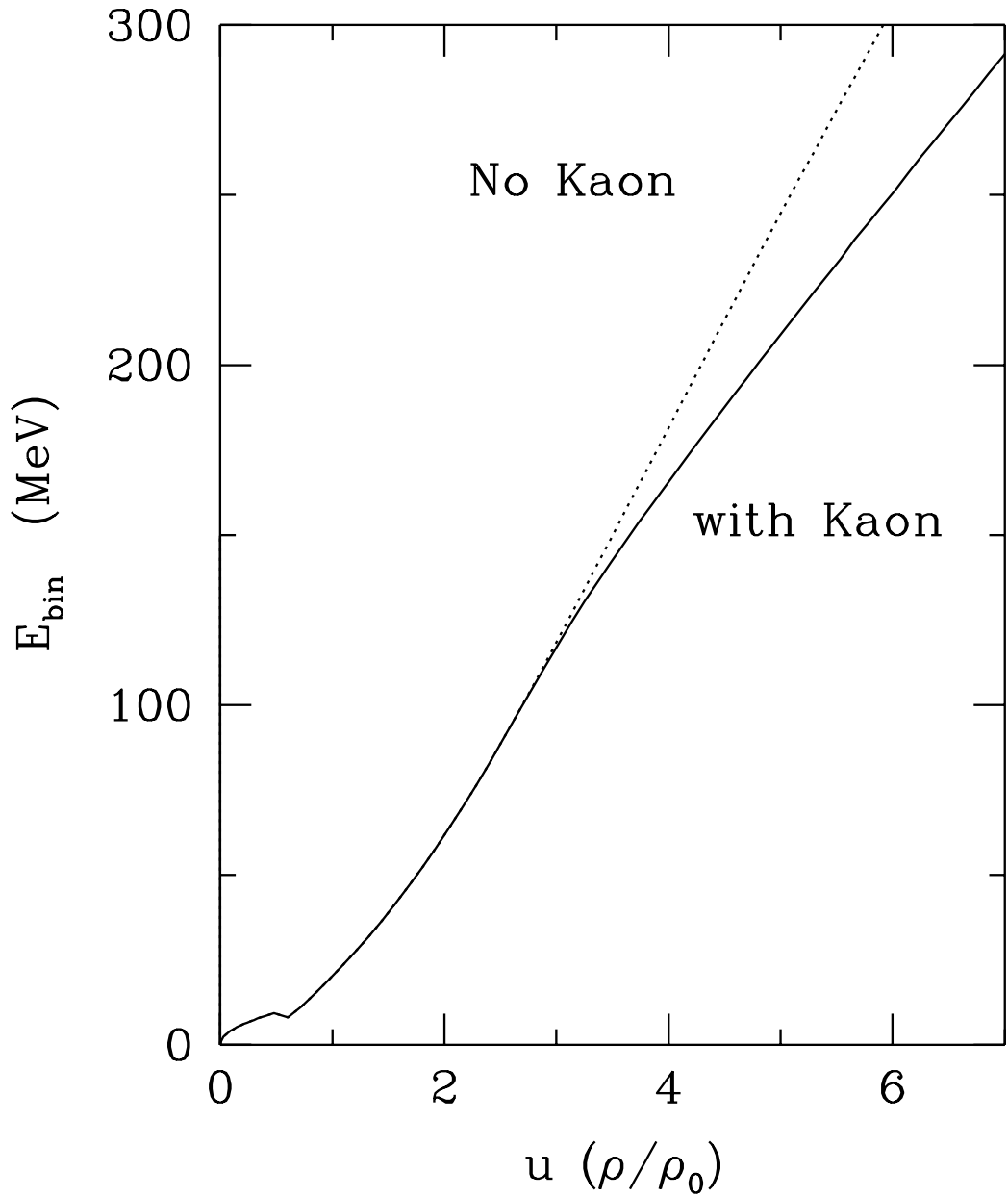,height=7.0in,width=7.0in}
\caption{Energy density as a function of
nucleon density. The solid and dotted curves
are obtained with and without kaons.
\label{ener}}
\end{center}
\end{figure}

\begin{figure}
\begin{center}
\epsfig{file=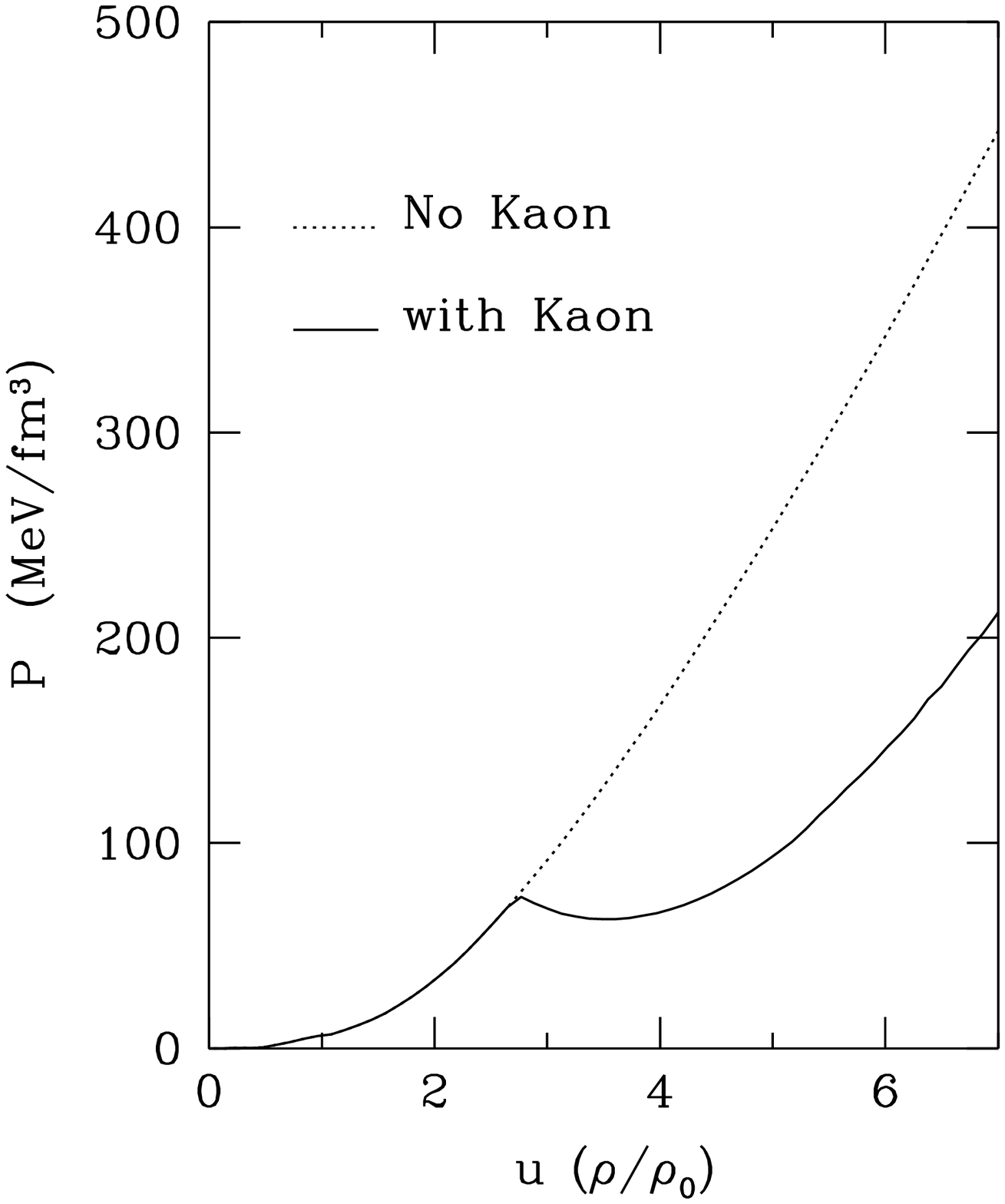,height=7.0in,width=7.0in}
\caption{Pressure as a function of
nucleon density. The solid and dotted curves
are obtained with and without kaons.
\label{pres}}
\end{center}
\end{figure}

\begin{figure}
\begin{center}
\epsfig{file=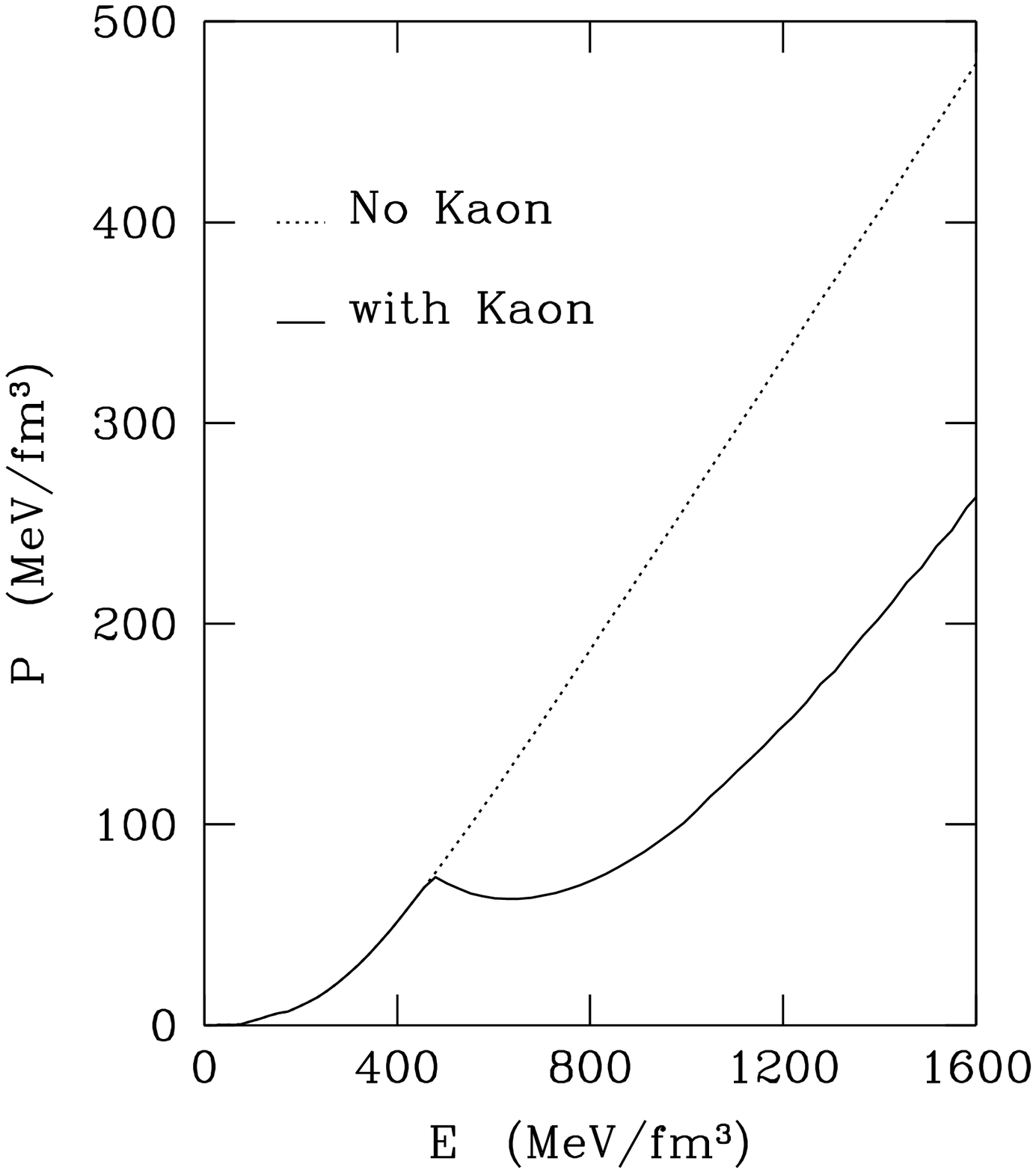,height=7.0in,width=7.0in}
\caption{Pressure as a function of
energy density. The solid and dotted curves
are obtained with and without kaons.
\label{enerpres}}
\end{center}
\end{figure}

In Figs. \ref{ener} and \ref{pres} we show the energy density and
pressure of system as a function of nucleon density, for both
cases with and with kaons. For densities lower than 0.08 fm$^{-3}$
we use the empirical equation of state as in Ref. \cite{thor94}.
The match between this and the equation of state of Furnstahl 
{\it et al.} produces the small dip near 0.5$\rho_0$. 
Both the energy density and pressure
are seen to be lowered once kaon condensation sets in. 
The effect is particularly strong in the pressure. Since
the properties of neutron stars also depend sensitively on the
ratio $P/\varepsilon$, we show in Fig. \ref{enerpres} the
pressure as a function of energy density. The softening of the
equation of state by kaon condensation is appreciable.

\begin{figure}
\begin{center}
\epsfig{file=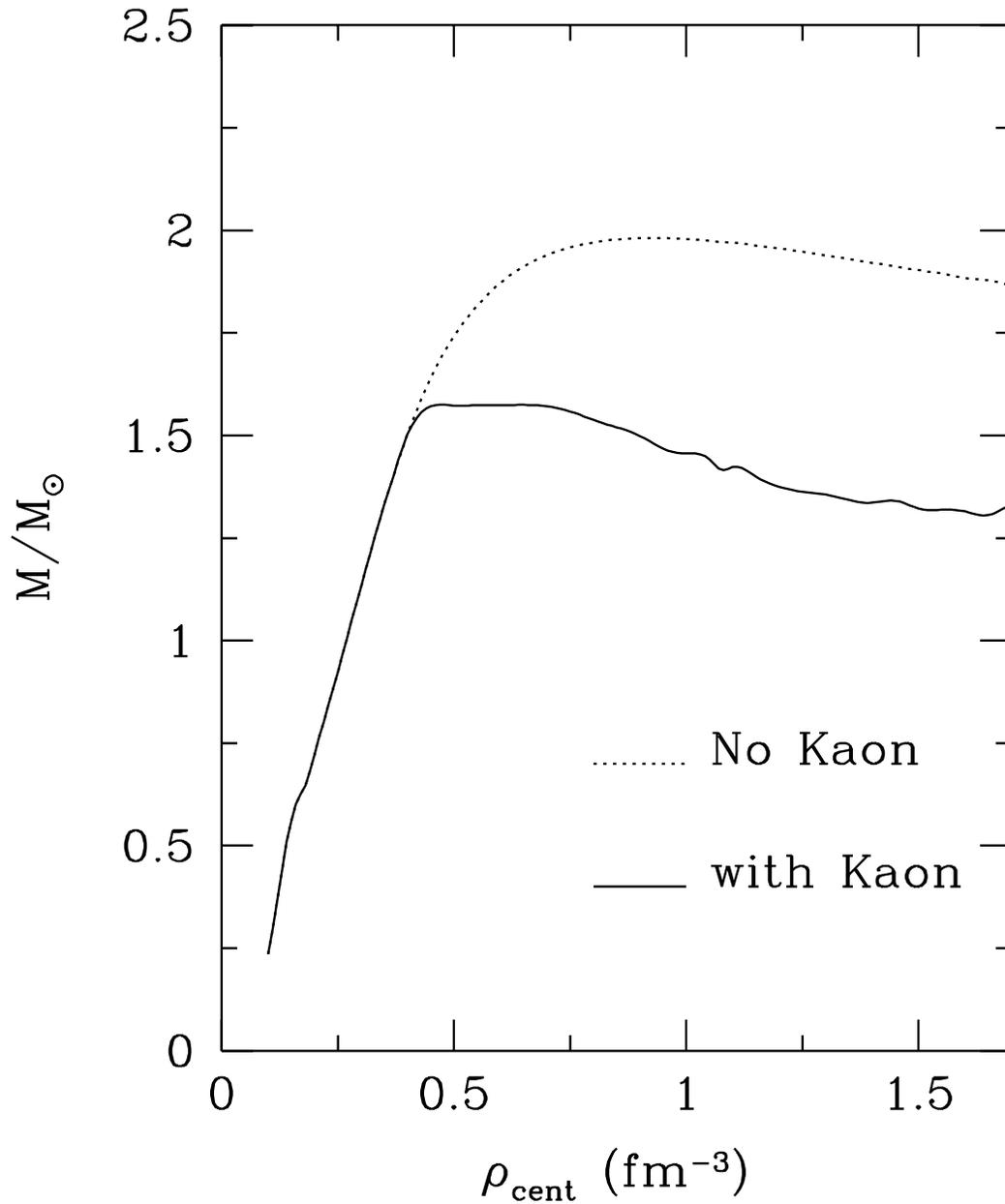,height=7.0in,width=7.0in}
\caption{neutron star mass as a function of
central density. The solid and dotted curves
are obtained with and without kaons.
\label{mass}}
\end{center}
\end{figure}

\begin{figure}
\begin{center}
\epsfig{file=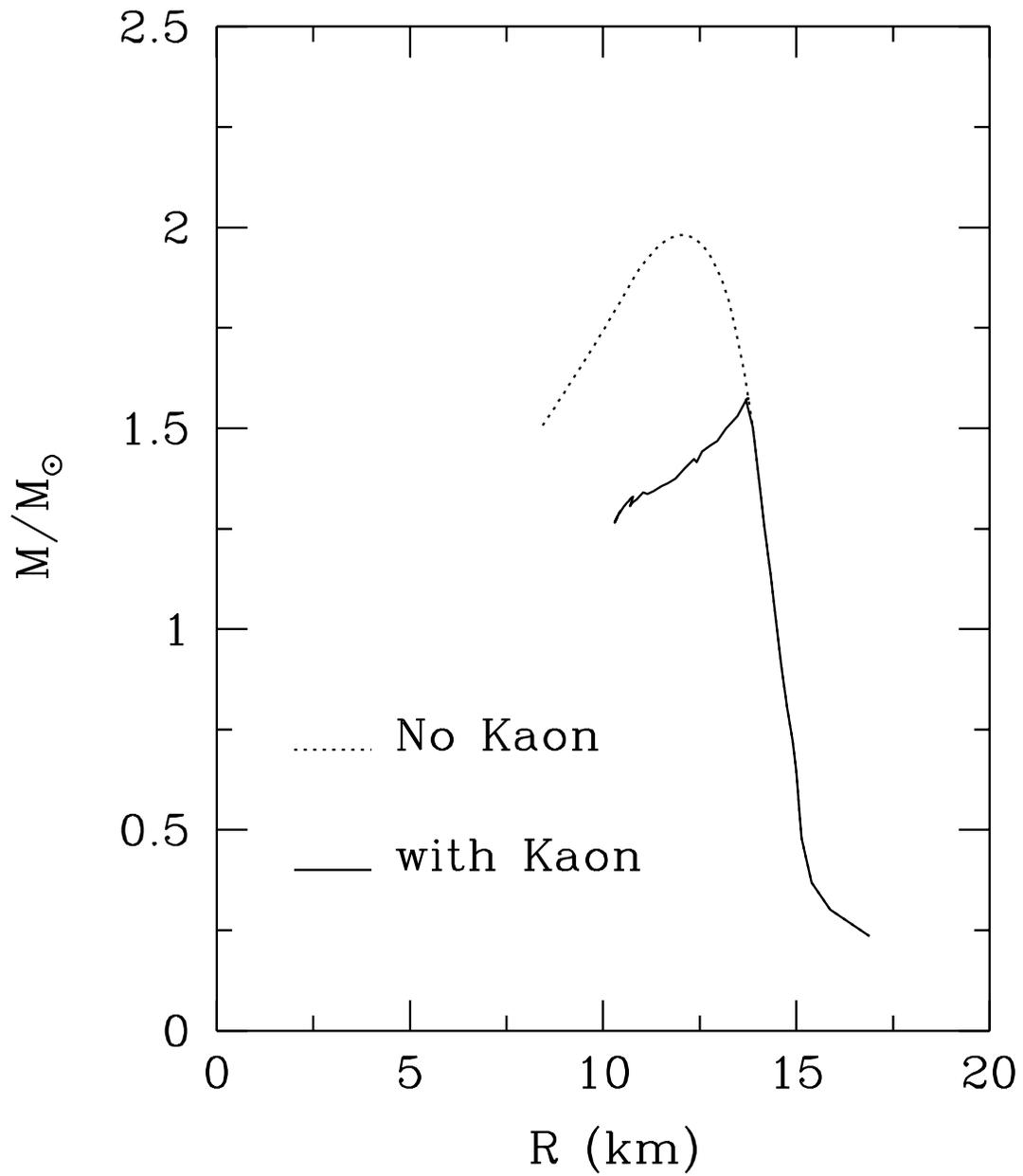,height=7.0in,width=7.0in}
\caption{neutron star mass as a function of
radius. The solid and dotted curves
are obtained with and without kaons.
\label{radius}}
\end{center}
\end{figure}

The results for neutron star mass as a function of central
density $\rho _{cent} $ are shown in Fig. \ref{mass}. It is seen
that, without kaons, the maximum neutron star mass in 
this model is about 2$M_\odot$. Similar 
conclusions, with neutron star mass in the range of 
2.1-2.3$M_\odot$, have been obtained in Ref. 
\cite{eng96,lee97} based on the nuclear equation of state
from the Dirac-Brueckner-Hartree-Fock approach. 
When kaons are included, the maximum mass of neutron 
stars is about 1.5$M_\odot$. The dependence of 
neutron star mass on radius is shown in Fig. \ref{radius}.
It is seen that with kaon condensation, the radius
of maximum mass neutron star increases. 

\section{summary and outlook}

In summary, we studied $K^+$ and $K^-$ production 
in Ni+Ni collisions at 1-2 AGeV, based on the relativistic 
transport model including the strangeness degrees of freedom. 
We found that the recent experimental data from the KaoS
collaboration are consistent with the predictions of the chiral
perturbation theory that the $K^+$ feels a weak repulsive potential
and $K^-$ feels a strong attractive potential in nuclear 
medium. Using the kaon in-medium properties constrained
by the heavy-ion data, we have studied neutron star properties
with and without kaon condensation. The maximum mass of 
neutron stars is found to be about 2.0$M_\odot$ based 
on conventional nuclear equations of state 
obtained from the effective Lagrangian of Furnstahl {\it et al.}. 
This can be reduced to about 1.5$M_\odot$,
once kaon condensation is introduced.
We have emphasized the growing interdependence
between hadron physics, relativistic heavy-ion physics
and the physics of compact stars in astrophysics.

The recent experimental data from the COSY-11 collaboration
\cite{cosy,cosy11} on near threshold kaon and antikaon 
production cross section in proton-proton collisions 
provide much needed information for the parameterizations
of elementary cross sections. It is realized that
both the Randrup-Ko parameterization for $K^+$
production cross section and Sch\"urmann-Zwermann
parameterization for the $K^-$ production cross section
overestimate the data near threshold by a large factor.
Use of the new parameterizations that incorporate 
the latest COSY-11 data reduces the contributions
to $K^+$ and $K^-$ yields in heavy-ion collisions
from baryon-baryon interactions \cite{cass97a,cass97b}.
As a result, contributions from pion-baryon and pion-hyperon
interactions, that were previously found to be less
important than those from baryon-baryon interaction,
turn out to be important, and in some cases even become
dominant.

Our knowledge on elementary cross sections has been
improved over the last a few years, mainly because of the
new experimental data near the threshold, and also because
of the development in the theoretical study of the
elementary cross sections. Nevertheless, to extract kaon
medium effects from their yields in heavy-ion collisions
is still a delicate task. On the other hand, the ratio,
$K^+/K^-$, and especially the shapes of its transverse 
mass spectra and rapidity distribution, is less 
sensitive to the elementary cross sections, and
should be able to provide more definite information
about kaon in-medium properties. Also, the collective 
flow signals of $K^+$ and $K^-$ \cite{likoli95,ritman,liko96}
in heavy-ion collisions are very useful probes of
kaon medium effects.

\vskip 1cm
We are grateful to C.M. Ko, T.T.S. Kuo, M. Prakash, and M.
Rho for useful discussions. We thank N. Herrmann and P. 
Senger for sending us their data files and for useful
communications. We also thank W. Weise for a careful
reading of the manuscript and for critical comments.
This work is supported in part by the 
Department of Energy under Grant No. DE-FG02-88ER40388.

\newpage

\newpage

\newpage

\newpage

\newpage

\newpage

\newpage

\newpage

\newpage

\newpage

\newpage

\newpage

\newpage

\newpage

\newpage

\newpage

\newpage

\newpage

\newpage

\newpage

\newpage

\newpage

\newpage

\newpage

\newpage

\newpage

\newpage

\newpage

\newpage

\newpage

\newpage

\newpage

\newpage

\newpage

\newpage

\newpage

\newpage

\newpage

\newpage

\newpage

\newpage

\newpage

\newpage

\newpage

\newpage

\newpage

\newpage

\newpage

\newpage

\newpage

\newpage

\newpage

\newpage

\newpage

\end{document}